\documentclass[a4paper,fleqn,usenatbib]{mnras}

\usepackage{newtxtext,newtxmath}
\usepackage[T1]{fontenc}
\usepackage{ae,aecompl}

\usepackage{graphicx}	% Including figure files
\usepackage{amsmath}	% Advanced maths commands
\usepackage{amssymb}	% Extra maths symbols
\usepackage{subcaption}
\usepackage{booktabs}
\usepackage[export]{adjustbox}
\title[Starburst-AGN mixing: NGC 1365, NGC 1068]{Starburst-AGN mixing: TYPHOON observations of NGC 1365, NGC 1068, and the effect of spatial resolution on the AGN fraction}

\author[J. J. D'Agostino et al.]{
Joshua J. D'Agostino,$^{1,2}$+\thanks{E-mail: joshua.dagostino@anu.edu.au}
Henry Poetrodjojo,$^{1,2}$
I-Ting Ho,$^{4}$
Brent Groves,$^{1,2}$
\newauthor
Lisa Kewley,$^{1,2}$
Barry F. Madore,$^{3}$
Jeff Rich,$^{3}$
Mark Seibert$^{3}$
\\
$^{1}$Research School of Astronomy and Astrophysics, The Australian National University, Cotter Road, Weston, ACT 2611, Australia \\
$^{2}$ARC Centre of Excellence for All Sky Astrophysics in 3 Dimensions (ASTRO 3D)\\
$^{3}$Observatories of the Carnegie Institution of Washington, 813 Santa Barbara St, Pasadena, CA 91101, USA\\
$^{4}$Max Planck Institute for Astronomy, K\"{o}nigstuhl 17, 69117 Heidelberg, Germany
}

\date{Accepted XXX. Received YYY; in original form ZZZ}

\pubyear{2018}

\begin{document}
\label{firstpage}
\pagerange{\pageref{firstpage}--\pageref{lastpage}}
\maketitle

\begin{abstract}
We demonstrate a robust method of resolving the star-formation and AGN contributions to emission lines using two very well known AGN systems: NGC 1365, and NGC 1068, using the high spatial resolution data from the TYPHOON/PrISM survey. We expand the previous method of calculating the AGN fraction by using theoretical-based model grids rather than empirical points. The high spatial resolution of the TYPHOON/PrISM observations show evidence of both star formation and AGN activity occurring in the nuclei of the two galaxies. We rebin the data to the lower resolutions, typically found in other integral field spectroscopy surveys such as SAMI, MaNGA, and CALIFA. The results show that when rebinned from the native resolution of TYPHOON ($< 200$ pc/pixel) to 1 kpc/pixel, the effects include a ${\sim} 3$ kpc increase in the radius of measured AGN activity, and a factor of $2 - 7$ increase in the detection of low surface brightness features such as shocks. All of this information is critical, because information on certain physical processes may be lost at varying resolutions. We make recommendations for analysing data at current IFU survey resolutions.
\end{abstract}


%%%%%%%%%%%%%%%%%%%%%%%%%%%%%%%%%%%%%%%%%%%%%%%%%%

%%%%%%%%%%%%%%%%% BODY OF PAPER %%%%%%%%%%%%%%%%%%

\section{Introduction}
% Starburst-AGN mixing -- mystery of link between two processes (Rebecca's paper has good intro)
% Heavily reference Rebecca's paper for original technique due to 'invention' of IFU data
% Mention basics of paper -- TYPHOON, spatial resolution

%Star formation and emission from active galactic nuclei (AGN) are two processes which are common in galaxies across the universe. However, despite the prevalence of the two processes co-existing in the same galaxy and causing a mechanism known as `starburst-AGN mixing', the exact nature of the relationship between the two processes is largely still a mystery.

Much work has been undertaken to establish the relationship between the accretion activity of supermassive black holes (SBH) and the evolution of their host galaxies \citep{CHR1999,HK2000,Fernandez2001,Kauffmann2003,Granato2004,LaMassa2012}.
We now know that the mass of the SBH correlates with other properties of the galaxy, such as the velocity dispersion \citep[$M$-$\sigma$ relation; e.g.][]{FM2000,Gebhardt2000,Tremaine2002,Gultekin2009,MM2013}, the stellar mass in the bulge \citep[$M_{BH}$-$M_{*}$ relation; e.g.][]{Magorrian1998,MH2003,Bennert2011,MM2013}, and the luminosity of the bulge \citep[$M_{BH}$-$L$; e.g.][]{MH2003,Gultekin2009,MM2013}. Therefore, one might expect that black hole activity correlates with star formation. Yet, all current theories that link these processes, such as mergers, starburst-driven winds or AGN-driven outflows \citep[e.g.][]{YKS2010,Rafferty2011} have remained unconvincing. No theoretical model has successfully been able to explain the relationship between star formation and AGN \citep[see the review by][]{AH2012}.

Emission lines such as H$\alpha$ are one of the key measures for the growth of the stellar mass and SBH in galaxies across cosmic time. However such lines
can arise from both AGN and star-forming processes, making the determination of the relative rate of growths difficult when both are present. This confusion
presents one of the difficulties in understanding the link between galaxy and SBH growth. Line ratio diagnostic diagrams, such as suggested by \citet{BPT1981,VO1987,Kewley2001,Kewley2006} can be used to determine the dominant power source within a galaxy. The flux from hydrogen recombination lines such as H$\alpha$ and H$\beta$ is proportional to the number of ionising photons emitted by a stellar cluster, which is proportional to the birthrate of massive stars \citep[][and references therein]{Dopita2002a}. Hence, the flux of H$\alpha$ specifically (rather than H$\beta$ which typically has a smaller flux due to the process that governs the two emission lines, as well as H$\beta$ being more prone to extinction and stellar absorption) is typically used as a star formation indicator. The [N \textsc{ii}]$\lambda$6584 and [O \textsc{iii}]$\lambda$5007 lines (hereafter simply [N \textsc{ii}] and [O \textsc{iii}] unless specified) are strong, forbidden lines and are easily measurable. These lines are AGN indicators because the extreme ultraviolet (EUV) radiation field from the accretion disk of the central AGN is harder than the radiation field from star formation \citep{Groves2004,Kewley2006,Kewley2013a}. 

% Here, first sentence: with the advent of IFS
Work on starburst-AGN mixing was first explored by \citet{Kewley2001}, using previous spectroscopy techniques. These previous spectroscopy techniques (e.g. long-slit) obtained a spectrum from a galaxy as a whole. As a result, galaxies were typically classified on the BPT diagram as either star-forming or from AGN, with the exception being when their line ratios placed them in the region of the BPT between the Kewley and Kauffmann demarcation lines \citep[see][respectively]{Kewley2001,Kauffmann2003}. These two lines are theoretical and empirical upper bounds of star formation in galaxies respectively. The region between these two lines is interpreted to imply mixing of both star formation and an additional hard component (from e.g. AGN, shocks, or diffuse ionised gas) in galaxies \citep{Yuan2010}. Through their work with starburst galaxy modelling, \citet{Kewley2001} demonstrated with their SDSS sample that a continuum of ionising sources (a `mixing sequence') exists in galaxies. This notion was taken further by \citet{Kewley2006}, who demonstrated that contributions from various ionising sources could be calculated and extracted from galaxies which show multiple sources of ionisation.

It is anticipated that the study of starburst-AGN mixing and uncovering the relationship between the two processes will become clearer with the advent of integral field spectroscopy (IFS). IFS, unlike previous methods, obtains spatially resolved spectra from individual spaxels across a galaxy, rather than a single spectrum that encompasses the galaxy as a whole. This allows both star formation and AGN processes in a single galaxy to be separated, and hence integral field unit (IFU) data is central to the study of starburst-AGN mixing. Starburst-AGN mixing using IFU data was first investigated by \citet{Davies2014a,Davies2014b}, who calculated the `AGN fraction' for NCG 7130 from the Great Observatory All-Sky LIRG Survey \citep[GOALS;][]{Armus2009}, and four galaxies (NGC 6394, NGC 2410, IC 0540, NGC 6762) from the Calar Alto Legacy Integral Field Area \citep[CALIFA;][]{Sanchez2012,Husemann2013}. They make use of the concept of `star-forming distance' defined in \citet{Kewley2006} to empirically fit each galaxy's mixing sequence from the star-forming to the AGN regions on the BPT diagram. We expand on the work conducted by \citet{Davies2014a,Davies2014b} by calculating the AGN fraction with new theoretical models, rather than the \citet{Davies2014a,Davies2014b} empirical method. 
% Rebecca's work

Using data obtained from the TYPHOON/Progressive Integral Step Method (PrISM) survey, we analyse two galaxies: NCG 1365 and NGC 1068, described in Section~\ref{sec:sample}. TYPHOON/PrISM is a survey of 100 of the nearest and largest spiral galaxies, resulting in an incredibly highly-resolved dataset. The survey and dataset is described further in Section~\ref{sec:obs}. The high resolution of the dataset gives a clean separation of both star formation and AGN processes in each galaxy, allowing very accurate calculations of the contribution to emission lines from both star formation and AGN processes. We calculate the AGN fraction for both NGC 1365 and NGC 1068 using new theoretical models. The models are described in Section~\ref{sec:grids}, and the new AGN fraction calculation is showcased in Section~\ref{sec:agnfrac}. Further, we explore systematically the impact of resolution on determining the true contributions to both processes in Section~\ref{sec:spatres} by rebinning the data to lower spatial resolutions and repeating the analysis. 

% Rigby 2006: The axis ratios of their host galaxies argue that extinction by host galaxies plays a key role in hiding nuclear emission lines

\section{Sample selection}
\label{sec:sample}

\subsection{NGC 1365}
\label{sec:1365data}

NGC 1365 is a large barred spiral galaxy \citep[SB(s)b;][]{dV1991} located at a distance of roughly 18 Mpc in the constellation Fornax. NGC 1365 is roughly four times bigger than the Milky Way, with a semi major axis of ${\sim} 32$ kpc. NGC 1365 has a Seyfert 1.8 nucleus \citep{VCV2006} with a rapidly rotating supermassive black hole of mass $M_{\mathrm{BH}} \approx 2 \times 10^6 M_\odot$ and spin parameter $a > 0.84$ \citep{Reynolds2013,Risaliti2013}.

Large-scale outflows from NGC 1365 have been reported since the middle of the 20$^{\mathrm{th}}$ century \citep[for a detailed discussion on the history of reported outflows, see][and references therein]{Veilleux2003}. \citet{Veilleux2003} confirm previous studies by measuring a biconical outflow south-east and north-west from the nucleus of NGC 1365. The opening angle of the biconical outflow has been suggested to be ${\sim} 100^{\rm{o}}$, with a position angle of $126^{\rm{o}}$ \citep{HL1996}. \citet{Veilleux2003} and \citet{SBH2010} also show high [N \textsc{ii}]/H$\alpha$ and [O \textsc{iii}]/H$\beta$ ratios particularly along the axis of the biconical outflow, up to values of 1.5 and 3 for each of the respective ratios \citep[figures 6d and 7d respectively from][]{Veilleux2003}. The high emission line ratios suggest photoionisation from the central AGN. The biconical outflow structure is roughly aligned with a radio jet-like feature at a position angle of $125^{\rm{o}}$, observed in \citet{SJL1995} and \citet{Morganti1999}.

\subsection{NGC 1068}
\label{sec:1068data}

As the prototypical Seyfert galaxy, NGC 1068 has been the subject of much work in the astronomical community, leading to it being arguably the most-studied active galaxy in the local universe. A Seyfert 2 \citep{OM1993} (R)SA(rs)b \citep{dV1991} galaxy, it is located at a distance of roughly 12.5 Mpc in the constellation Cetus. The galaxy's broad-line region is known to be obscured by an optically-thick torus \citep[e.g.][]{MA1983,AM1985,MGM1991,GB2016}, and hence has largely been hidden from view due to extinction in the past \citep{Rigby2006}. However, recent work such as \citet{Marinucci2016} and \citet{GB2016} has begun to resolve the nucleus and circumnuclear disk of NGC 1068.

NGC 1068 is also observed to have a large-scale biconical outflow structure \citep[e.g.][]{CBT1990,AMG1996,Pogge1988,CK2000}. Material within this bicone has been observed in the [O \textsc{iii}]$\lambda$5007 line to be moving outward from the nucleus at velocities of ${\sim} 3000$km s$^{-1}$ at a position angle of ${\sim} 32^{\rm{o}}$, believed to be radiatively accelerated by the AGN at the galaxy's centre \citep{Pogge1988,Cecil2002,Dopita2002b}. The outflow in [O \textsc{iii}] can be seen in Figure~\ref{fig:nIIHaoIIIHbb} with log([O \textsc{iii}]/H$\beta$) values near unity, extending from the centre of the galaxy to the north-east. Material with high values of log([O \textsc{iii}]/H$\beta$ are also seen to the south-west of the galaxy, partially obscured by the disk of the galaxy. This ejecta is aligned with the radio jet seen from this galaxy, which has a measured position angle of $34^{\rm{o}}$ \citep[e.g.][]{WU1983,WU1987,Pogge1988,Gallimore1996}. The models from \citet{Das2006} (especially their Figure~10) which attempt to model the bicone give an opening angle of ${\sim} 80^{\rm{o}}$ and position angle of ${\sim} 30^{\rm{o}}$. Values of parameters relating to the geometry of the bicone measured by \citet{Das2006} agree closely with those calculated by \citet{CK2000}, with the exception of the value for the bicone position angle for which \citet{CK2000} suggest $15^{\rm{o}}$. \citet{PSP2008} also measure a value for the bicone orientation using emission lines in the infrared which is in closer agreement to that suggested by \citet{CK2000}. \citet{PSP2008} suggest a bicone orientation with a position angle of ${\sim} 10^{\rm{o}}$. The position angle suggested by \citet{PSP2008} is in better agreement with the \textit{HST} [O \textsc{iii}] image results from \citet{Evans1991}. \citet{PSP2008} do however agree with the opening angle of the bicone calculated by \citet{Das2006}, finding an opening angle of ${\sim} 82^{\rm{o}}$. However, as mentioned by \citet{GM2011}, derivation of the actual geometry of the ionisation bicone close to the centre is not straightforward when observations have large spatial scales. Instead, they propose X-ray polarimetry to constrain the geometry of the innermost outflows. 

%Saturation at the centre of the NGC 1068 datacube due to the AGN resulted in several 'NaN' values, impairing analysis of the starburst-AGN fraction at the centre of the galaxy. This was rectified by co-author J. Rich, in a process listed in Section \ref{sec:1068fix}.

\section{Observations and data reduction}
\label{sec:obs}

\subsection{Observations}

TYPHOON is a survey which uses the Progressive Integral Step Method (PrISM), also known as the `step-and-stare' or `stepped-slit' method to construct 3D datacubes of 100 of the closest and largest galaxies in the southern hemisphere. The survey is being undertaken on the 2.5m du Pont telescope at the Las Campanas Observatory in Chile. Whilst not using an IFU \textit{per se}, the observations are made by progressively stepping a long-slit aperture across a galaxy to mimic the effect of an IFU. Thus, the end product is a high spectral resolution datacube. The datacubes produced from the TYPHOON/PrISM survey are highly resolved, with a spatial resolution of ${\sim}$ 4-5 pc at best \citep{SM2012}. Further details about the TYPHOON/PrISM survey will be explained in a forthcoming publication (Seibert et al. in prep.)

NGC 1365 was observed using the Wide Field reimaging CCD (WFCCD) imaging spectrograph on the du Pont telescope. The WFCCD has a field of view of 25' and we construct the 3D data cube using a custom long-slit ($18' \times 1.65''; 0.5\;\rm{arcmin}^2$) which was placed along the north-south direction and progressively scanned across the galaxy through stepping and staring. Each pointing position was integrated for 600 seconds before the slit was moved by one slit width in the east-west direction for the next integration.  This procedure was repeated until the optical disk of NGC 1365 was covered. In total, 223 observations covering an area of approximately $6' \times 18'$ were taken during 15 nights over four observing runs in November 2011, January 2016, February 2016  and  August 2016. NGC 1068 was also observed using the WFCCD imaging spectrograph on the du Pont telescope, with 63 observations taken between the 8$^\mathrm{th}$ and 11$^\mathrm{th}$ of October 2012 inclusive. The observations cover an area of $1.73' \times 18'$, centered on the nucleus and inner star-forming ring. Each exposure was also integrated for 600 seconds. Observations were performed only when the seeing was less than the slit-width of $1.65''$. Spectrophotometric flux standards were observed each night.

\subsection{Data reduction}

\begin{figure*}
\centering
\begin{subfigure}{0.85\textwidth}
\includegraphics[width=\linewidth]{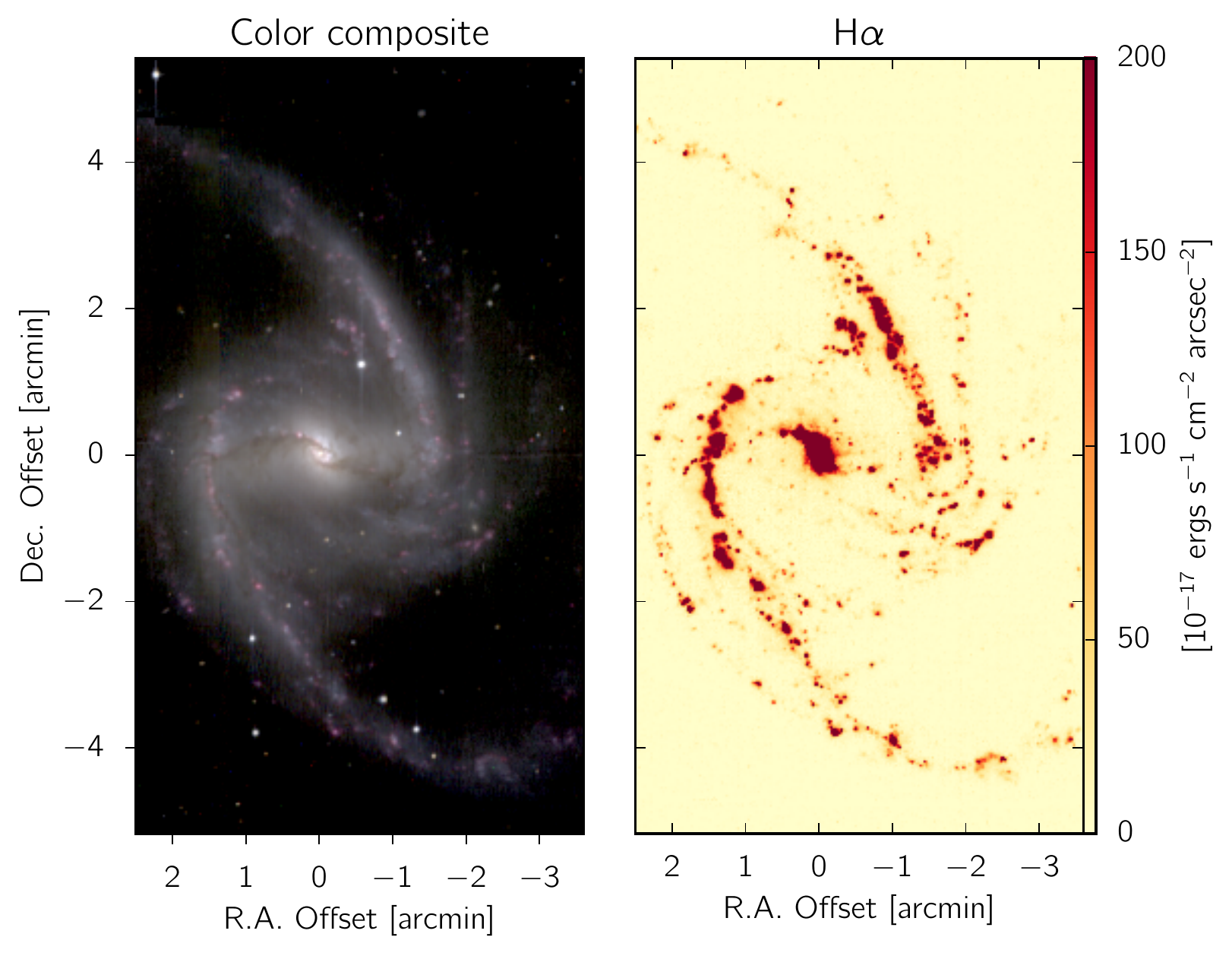}
\caption{NGC 1365}
\label{fig:mapsa}
\end{subfigure}\hspace{0.2\textwidth}
\begin{subfigure}{0.85\textwidth}
\includegraphics[width=\linewidth]{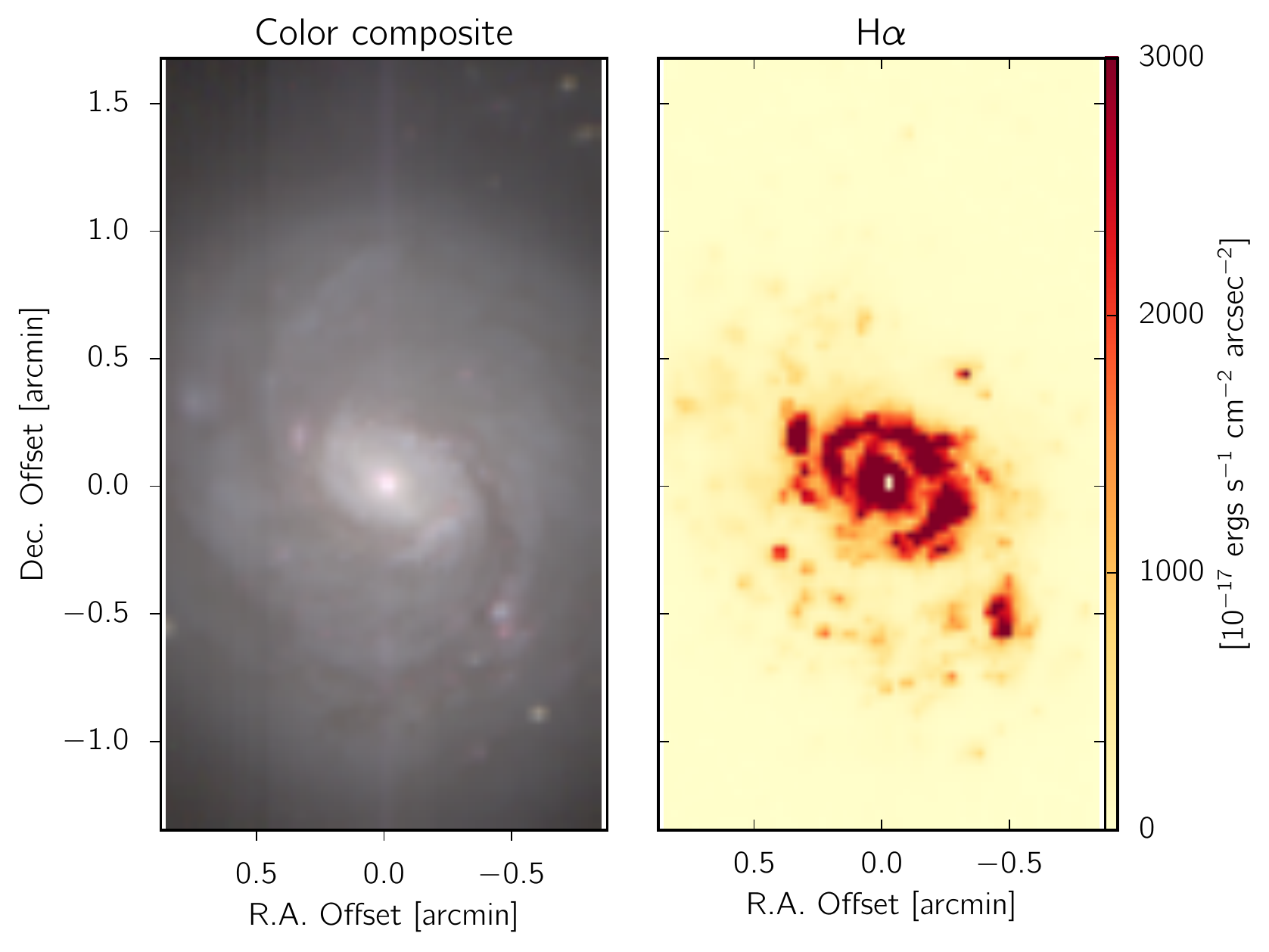}
\caption{NGC 1068}
\label{fig:mapsb}
\end{subfigure}
\caption{Colour composite (\textit{BVR}) and H$\alpha$ maps for NGC 1365 and NGC 1068 from the TYPHOON datacubes. For both images (and all hereafter), north is vertically upwards, and east to the left.}
\label{fig:maps}
\end{figure*}

\begin{figure*}
\centering
\begin{subfigure}{0.41\textwidth}
\includegraphics[width=\linewidth]{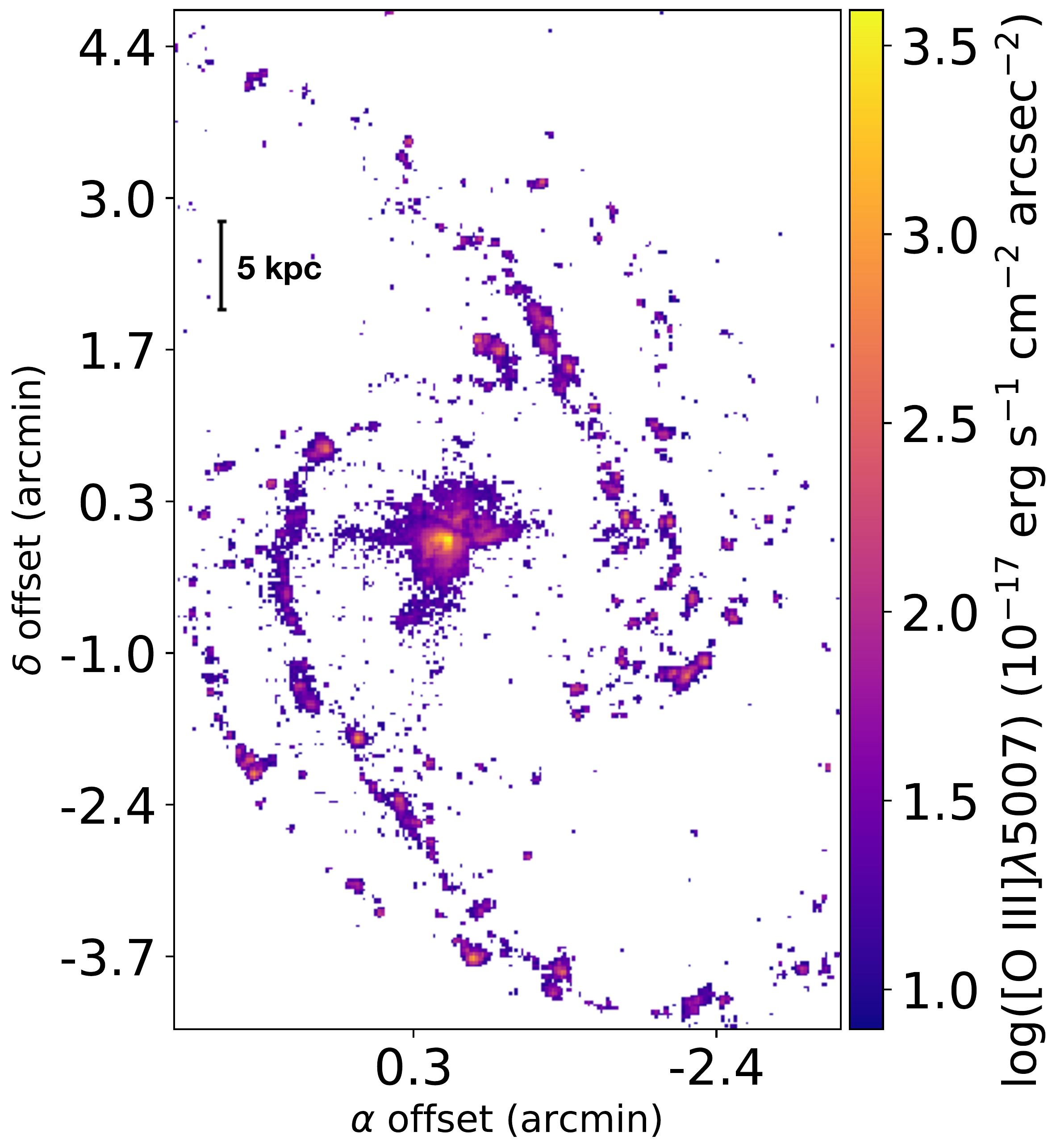}
\caption{NGC 1365}
\label{fig:oIIImapsa}
\end{subfigure}\hspace{0.1\textwidth}
\begin{subfigure}{0.3\textwidth}
\includegraphics[width=\linewidth]{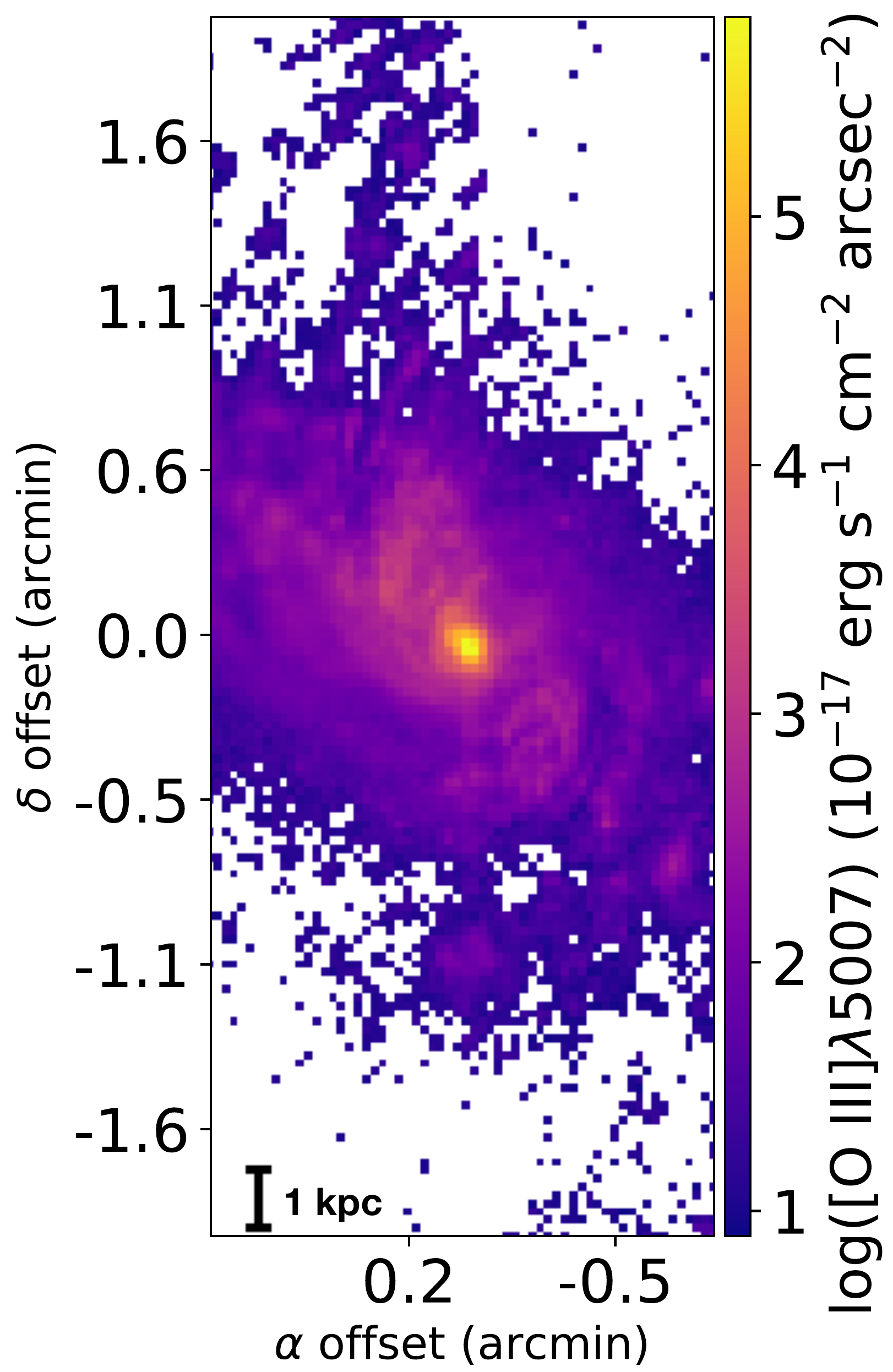}
\caption{NGC 1068}
\label{fig:oIIImapsb}
\end{subfigure}
\caption{[O \textsc{iii}] distribution maps for NGC 1365 and NGC 1068.}
\label{fig:oIIImaps}
\end{figure*}

\begin{figure*}
\centering
\begin{subfigure}{0.85\textwidth}
\includegraphics[width=\linewidth]{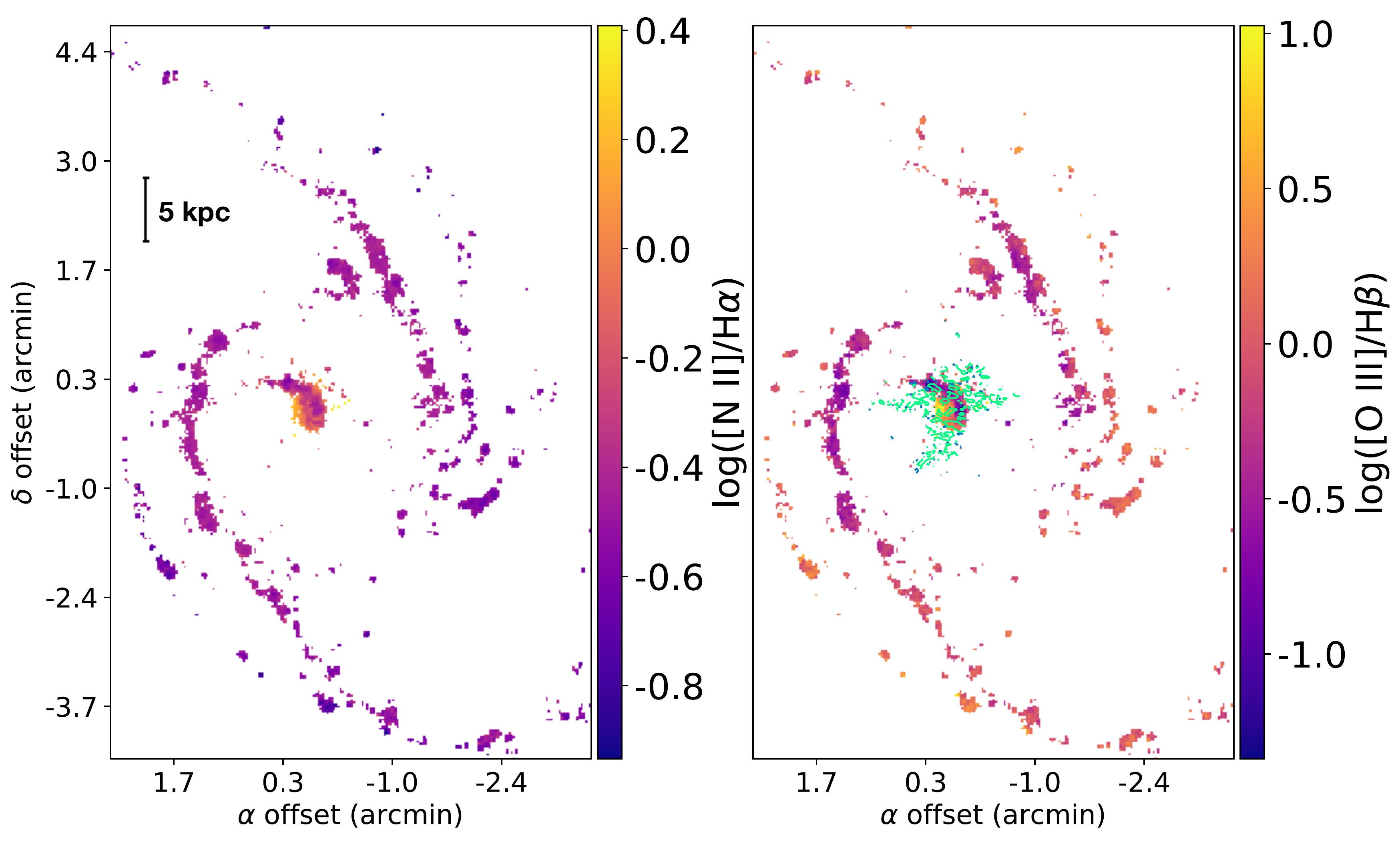}
\caption{NGC 1365}
\label{fig:nIIHaoIIIHba}
\end{subfigure}\hspace{0.2\textwidth}
\begin{subfigure}{0.85\textwidth}
\includegraphics[width=\linewidth]{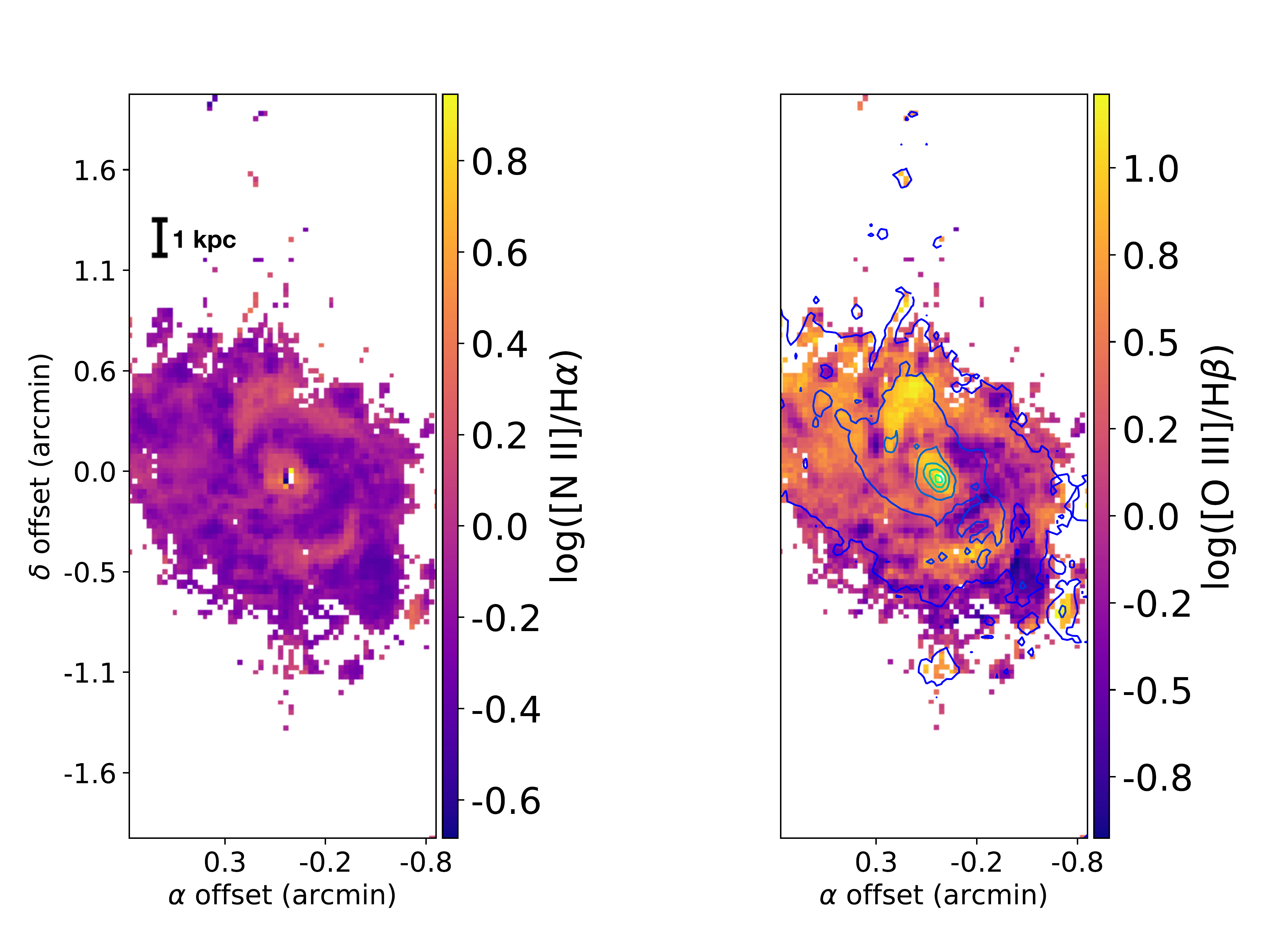}
\caption{NGC 1068}
\label{fig:nIIHaoIIIHbb}
\end{subfigure}
\caption{[N \textsc{ii}]/H$\alpha$ and [O \textsc{iii}]/H$\beta$ distribution maps for NGC 1365 and NGC 1068. Contours of [O \textsc{iii}] with a signal-to-noise ratio greater than 3 are included on the [O \textsc{iii}]/H$\beta$ maps. The [O \textsc{iii}] contours are shown for the inner 7 kpc of NGC 1365, ranging from log([O \textsc{iii}]) = 1.5 (light green) to log([O \textsc{iii}]) = 0.5 (deep blue) $10^{-17}$ erg s$^{-1}$ cm$^{-2}$ arcsec$^{-2}$ in increments of 0.5 dex. For NGC 1068, contours for the whole map are shown, ranging from log([O \textsc{iii}]) = 5.6 (light green) to log([O \textsc{iii}]) = 1.6 (deep blue) $10^{-17}$ erg s$^{-1}$ cm$^{-2}$ arcsec$^{-2}$ in increments of 0.8 dex.}
\label{fig:nIIHaoIIIHb}
\end{figure*}

The data are reduced using standard long-slit data reduction techniques. The wavelength  calibration  has  a  typical root-mean-square value 0.05\AA\; for the entire data set.  Flux calibration  is  accurate  to  2\%  at  the  spaxel  scale  over  the range of 4500 to 7500\AA. The reduced long-slit 2D spectra are tiled together to form the 3D data cube. The final reduced data cube covers a wavelength range of 3650 to 8150\AA, with spectral  and  spatial  samplings  of  1.5\AA\;  and  $1.65''$, respectively. From fitting Gaussians to field stars in the reduced data cube, we estimate that the full-width half maximum of the point spread function is approximately $2''$, which corresponds to approximately 169 pc at the assumed distance of roughly 18 Mpc. The instrumental dispersion is approximately 3.5\AA\; ($\sigma$; correspond to $R \approx 850$ at 7000\AA).

The colour-composite images extracted from the reduced datacube are shown in the left-hand panels of Figure~\ref{fig:maps} for both galaxies NGC 1365 and NGC 1068. The H$\alpha$ flux maps derived from the datacube for both galaxies are shown in the right-hand panels of Figure~\ref{fig:maps}. We use the emission line fitting tool \textsc{LZIFU} \citep{lzifu,lzifu2} to measure emission line fluxes from the raw data, following the method outlined in \citet{Ho2017}. The flux of H$\alpha$ emission peaks at the centre of the galaxy for both galaxies. Hence we use the spaxel containing the highest value of H$\alpha$ flux as the centre of the galaxy, from which we measure our radial offsets. 

%Whilst both galaxies have been known to display clear AGN signatures such as nuclear outflows, the distribution of H$\alpha$ is seen to follow the spiral arms in Figure~\ref{fig:maps}. 

The large-scale distribution of H$\alpha$ is seen to follow the spiral arms in both galaxies. However, the centres of both NGC 1365 and NGC 1068 have been shown to be dominated by harder-ionising processes, such as AGN and shocks (see Section~\ref{sec:sample}), resulting in outflows seen in emission lines such as [O \textsc{iii}]$\lambda$5007. Their centres also contain significant star-formation, such as the nuclear 1kpc diameter starburst ring in NGC 1068 \citep[e.g.][]{Wilson1991}. Therefore, while some clear AGN driven structures can be detected using single lines like the biconical outflow in the [O \textsc{iii}] map of NGC 1365 in Figure \ref{fig:oIIImaps}, most lines towards the centre arise from some combination of star formation and AGN affected gas. This is most clearly seen in Figure~\ref{fig:nIIHaoIIIHb}, which shows increases towards the nucleus in both the [N \textsc{ii}]/H$\alpha$ and [O \textsc{iii}]/H$\beta$ ratios. We note that the line ratio maps in Figure~\ref{fig:nIIHaoIIIHb} do not show evidence of a biconical outflow in NGC 1365 due to low signal-to-noise in the H$\beta$ line. Hence, we make use of the BPT diagram and each galaxy's mixing sequence to separate star formation and AGN activity. For this work, we apply a consistent signal-to-noise cut of 3 on all emission lines in the BPT diagram (H$\alpha$, H$\beta$, [O \textsc{iii}], [N \textsc{ii}]).

\subsubsection{Correcting for saturation in NGC 1068}
\label{sec:1068fix}

The emission lines affected by saturated pixels include [O \textsc{iii}]$\lambda\lambda$4959,5007, H$\alpha$ and [N \textsc{ii}]$\lambda\lambda$6543,6584. In the pipeline, pixels that are saturated are flagged and carried through as `NaN' pixels. We reproduce the line profile by performing a fit to the unsaturated emission line(s). For spaxels with saturated lines, we use \textsc{idl/mpfit} to fit the H$\beta$ profile with a two-component gaussian.

\begin{figure}
\centering
\includegraphics[width=\columnwidth]{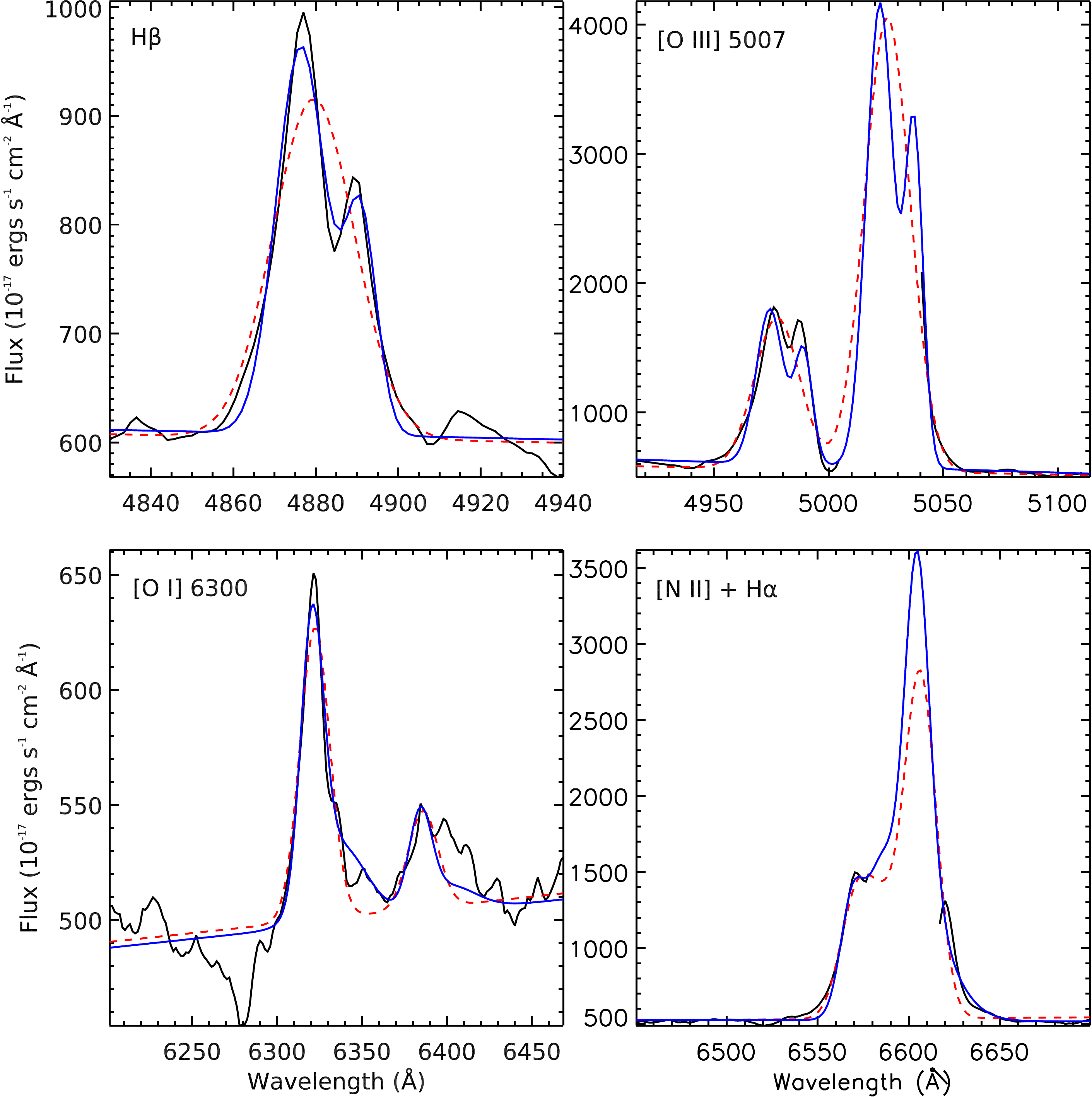}
\caption{The fitting correction applied to the NGC 1068 datacube for NaN spaxels at the centre. Figure shows corrections to the H$\beta$, [O \textsc{iii}], [O \textsc{i}], [N \textsc{ii}], and H$\alpha$ emission lines. The one-component gaussian fits are shown in red, whilst the two-component fits are shown in blue.}
\label{fig:1068fix1}
\end{figure}

The [O \textsc{iii}]$\lambda\lambda$4959,5007 profile with saturated (NaN) pixels is fit with \textsc{idl/mpfit} using a two-component gaussian with the redshift and velocity width of the gaussian components fixed to the values derived from the H$\beta$ fit. The line peaks, continuum level and continuum slope are allowed to vary (the [O \textsc{iii}]$\lambda\lambda$4959,5007 ratio is held fixed at 1:3), using the Gaussian shape and the unsaturated wings of the line to constrain the missing data.  Any `NaN' pixels in the region of [O \textsc{iii}] are replaced with the value from the resulting fit. This process is repeated using the [O \textsc{i}]$\lambda\lambda$6300,6363 doublet to generate a two-component gaussian used to fit the [N \textsc{ii}]$\lambda\lambda$6543,6584 + H$\alpha$ region in the same fashion (redshift and width are fixed, the [N \textsc{ii}]$\lambda\lambda$6543,6584 ratio is fixed to 1:3, the peaks of the [N \textsc{ii}] and H$\alpha$ lines are allowed to vary). The example in Fig.~\ref{fig:1068fix1} shows the H$\beta$ (top left) and [O \textsc{i}] (bottom left) fits - a single component fit is shown in red and the two-component fit is shown in orange. The fits to [O \textsc{iii}] (upper right) and [N \textsc{ii}]+H$\alpha$ (bottom right) using the same gaussian components are shown on the right.

\begin{figure}
\centering
\includegraphics[width=\columnwidth]{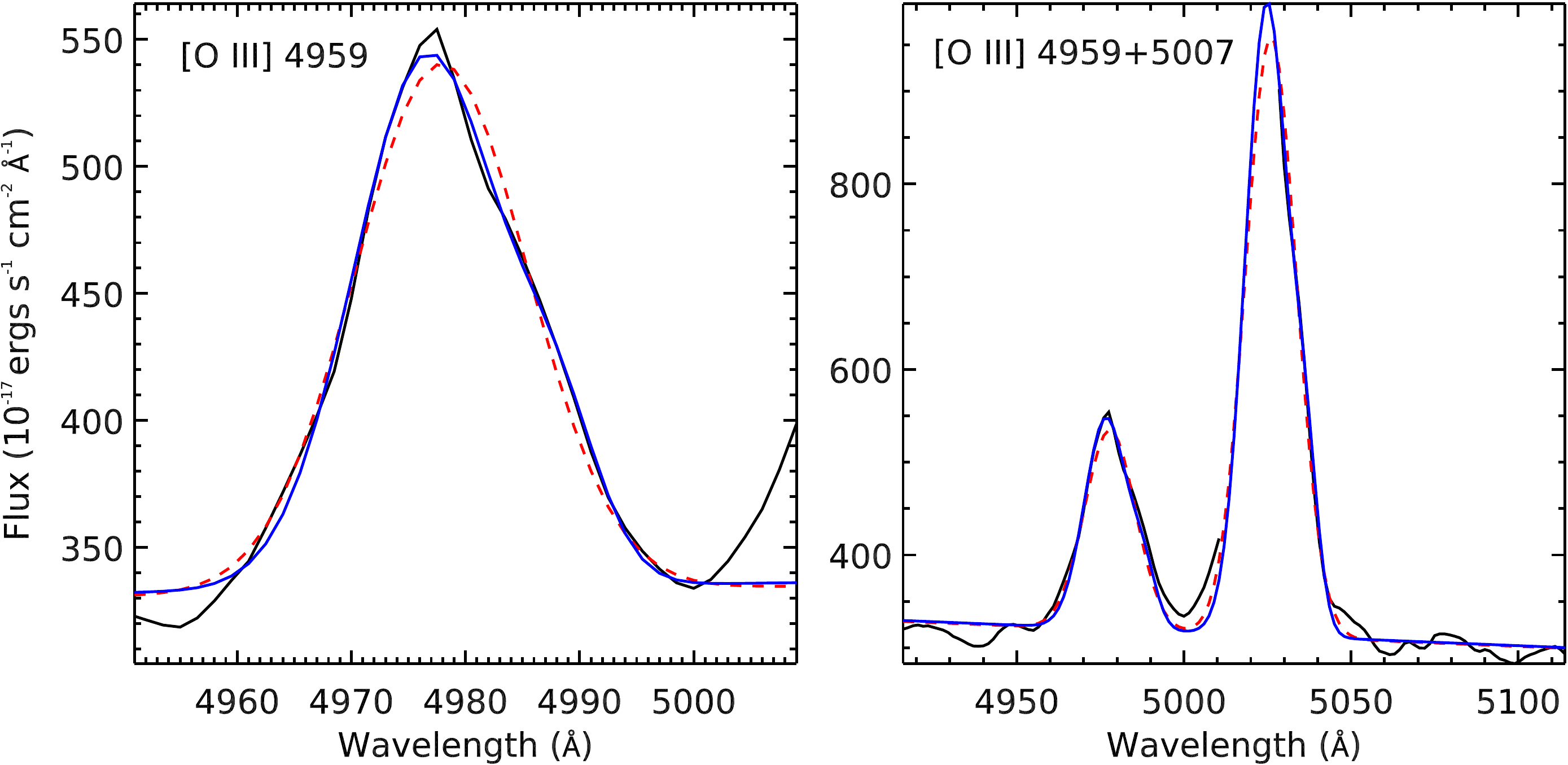}
\caption{Figure showing the fit to the [O \textsc{iii}]$\lambda$4959, used to fit the [O \textsc{iii}]$\lambda \lambda$4959,5007 when the signal-to-noise in the H$\beta$ profile is too low to provide reliable fit parameters. The one-component gaussian fits are shown in red, whilst the two-component fits are shown in blue.}
\label{fig:1068fix2}
\end{figure}

Finally, there are some spaxels where the signal-to-noise in the H$\beta$ profile is too low to provide a reliable fit to be used for the [O \textsc{iii}] doublet. For these four spaxels, we fit the [O \textsc{iii}]$\lambda$4959 line to generate the two-component gaussian widths and redshifts, and apply the results to the [O \textsc{iii}]$\lambda\lambda$4959,5007 simultaneously with the resulting widths and redshifts fixed. In Fig.~\ref{fig:1068fix2}, the left panel shows the fit to [O \textsc{iii}]$\lambda$4959 and the right panel shows the combined [O \textsc{iii}]$\lambda\lambda$4959,5007 fit using the parameters from the [O \textsc{iii}]$\lambda$4959 fit. Again a single component gaussian is shown in red and a two-component gaussian is shown in orange. 

The spaxels with these corrections applied are located at the following ($\alpha$,$\delta$) arcminute offsets from the centre of the galaxy):
(-0.028,-0.054), 
(-0.055,-0.054), 
(-0.028,-0.027), 
(-0.055,-0.027), 
(-0.028,0.000), 
(-0.055,0.000), 
(0.000,0.027), 
(-0.028,0.027), 
(-0.055,0.027), 
(0.000,0.054), 
(-0.028,0.054), 
(-0.055,0.054), 
(-0.028,0.081), 
(-0.055,0.081). 

In total, when considering spaxels with a signal-to-noise cut greater than 3 in all the emission lines on the BPT diagram, the percentage of spaxels with a correction applied is only 0.5\%.

%For this work, we selected two galaxies (NGC 1365 and NGC 1068) which both show confirmed AGN activity
\pagebreak
\section{Photoionisation Model Grids}
\label{sec:grids}

The photoionisation grids used during this work were created using the photoionisation code \texttt{MAPPINGS V.1}. We compute separate model grids for both the star-formation region and the AGN region of the BPT diagram, described subsequently.

\subsection{Starburst model grid}

The parameters involved in the creation of the starburst model grid are based on the findings and recommendations given by \citet{gridpaper}. We consider all aspects of the H \textsc{ii} region, including a stellar cluster and physical structure of the region. 

\subsubsection{Ionising stellar spectrum}

To synthesise the ionising stellar spectrum, we use the stellar population synthesis code Stochastically Lighting Up Galaxies \citep[SLUG;][]{daSilva2012, Krumholz2015}. The stellar spectrum is created using the initial mass function of \citet{Kroupa2002}, the Geneva "High" mass-loss (HIGH) stellar evolutionary tracks published in \citet{Meynet1994}, and the stellar atmospheres of \citet{Lejeune1997} with the addition of updated Wolf-Rayet (WR) and OB star atmospheres from \citet{HM1998} and \citet{Pauldrach2001} respectively. We favour the use of certain stellar models based on updates to the input physics from previous models. The Geneva group's HIGH stellar tracks include a correction to the definition of stellar effective temperature at the WR phase. Similarly, specialised modelling of WR stars by \citet{HM1998} and OB stars by \citet{Pauldrach2001} include updated physics to model both processes of line blanketing and line blocking, as well as a revised EUV and X-ray radiation model as a result of shock cooling zones in the OB stellar winds. We assume a stellar population at 10 Myr, undergoing constant star formation at a rate of $1 M_\odot \;\rm{yr}^{-1}$. We use an age of 10 Myr, as 10 Myr is the age at which 99.9\% of ionising photons have been released for a single stellar generation undergoing constant star formation \citep[e.g.][]{CL2001,Feltre2016}. Hence, the shape of the stellar spectrum varies negligibly beyond this point. Despite the fact that we resolve individual H \textsc{ii} regions harbouring a single stellar cluster in galaxies using TYPHOON data, we use models which assume a continuous star-formation history. As we resolve individual H \textsc{ii} regions across the entire galaxy (see Figure~\ref{fig:hiiregs}), it is fair to assume the initial conditions and rates of star formation within each H \textsc{ii} region differs throughout the galaxy, especially considering galaxies containing an AGN with a rich history of large-scale outflows. Hence, our models consider radiation from stellar clusters of all ages up to and including 10 Myr.

\subsubsection{H \textsc{ii} region structure}
\label{sec:sbmodel_hii}

We assume a spherical geometry for our H \textsc{ii} region, and adopt the elemental abundance solar reference of \citet{AG1989}, and metallicity scaling prescription of \citet{Nichollsabund}, matching the stellar abundance to the overall metallicity of the stellar tracks. Our model includes dust, using the elemental depletion pattern of \citet{Jenkins2014}, with a fraction of 0.3 of the carbon dust in the form of polycyclic aromatic hydrocarbons (PAHs). We use five metallicities in our starburst grid, with metallicities of $Z = 0.001, 0.004, 0.008, 0.020$, and 0.040. $Z$ here refers to the mass fraction of heavy elements (elements heavier than helium), defined by the relationship $X (\rm{H}) + Y (\rm{He}) + Z = 1$. The range of metallicities sufficiently covers the metallicity range measured in both galaxies NGC 1365 and NGC 1068. We use the $R_{23}$ metallicity diagnostic from \citet{KK04} to measure the range of metallicity in both galaxies. This is discussed further in Section~\ref{sec:use}. We include a range of the ionisation parameter $Q(N)$, defined as the number of  ionising photons emitted from the source relative to the number density of all ions in the nebula, with values of log($Q(N)\;\mathrm{cm}^{-2} \mathrm{s}^{-1}$) from 6.5 to 8.5 inclusive, in increments of 0.25 dex. The range of ionisation parameter is comparable to that used by \citet{Kewley2001}. We compute our model in isobaric conditions, setting the initial pressure of the H \textsc{ii} region to be $P/k = 8 \times 10^{5}\;\mathrm{K}\;\mathrm{cm}^{-3}$. Under the assumption of an H \textsc{ii} region initial temperature of 8000K, this approximately corresponds to an electron density of $n = 100 \;\rm{cm}^{-3}$. Figure~\ref{fig:edens} shows the electron density distribution across both NGC 1365 and NGC 1068 calculated using the [S \textsc{ii}] ratio, showing an electron density of $n = 100 \;\rm{cm}^{-3}$ to be a reasonable assumption in the star-forming regions of both galaxies. Our model is assumed to be radiation-bounded, and hence the model is terminated once 99\% of H \textsc{ii} recombination has occurred

\subsection{AGN model grid}

Our AGN models are computed assuming a plane-parallel geometry, and use the same definitions for the abundance pattern and metallicity scaling, and ionisation parameter as our starburst model described in Section~\ref{sec:sbmodel_hii}. We use six metallicities when computing our AGN model grids, with values of $Z$ = 0.001, 0.004, 0.008, 0.020, 0.040 and 0.060. However, the low metallicities used in these grids are arguably redundant, as low-metallicity AGN are rare \citep[e.g.][]{GHK2006}. In increments of 0.25 dex, the ionisation parameter varies from log($Q(N)\;\mathrm{cm}^{-2} \mathrm{s}^{-1}$) = 7.0 to 11.0 inclusive.

We compute the AGN models also in isobaric conditions, however the value of initial pressure differs between the models used for NGC 1365 and NGC 1068. For NGC 1365, we use an initial pressure of $P/k = 2 \times 10^7$, and for NGC 1068, $P/k = 2 \times 10^8$. Under the assumption of a narrow-line region (NLR) initial temperature of 20,000K, this corresponds to electron densities of $n = 1000 \;\rm{cm}^{-3}$ and $n = 10,000 \;\rm{cm}^{-3}$ for NGC 1365 and NGC 1068 respectively, supported through the density calculation from the [S \textsc{ii}] ratio in Figure~\ref{fig:edens}. 

%Spherical geometry, the elemental abundance pattern of \citet{Nichollsabund}, and depletion pattern of \citet{Jenkins2014} \citep[see][for details]{gridpaper} are common to both our starburst and AGN grids, as is the use of polycyclic aromatic hydrocarbons (PAHs), with a carbon dust depletion fraction of 0.3. Both grids are also computed in isobaric conditions, however the pressure assigned to each varies:
%assuming an initial temperature of 8,000K for the starburst grids, and 20,000K for the AGN grids, the pressure in $P/k$ units for each grid was calculated using $P/k = nT$, where $n$ refers to the initial electron density. The electron density across both galaxies can be seen in Fig.~\ref{fig:edens}. As a result, the electron densities chosen for the starburst and AGN grids for NGC 1365 are 100 cm$^{-3}$ and 1,000 cm$^{-3}$ respectively, and for NGC 1068 are 100 cm$^{-3}$ and 10,000 cm$^{-3}$ respectively.

\begin{figure*}
\centering
\begin{subfigure}{0.3\textwidth}
\includegraphics[width=\linewidth]{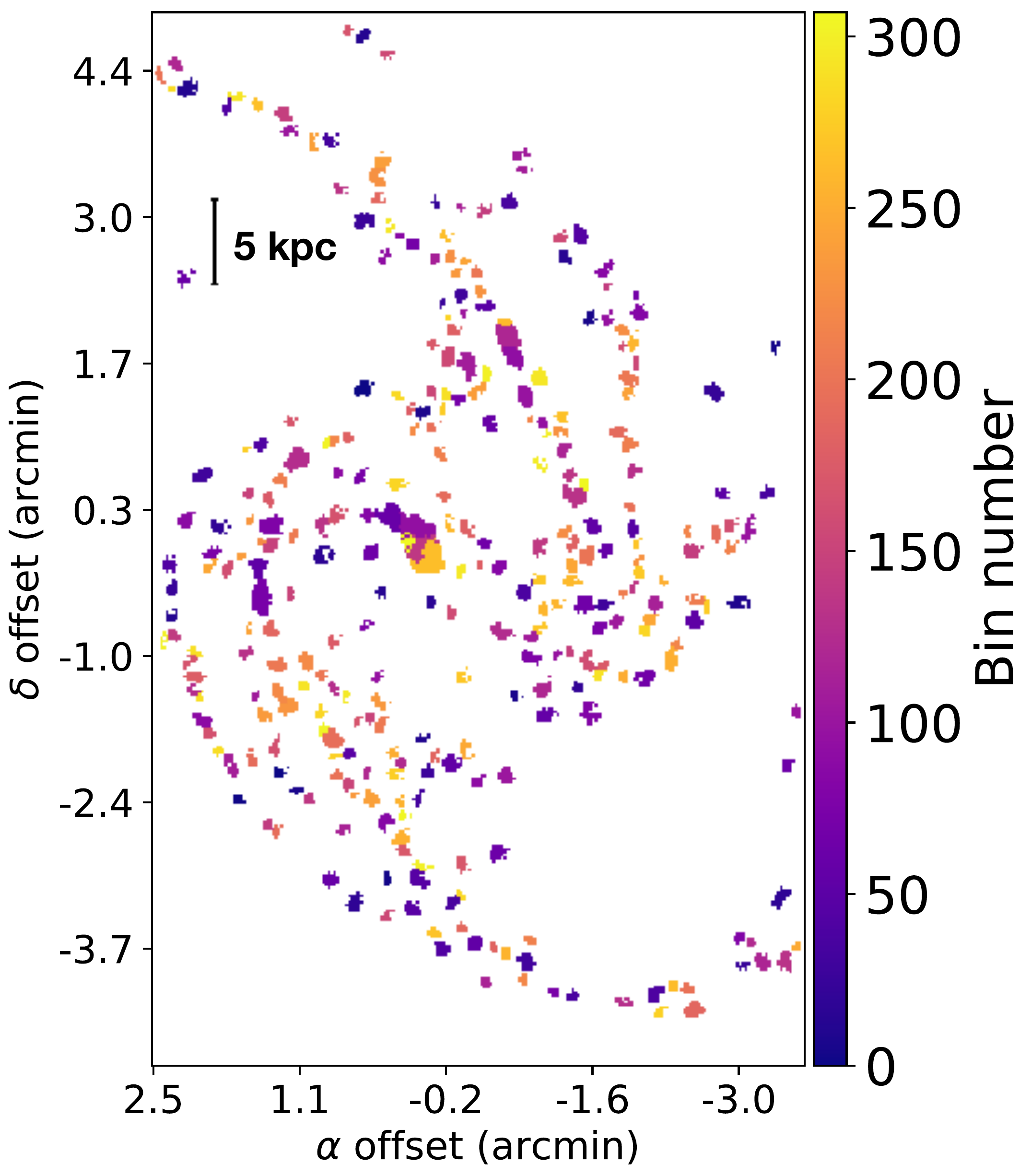}
\caption{NGC 1365}
\label{fig:hiiregsa}
\end{subfigure}\hspace{0.1\textwidth}
\begin{subfigure}{0.21\textwidth}
\includegraphics[width=\linewidth]{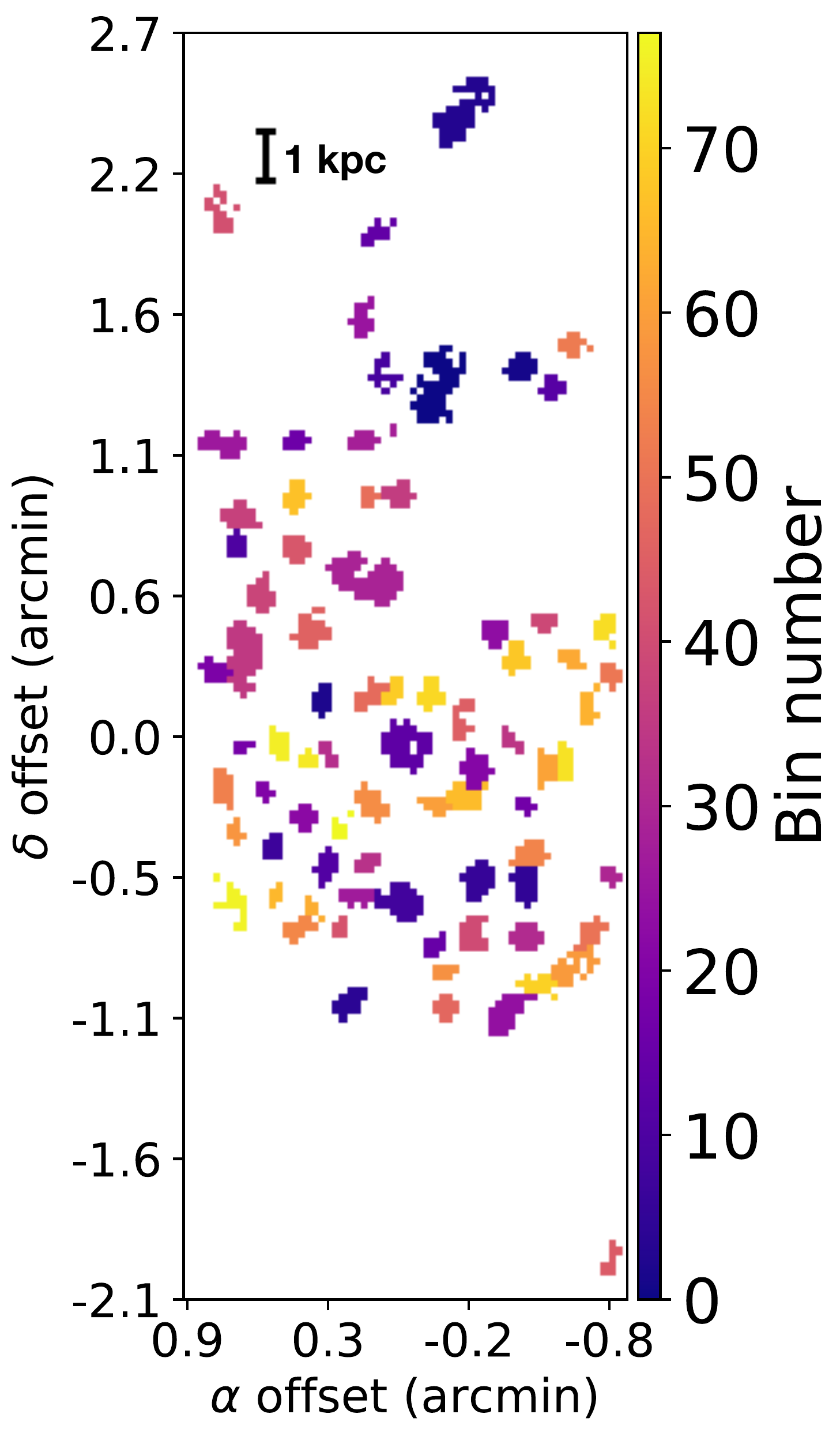}
\caption{NGC 1068}
\label{fig:hiiregsb}
\end{subfigure}
\caption{Figure showing location of H \textsc{ii} regions in NGC 1365 in (a) and NGC 1068 in (b). These H \textsc{ii} regions were identified using \textsc{hiiphot}, developed by \citet{Thilker2000}.}
\label{fig:hiiregs}
\end{figure*}

\begin{figure*}
\centering
\begin{subfigure}{0.3\textwidth}
\includegraphics[width=\linewidth]{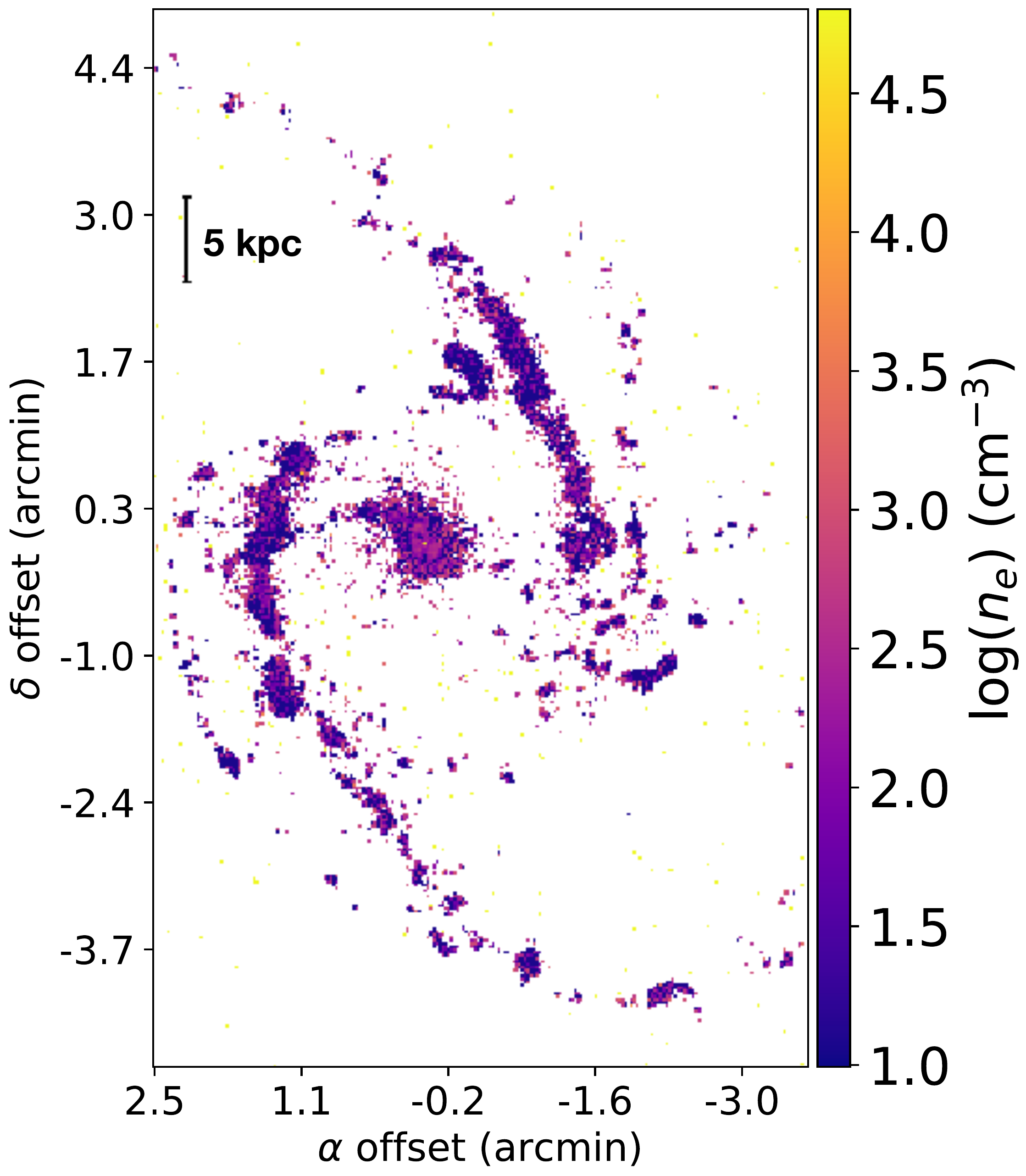}
\caption{NGC 1365}
\label{fig:edensa}
\end{subfigure}\hspace{0.1\textwidth}
\begin{subfigure}{0.27\textwidth}
\includegraphics[width=\linewidth]{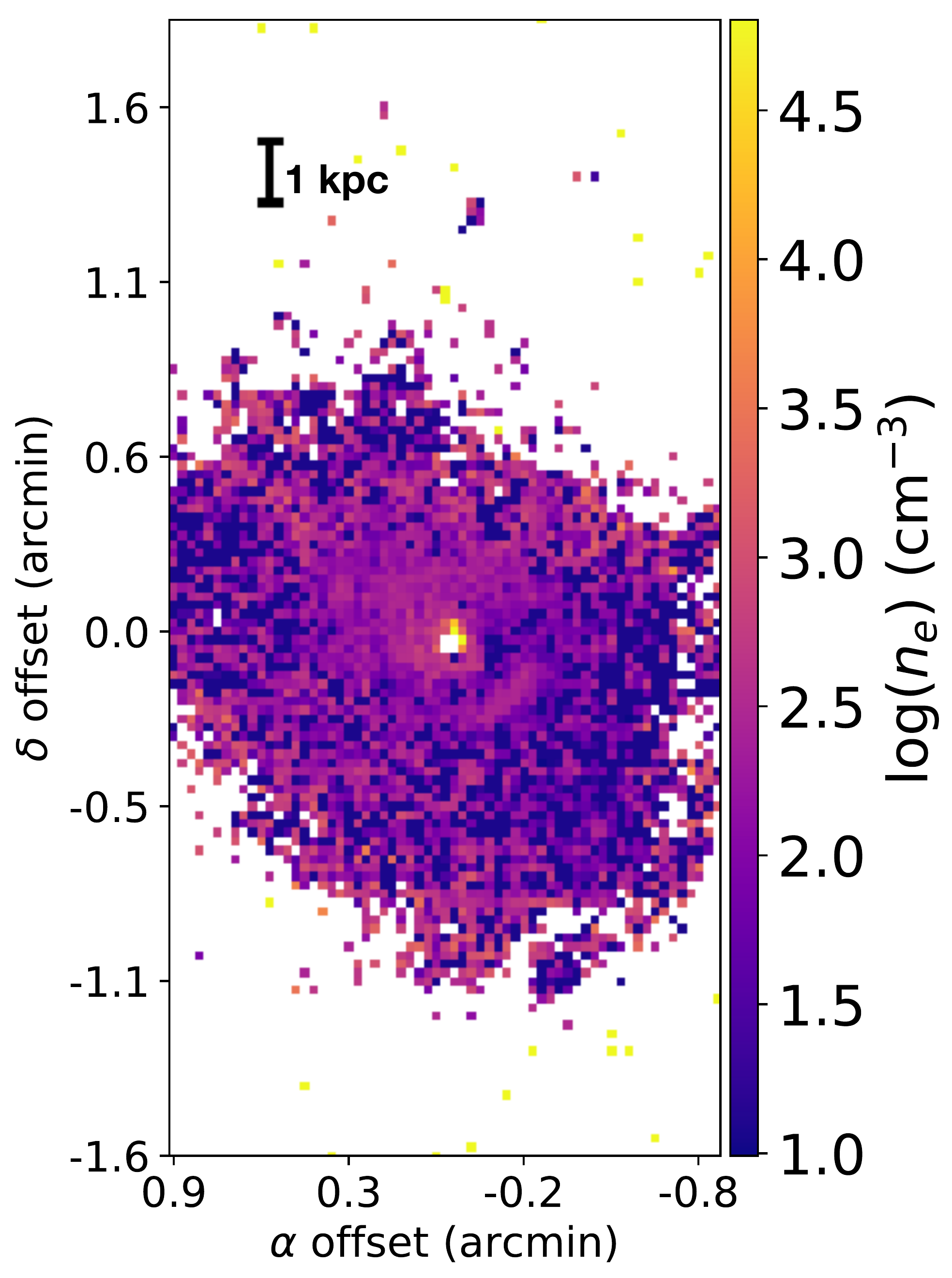}
\caption{NGC 1068}
\label{fig:edensb}
\end{subfigure}
\caption{Figure showing range of electron density across both NGC 1365 in (a) and NGC 1068 in (b).}
\label{fig:edens}
\end{figure*}

%Our starburst grid models adopt the Binary Population and Spectral Synthesis models \citep[BPASS; see][]{Eldridge2008, ES2009, ES2012, Stanway2016} to provide the ionising spectrum. A continuous binary stellar population at an age of 10 Myr is used, in favour of a single stellar cluster. \citet{Levesque2010} showed a continuous star formation history (SFH) produces a harder far-ultravoilet (FUV) ionising spectrum, and showed better agreement with emission line ratios found in their galaxy sample than for an instantaneous burst SFH at the same age. \citet{Stanway2016}, amongst others, find a boosted (50-60\%) Hydrogen ionising flux in stellar populations at low metallicities ($0.05Z_\odot \leq Z \leq 0.3Z_\odot$), and a more modest 10-20\% increase in the flux at higher (near-solar) metallicities, compared to single-star stellar populations. The BPASS model used in particular contains two power law segments, with indices of $\alpha_1 = 1.3$ between 0.1$M_\odot$ and 0.5$M_\odot$, and $\alpha_2 = -2.0$ between 0.5$M_\odot$ and 300$M_\odot$.

%Our photoionisation models are radiation-bounded when 99\% of H \textsc{ii} recombination has taken place. We use five metallicities constrained by the stellar tracks in our stellar population synthesis models: $Z = 0.001, 0.004, 0.008, 0.020, 0.040$. We select models with ionisation parameter in units of log($Q(N)$) from 6.5 to 8.5 inclusive, in increments of 0.25 dex.

The AGN region models are computed with a power-law spectrum of the form $c \nu^{\alpha}$ \citep{Groves2004}, fixing $\alpha$ = -1.2. Lower values of $\alpha$ have been historically favoured \citep[${\sim} -1.4$ to ${\sim} -2.0$; e.g.][]{Groves2004,BWH2006}, however we find lower values of $\alpha$ fail to completely explain the TYPHOON data on the BPT diagram. Further justification on the use of $\alpha = -1.2$ is dicussed in Appendix~\ref{sec:alpha}. The value of $\alpha$ however has been known to vary wildly amongst AGN \citep[e.g. see detailed catalogue of AGN by][]{Ackermann2011}. %However, this range of ionisation parameter values was deemed insufficient to cover the spread of data points when analysing the [O \textsc{iii}]/H$\beta$ vs [S \textsc{ii}]/H$\alpha$ diagram. As a result, one extra value of the ionisation parameter (log($Q(N)$) = 6.75) was added to the AGN photoionisation grid for the aforementioned diagnostic diagram.

\section{The starburst-AGN mixing sequence}
\label{sec:agnfrac}
% New technique, updated from original star-forming distance technique from Davies et al. (Mention shortcomings)
% Henry's code helps constrain basis points
% I-Ting's work helps select bins to use
% Metallicity gradient, but no ionisation parameter gradient

\begin{table*}
\centering
\begin{tabular}{| c | c | c | c | c | c | c | c |}
%\begin{center}
\toprule
\textbf{Galaxy} & \multicolumn{2}{c}{\textbf{Metallicity ($Z_\odot$)}} & \multicolumn{5}{c}{\textbf{Basis Point Coordinates (log([N \textsc{ii}]/H$\alpha$, log([O \textsc{iii}]/H$\beta$)}} \\
\midrule
& \textbf{Edge} & \textbf{Centre} & \textbf{SB$_\mathrm{1}$} & \textbf{SB$_\mathrm{2}$} & \textbf{SB$_\mathrm{3}$} & \textbf{AGN$_\mathrm{1}$} & \textbf{AGN$_\mathrm{2}$} \\
\midrule
NGC 1365 & 0.55 & 1.42 & (-1.01, 0.39) & (-0.64, -0.86) & (-0.42, -1.27) & (0.24, 1.14) & (0.46, 0.62)  \\
NGC 1068 & 1.02 & 1.41 & (-0.52, -0.19) & (-0.48, -0.99) & (-0.38, -1.49) & (0.22, 1.31) & (0.49, 0.71)  \\
\bottomrule
%\end{center}
\end{tabular}
\caption{Derived quantities for NGC 1365 and NGC 1068 for use on the BPT diagram.}
\label{tab:derived}
\end{table*}

As demonstrated by \citet{Kewley2001} and \citet{Kauffmann2003}, star-forming galaxies lie along a tight curve on the BPT diagram. However, as an AGN increases in activity until it completely dominates over emission from star formation, the galaxy moves off this star-forming sequence onto a `mixing sequence' \citep[e.g.][]{Kewley2006}. As discussed by \citet{Kewley2001} and demonstrated empirically by \citet{Davies2014a,Davies2014b}, the relative fractional contribution of
star-formation and the AGN to the emission line luminosity can be determined from this mixing sequence.

\citet{Davies2014a,Davies2014b} used `basis points' to define empirical 100\% starburst and 100\% AGN positions on the BPT diagram. The starburst and AGN basis points were assigned to the spaxels which contained the lowest and highest [O \textsc{iii}]/H$\beta$ ratio respectively. These basis points corresponded to 0\% and 100\% AGN activity, respectively. All other spaxels were assigned an AGN fraction between 0\% and 100\% corresponding to their star-forming distance \citep[$d_{\mathrm{SF}}$;][]{Kewley2006} along the line between the two basis points. The star-forming distance $d_{\mathrm{SF}}$ is a measure of a spaxel's (or galaxy's) offset in its [N \textsc{ii}]/H$\alpha$ and [O \textsc{iii}]/H$\beta$ ratios from the star-forming sequence on the BPT diagram, mapped out by pure star-forming galaxies in SDSS \citep[see][]{Kauffmann2003}. The starburst basis point was defined as having a $d_{\mathrm{SF}} = 0$. The complement to the AGN fraction with respect to 100\% is considered to be the relative fraction of emission from star formation. Using data from the TYPHOON survey, we extend this method, described below.

The result of applying the method described by \citet{Davies2014a,Davies2014b} to NGC 1365 from the TYPHOON survey is shown in Figure~\ref{fig:agnfrac_old}. The method from \citet{Davies2014a,Davies2014b} when applied to IFU data shows AGN fractions of up to ${\sim} 50$\% for several spaxels along the star forming sequence of NGC 1365. These spaxels are found below the Kauffmann line, which is considered to be the upper limit of pure star formation.

\subsection{Use of photoionisation grids}
\label{sec:use}

We use the photoionisation grids described in Section~\ref{sec:grids} to aid in the selection of the basis points. The basis points selected by \citet{Davies2014a,Davies2014b} for their starburst-AGN mixing work are purely empirical. Hence, the basis points contain no physical information about the galaxy. 
Through the use of photoionisation grids, star formation and AGN activity can be separated with the basis points containing physical information about the galaxy, such as the metallicities, ionisation parameters, pressures of both the H \textsc{ii} and AGN regions, and hardness of the AGN power-law radiation field (Thomas et al. in prep.). In addition, theoretical models also allow the prediction of spectra at other wavelengths. 

%By moving to a theoretical calculation of the starburst-AGN fraction through the use of photoionisation grids, the basis points selected detail physical properties about the galaxy, such as the metallicities, ionisation parameters and pressures of both the H \textsc{ii} and AGN regions. 

To constrain the basis points of both the AGN and star-forming grids we use the independently determined gas-phase metallicity. The $R_{23}$ metallicity diagnostic from \citet{KK04} (KK04) was used to diagnose metallicities from H \textsc{ii} regions within the galaxy. These H \textsc{ii} regions were identified using the \textsc{idl} routine \textsc{hiiphot}, developed by \citet{Thilker2000}. The H \textsc{ii} regions were found to be located throughout the galaxy, thus making possible the calculation of metallicity gradients from the central regions to the outskirts of both galaxies. The location of the H \textsc{ii} regions across NGC 1365 and NGC 1068 can be seen in Figure~\ref{fig:hiiregs}. The metallicities for the central and outer regions of both NGC 1365 and NGC 1068 can be found in Table~\ref{tab:derived}.

\begin{figure*}
\centering
\includegraphics[width=\textwidth]{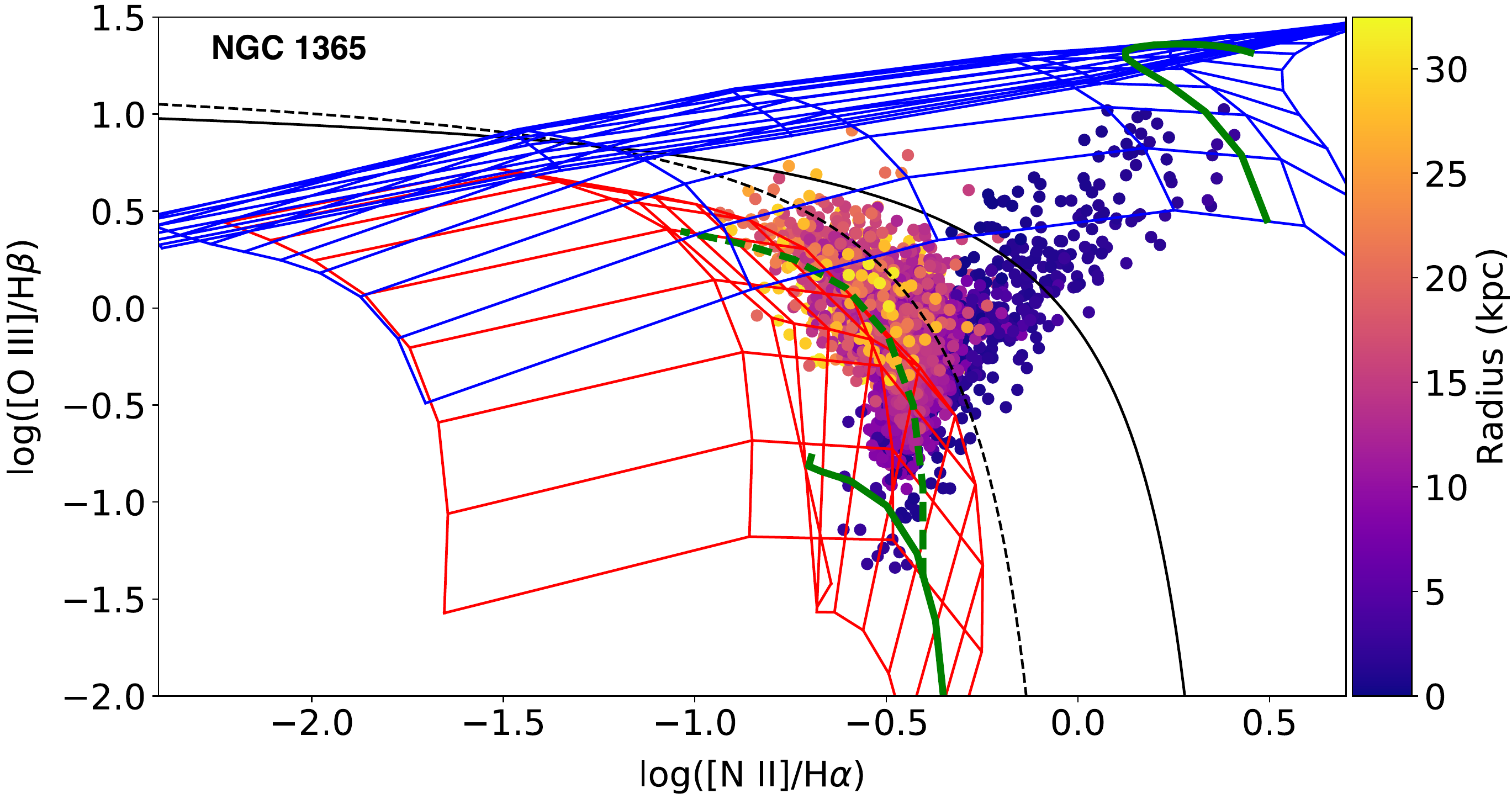}
\caption{The BPT diagram of NGC 1365, with each spaxel coloured according to radius from the centre of the galaxy. The black dashed curve is the \citet{Kauffmann2003} empirical maximum starburst line, and the black solid line is the \citet{Kewley2001} theoretical maximum starburst line. The H \textsc{ii} region grid is shown in red, and the AGN grid is shown in blue. The solid green grid lines on both grids represent the metallicity at the centre of the galaxy, and the dashed green grid line on the H \textsc{ii} region grid represents the metallicity at the edge of the galaxy. Both metallicity values are stated in Table~\ref{tab:derived}.}
\label{fig:1365_normal_radius}
\end{figure*}

Seen in Figure~\ref{fig:1365_normal_radius}, coloured by radius from the centre of the galaxy, is the presence of a large star-forming sequence in NGC 1365. This star-forming sequence extends from the nucleus to the outermost regions of the galaxy. The starburst grid shows a metallicity gradient within this star-forming sequence, beginning at a high metallicity within the nucleus, and extending to a low metallicity at the edges of the galaxy. 

\begin{figure*}
\centering
\includegraphics[width=\textwidth]{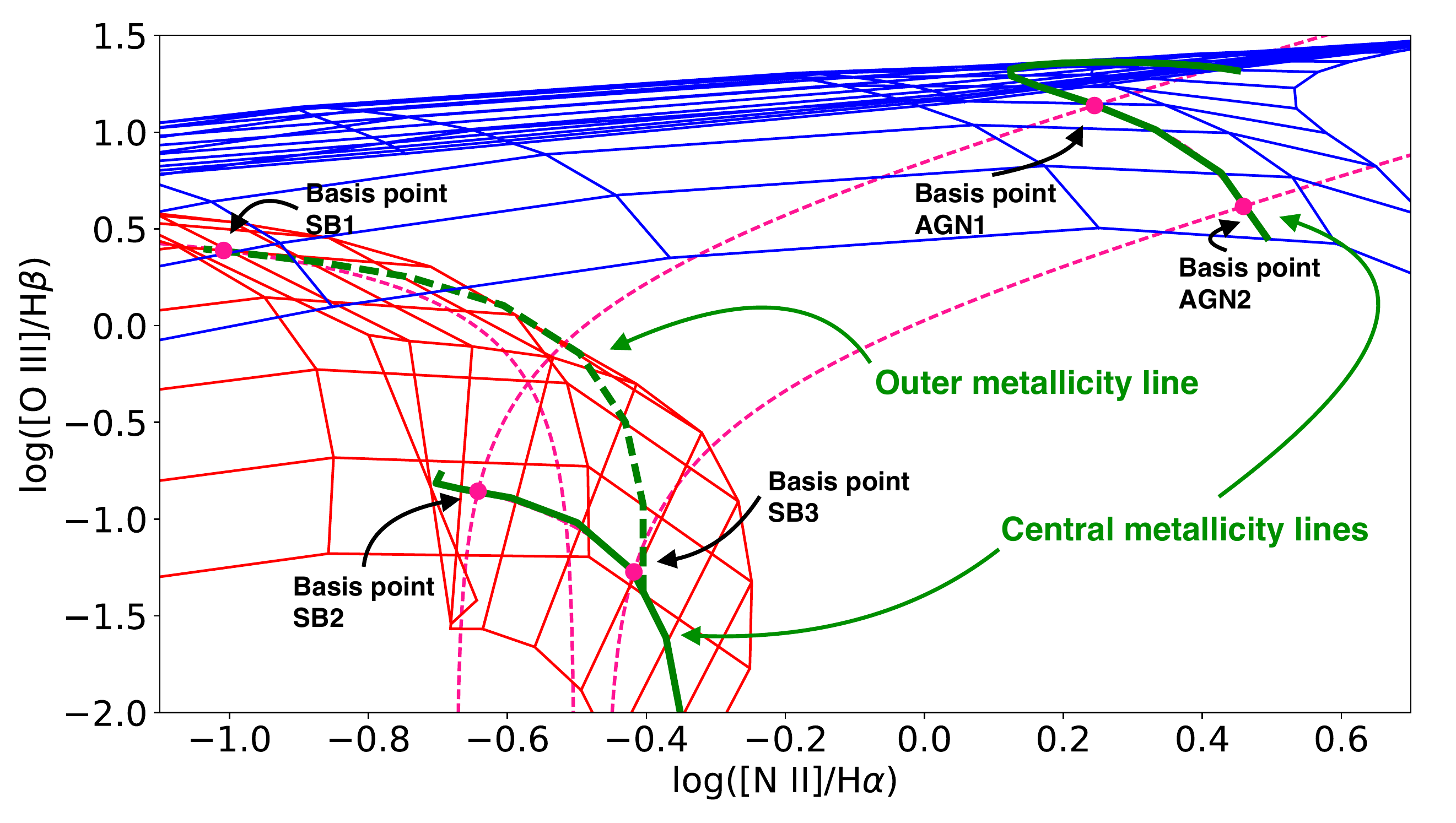}
\caption{The basis points exhibited on the photoionisation grids, with the location of each constrained using the metallicity measurements of NGC 1365. The starburst basis points SB1, SB2, and SB3 define the starburst basis line, while the basis points SB2, SB3, AGN1, and AGN2 together define the nuclear column.}
\label{fig:grids_toy}
\end{figure*}

We used the metallicity calculations to construct extra lines on the photoionisation grids, corresponding to the metallicities of the central and outer regions of the two galaxies. All basis points are positioned along these new grid lines, depending on the region of the galaxy for which they correspond. The positions of basis points SB1, SB2, and SB3 are defined by the starburst grid, whilst the positions of basis points AGN1 and AGN2 are defined by the AGN grid. Basis point SB1 is positioned along the line corresponding to the outer metallicity of the galaxy, at the edge of the star-forming sequence of spaxels. Basis points SB2 and SB3 are positioned along the line of central metallicity at the nuclear region of the star-forming sequence. Basis points AGN1 and AGN2 are also positioned along the line of central metallicity. The basis points are placed along the lines of constant metallicity such that they encompass the spread in ionisation parameter in the data (see Section~\ref{sec:nuc_column}). The starburst basis points SB1, SB2, and SB3 define the \emph{starburst basis line}, described in Section~\ref{sec:sb_basis_line}. Basis points SB2, SB3, AGN1, and AGN2 altogether define the \emph{nuclear column}, defined in Section~\ref{sec:nuc_column}. As an example, we show all five basis points located on the photoionisation grids in Figure~\ref{fig:grids_toy}, using the central and outer metallicities of NGC 1365 to determine their locations. Figure~\ref{fig:grids_toy} also demonstrates the basis points defining the starburst basis line, and the nuclear column.

\subsection{Starburst basis line}
\label{sec:sb_basis_line}

\begin{figure*}
\centering
\includegraphics[width=\textwidth]{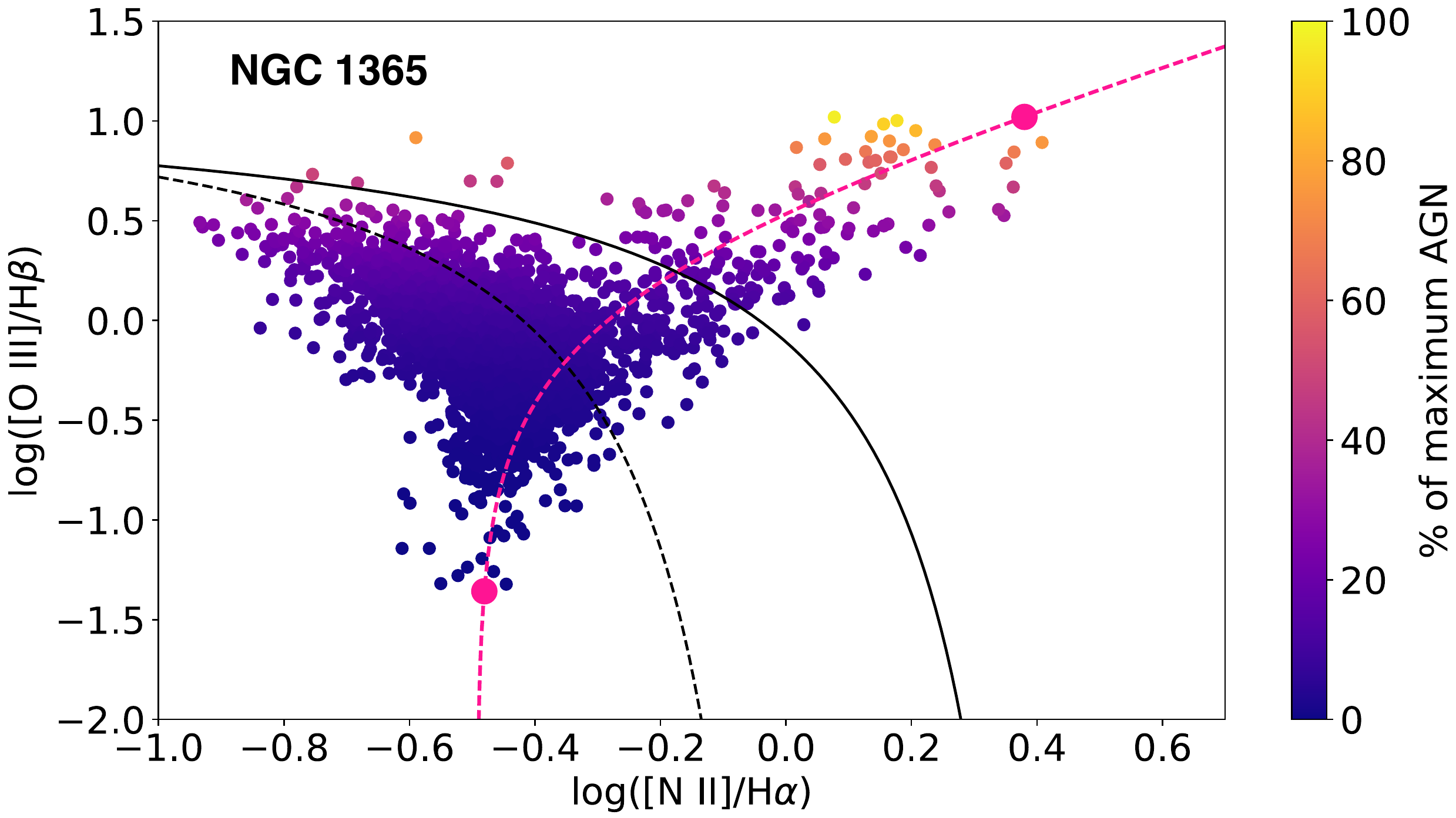}
\caption{The AGN fraction for NGC 1365, using the old calculation by \citet{Davies2014a,Davies2014b}.  The method from \citet{Davies2014a,Davies2014b} gives AGN fractions of up to ${\sim} 50$\% for some spaxels below the Kauffmann line -- a well-known empirical line which signifies the upper limits of pure star formation.}
\label{fig:agnfrac_old}
\end{figure*}

\citet{Davies2014a,Davies2014b} assign a star-forming distance of $d_{\mathrm{SF}} = 0$ to the high-metallicity end of the star-forming sequence, as the majority of AGN-dominated galaxies are shown to mix with high-metallicity H \textsc{ii} regions \citep{GHK2006}. The high spatial resolution of the TYPHOON IFU data shows a star-forming sequence present in each individual galaxy, following the path mapped out by the pure star-forming galaxies in SDSS \citep[see][]{Kauffmann2003}. By defining a single point as $d_{\mathrm{SF}} = 0$, the spaxels on the star-forming sequence which extend out to lower metallicities are given progressively higher values for their star-forming distance. Seen in Figure~\ref{fig:agnfrac_old}, lower-metallicity star-forming spaxels are designated larger AGN fractions than expected.

We extend the notion of a starburst basis point defined as $d_{\mathrm{SF}} = 0$, to a \emph{starburst basis line}. Any spaxel present along the line is defined as having $d_{\mathrm{SF}} = 0$. This line is indicated on Figure~\ref{fig:1365agnfrac_nogrids_annotated}, extending between the starburst basis point SB1, and the midpoint of the starburst basis points SB2 and SB3.

\subsection{Nuclear column}
\label{sec:nuc_column}

Interestingly, all spaxels seen in the mixing sequence of NGC 1365 in Figure~\ref{fig:1365_normal_radius} are located in the nucleus, showing both star formation and AGN activity occurring as nuclear processes. Seen along this mixing sequence in Figure~\ref{fig:1365_normal_radius} is a clear spread in the ionisation parameter. We further extend the method displayed by \citet{Davies2014a,Davies2014b} by increasing the number of basis points located at either end of the mixing sequence from one to two, in order to account for this spread in ionisation parameter ($\Delta \mathrm{log} Q(N) {\sim} 0.5$ dex). The result is defined as the \emph{nuclear column}, and indicated in Figure~\ref{fig:1365agnfrac_nogrids_annotated} as extending between the 100\% nuclear starburst line, and the 100\% measureable AGN line. The nuclear column ensures the star-forming distance in the spaxels along the mixing sequence remains consistent, despite the changes in ionisation parameter. The AGN fraction for each spaxel is then determined by each spaxel's distance along a unique line projected between the 100\% nuclear starburst line and the 100\% AGN line.

% by fitting a line to the mixing sequence, and calculating each spaxel's distance along the line in order to arrive at the starburst-AGN fraction. This was a purely empirical method, which used the concept of `basis points' to define the 100\% starburst and 100\% AGN points. These two basis points defined the minimum and maximum along the line of best fit, and were placed on the line at the minimum and maximum [O \textsc{iii}]/H$\beta$ ratios found in the mixing sequence. 

%Our initial method aimed to perform the same starburst-AGN fraction calculation, however using the photoionisation grids described in Section \ref{sec:grids} to aid in the basis point selection. By calculating the metallicity gradient of each galaxy using identified H \textsc{ii} regions \citep[identified through the use of the \textsc{idl} routine \textsc{hiiphot}, developed by][]{Thilker2000} as explained in \citet{Poetrodjojo}, we were able to use the grids to constrain the basis points of the starburst-AGN fraction by metallicity. During this attempt, the galaxy's outer metallicity was for use in the starburst grid, and the central metallicity applied to the AGN grid. Along these metallicity constraint lines, the basis points were placed at positions where they best fit the data. From there, once the basis points were in place, the starburst-AGN fraction calculation method was identical to that shown in \citet{Davies2014a,Davies2014b}. A resulting example of this calculation can be seen in Fig.~\ref{fig:agnfrac_old}, where the starburst-AGN fraction is shown in terms of fraction of AGN activity.

\begin{figure*}
\centering
\includegraphics[width=\textwidth]{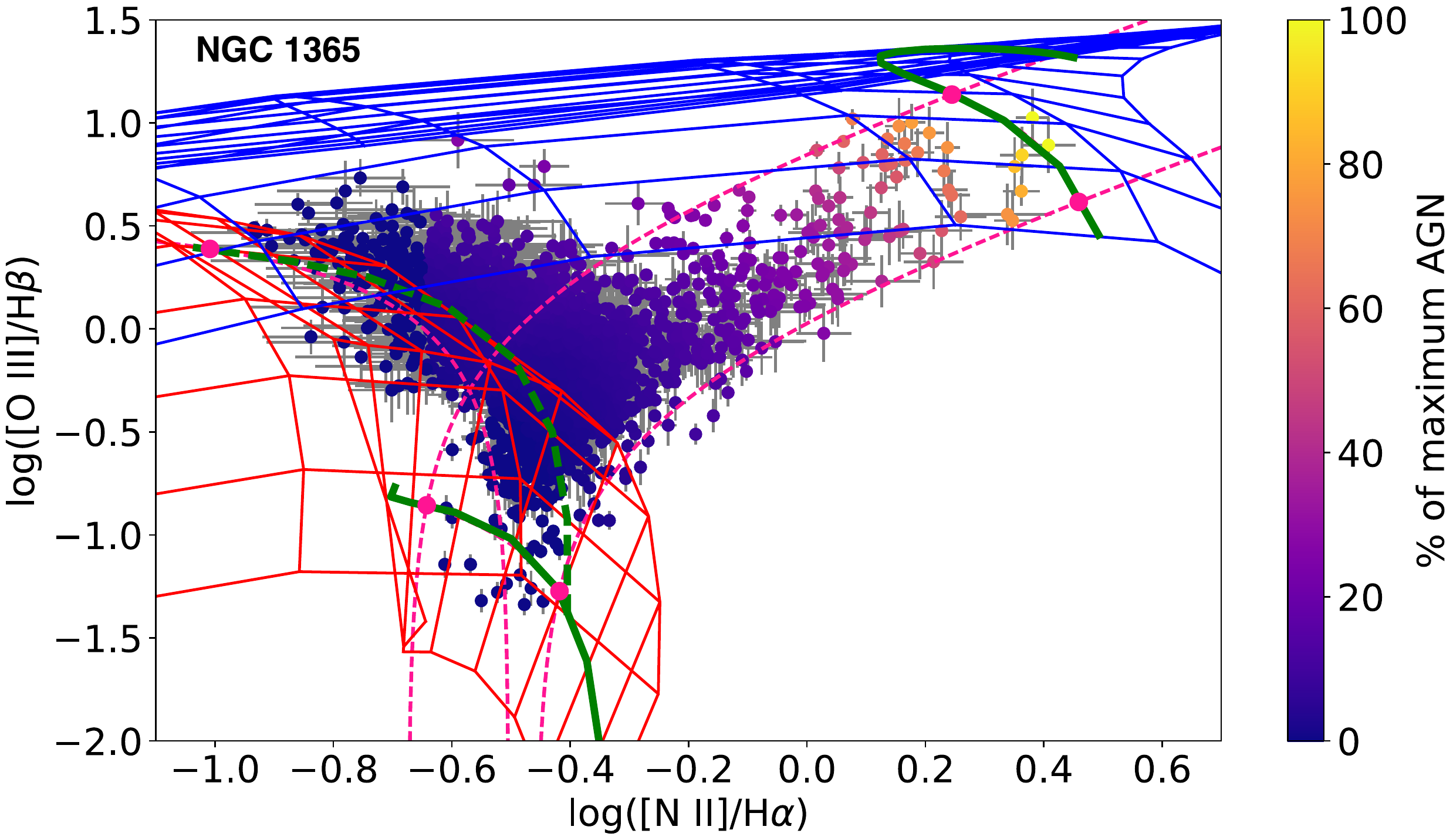}
\caption{AGN fraction [O \textsc{iii}]/H$\beta$ vs [N \textsc{ii}]/H$\alpha$ diagram for NGC 1365.}
\label{fig:1365BPTagnfrac}
\end{figure*}%\hspace{\textwidth}

\begin{figure*}
\centering
\includegraphics[width=\textwidth]{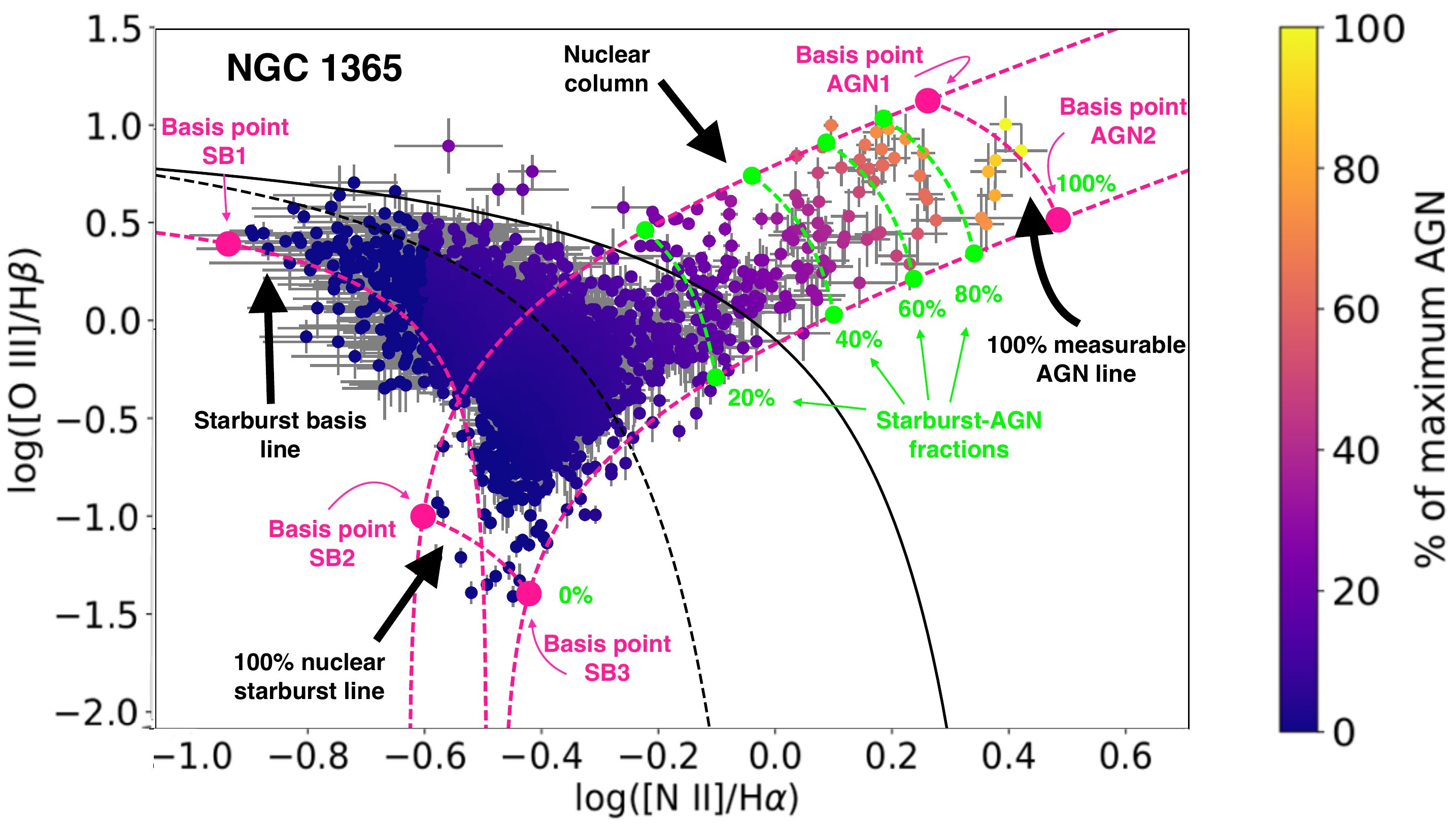}
\caption{The BPT diagram for NGC 1365 showing the fractional contribution of AGN excited gas to the H$\alpha$ line using our new calculation method. This figure also overlays the original \citet{Kauffmann2003} and \citet{Kewley2001} AGN diagnostic lines (black lines, Figure \ref{fig:1365_normal_radius}), the new basis points and mixing curves based on our model grids (pink points and lines, Figure \ref{fig:grids_toy} and Section \ref{sec:agnfrac}), and the resulting starburst-AGN mixing sequence (green curves).}
\label{fig:1365agnfrac_nogrids_annotated}
\end{figure*}

\subsection{Relative AGN fractions}

We calculate the relative AGN fractions for several emission lines for both NGC 1365 and NGC 1068, given as a percentage of the total emission. This calculation is identical to that displayed and described in \citet{Davies2014a}. The complement of the AGN fraction with respect to 100\% signifies the relative fraction of the total emission from star formation. The total luminosity of any emission line in a given galaxy can be calculated by the following:
\[ L_{\mathrm{Tot}} = \sum_{i = 1}^{n} L_i \]
where $L_i$ is the luminosity in spaxel $i$ for the emission line in question. The total luminosity of the emission line attributable to AGN activity can be calculated from
\[ L_{\mathrm{AGN}} = \sum_{i = 1}^{n} f_{i}^{\mathrm{AGN}} L_i \]
where $f_{i}^{\mathrm{AGN}}$ is the AGN fraction in spaxel $i$. It follows that the relative fraction of emission attributable to AGN activity for the given emission line is given by
\[ f_{\mathrm{Tot}}^{\mathrm{AGN}} = \frac{L_{\mathrm{AGN}}}{L_{\mathrm{Tot}}} \] 
Shown in Table~\ref{tab:agnfracs} are the results of the above calculations for several emission lines. We discuss the results for NGC 1365 and NGC 1068 below individually.

\subsubsection{NGC 1365}
\label{sec:1365}
% Need composite image and Ha map for both.

%Shown in Figure.~\ref{fig:1365} is the starburst-AGN fraction BPT diagram and map following our new method. The map of the starburst-AGN fraction distribution across the galaxy can be seen in Fig.~\ref{fig:1365BPTagnmap}. 

AGN activity from NGC 1365 has been inferred previously, with studies showing the nucleus of NGC 1365 to be luminous in hard x-rays \citep[e.g.][]{Risaliti2007,Risaliti2013}, radio wavelengths \citep[e.g.][]{SJL1982,SJL1995} and FUV radiation \citep[e.g.][]{PT1983}. Further, the nucleus of NGC 1365 has been shown to produce large fluxes of collisionally-excited emission lines, with its high flux of [O \textsc{iii}] very well documented \citep{PT1983,Veilleux2003}. Large [O \textsc{iii}]/H$\beta$ ratios in the nucleus of NGC 1365 \citep{PT1983}, as well as [O \textsc{iii}]/H$\alpha$ ratios much larger than unity in the nucleus \citep{Veilleux2003} have led previous authors to conclude the [O \textsc{iii}] is produced by photoionisation from the central AGN of NGC 1365. 

\begin{figure*}
\centering
\begin{subfigure}{\columnwidth}
\includegraphics[width=\textwidth]{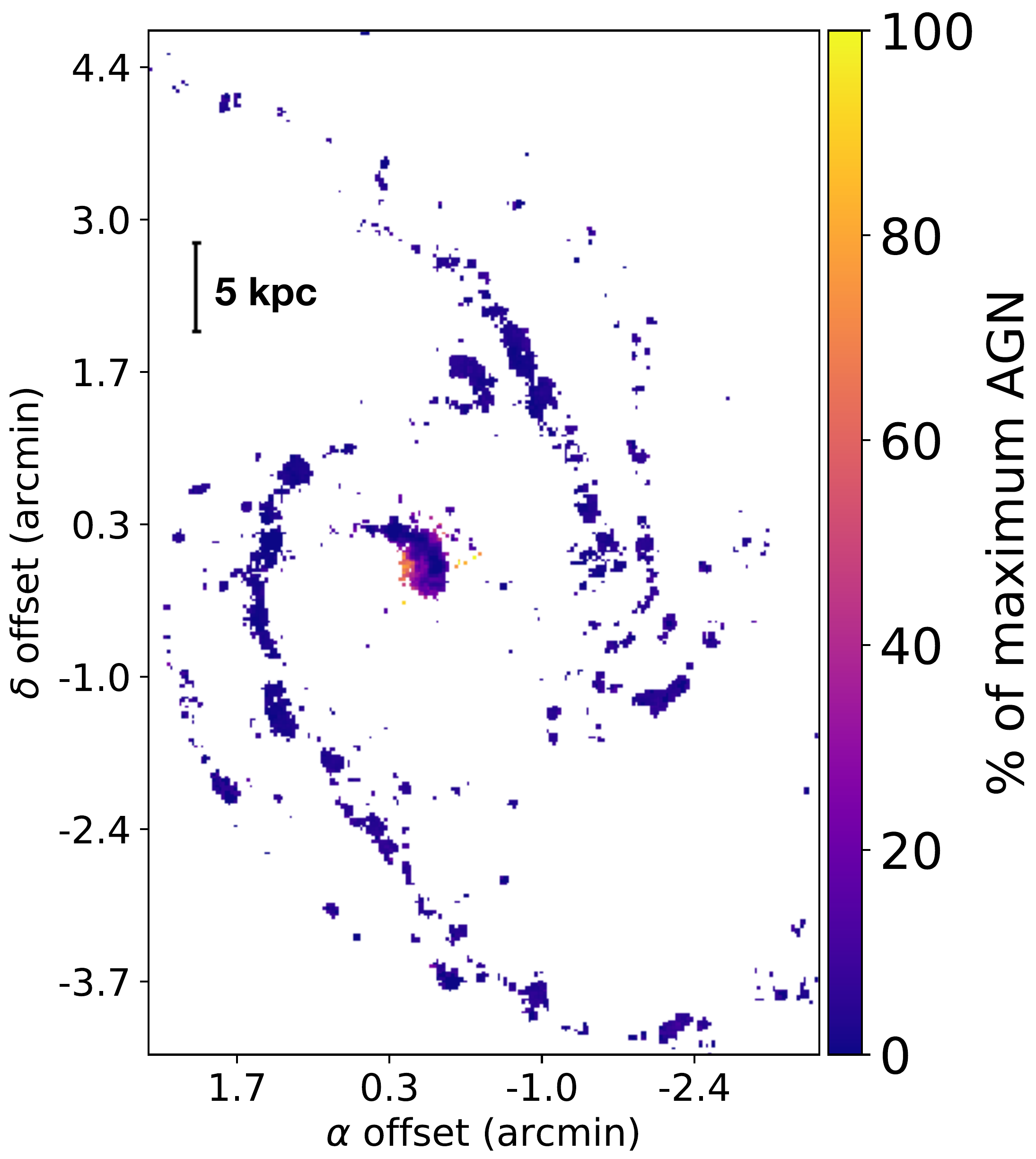}
\caption{}%[O \textsc{iii}]/H$\beta$ vs [N \textsc{ii}]/H$\alpha$ AGN fraction map.}
\label{fig:1365BPTagnmap}
\end{subfigure}%\hspace{0.1\columnwidth}
\begin{subfigure}{\columnwidth}
\includegraphics[scale=0.3]{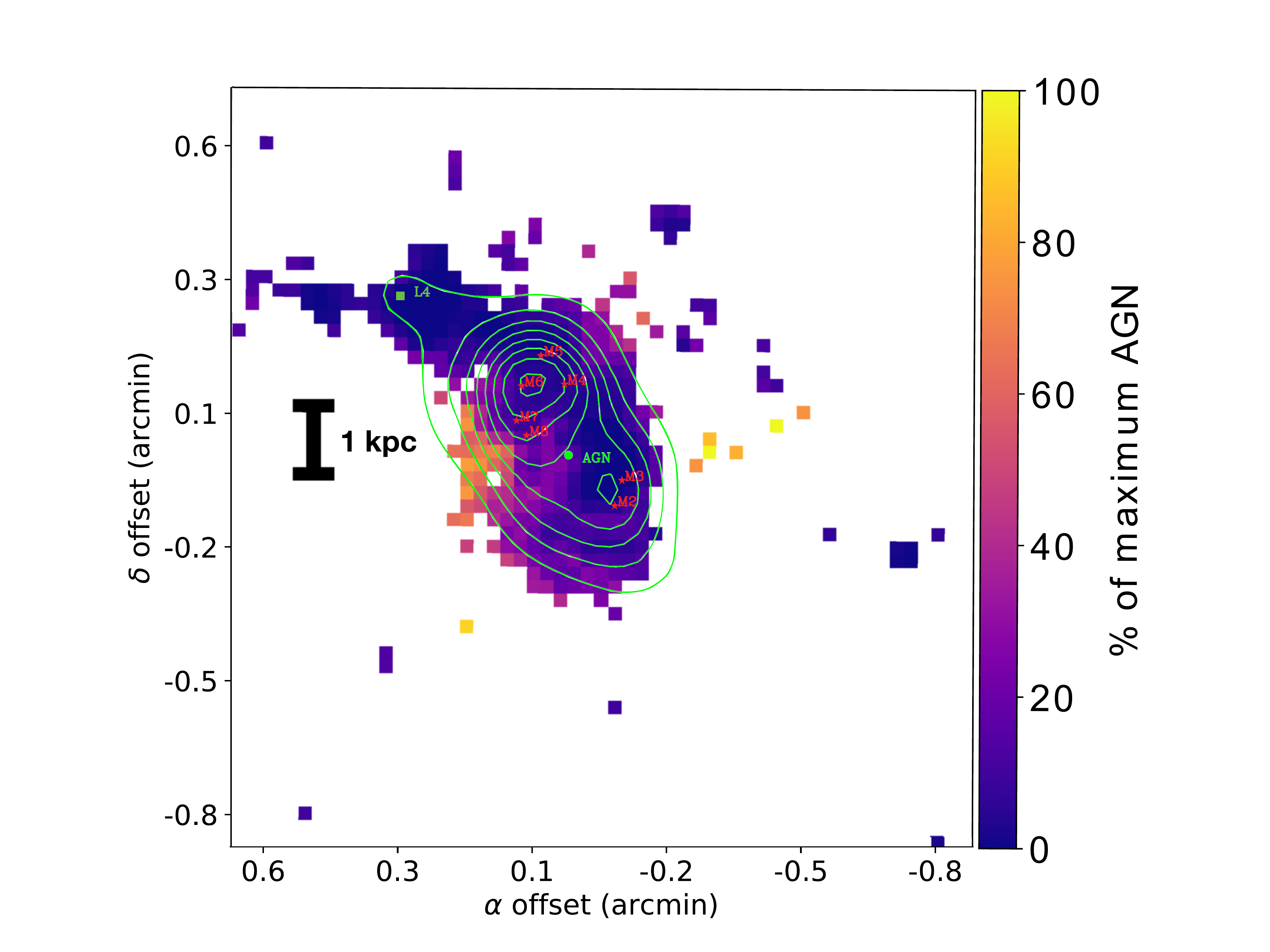}
\caption{}%[O \textsc{iii}]/H$\beta$ vs [N \textsc{ii}]/H$\alpha$ starburst-AGN fraction map of the centre of NGC 1365. A}
\label{fig:1365BPTagnmap_cz}
\end{subfigure}
%\begin{subfigure}{\columnwidth}
%\includegraphics[width=\columnwidth]{1365_SII_agnmap_normal.pdf}
%\caption{}%[O \textsc{iii}]/H$\beta$ vs [S \textsc{ii}]/H$\alpha$ starburst-AGN fraction map}
%\label{fig:1365SIIagnmap}
%\end{subfigure}
\caption{%The starburst-AGN fraction BPT diagram (a) and map (b) for NGC 1365 using the new fraction method. The basis points at the centre of the galaxy are constrained by the grids in both ionisation parameter and metallicity (green lines), whilst the basis point for the spaxels located at the edge of the galaxy is constrained solely by metallicity (orange line), and is located at the point where it fits the data.
Map of NGC 1365 (a), with spaxels coloured to the [O \textsc{iii}]/H$\beta$ vs [N \textsc{ii}]/H$\alpha$ AGN fraction. The nucleus of NGC 1365 is shown in (b), overlaid with an image from \citet{AH2012b} showing star-forming clusters identified by \citet{Galliano2005} (red stars, M2, ..., M8), the H$\alpha$ hotspot L4 from \citet{Alloin1981} and \citet{Kristen1997}, and light-green contours showing the 70$\mu$m flux distribution from Herschel PACS \citep[square-root flux scaling, flux density of 85.5 Jy in the inner 15 arcseconds;][]{AH2012b}. The contours are in a square-root intensity scale, with flux of 70$\mu$m mid-IR emission.}
\label{fig:1365map}
\end{figure*}

Our results of starburst-AGN mixing are in agreement with the AGN activity in the nucleus of NGC 1365. Seen in Figures~\ref{fig:1365BPTagnfrac} and \ref{fig:1365agnfrac_nogrids_annotated} is the BPT diagram for NGC 1365, showcasing our new AGN fraction calculation. We note the presence of spaxels found below the 100\% star formation lines, and spaxels found above the 100\% AGN line. This indicates the uncertainties found within the photoionisation models, and may arise due to systematic uncertainties in the input ionising spectrum parameters (see D'Agostino et al. in prep. for a detailed discussion on the systematic uncertainties in input ionising stellar spectrum parameters in H \textsc{ii} region modelling). Nevertheless, all spaxels which lie below the 100\% star formation lines are set to 100\% star formation, and all spaxels found above the 100\% AGN line are set to 100\% AGN. The map of the distribution of AGN fractions in NGC 1365 is shown in Figure~\ref{fig:1365BPTagnmap}. Seen in Figure~\ref{fig:1365BPTagnmap_cz} is the nucleus of NGC 1365, coloured by AGN fraction. The nucleus of NGC 1365 shows regions of high AGN activity, corresponding to high [O \textsc{iii}]/H$\beta$ and [N \textsc{ii}]/H$\alpha$ emission line ratios. However, regions of high star formation (low AGN fractions) are also seen in the nucleus, with a clean mixing evident between regions of low and high AGN fractions. Star formation in the nucleus of NGC 1365 is also confirmed in the BPT diagram of NGC 1365 coloured by radius from the centre of the galaxy, shown in Figure~\ref{fig:1365_normal_radius}. Nuclear star formation in NGC 1365 has also been found previously \citep[e.g.][and references therein]{PT1983,FN1998,Galliano2008,AH2012b}. In Figure~\ref{fig:1365BPTagnmap_cz}, we show the position of several star-forming clusters shown in \citet{AH2012b}, coinciding with regions of low AGN fractions (high star-forming emission) in the nucleus of NGC 1365. Contours of 70$\mu$m IR emission are also shown in Figure~\ref{fig:1365BPTagnmap_cz}, taken from \citet{AH2012}. The detection of mid-IR emission at 70$\mu$m \citep[and other wavelengths; see][]{AH2012} in the nucleus of NGC 1365 suggests an indirect detection of nuclear star formation, as a prominent dust lane surrounds the nuclear region of NGC 1365, obscuring a significant fraction of optical emission from our field of view \citep{AH2012}.

%\begin{figure*}%{\columnwidth}
%\includegraphics[width=\textwidth]{1365_agnfrac_normal_nogrids_pcvals_SLUG.eps}
%\caption{Starburst-AGN fraction [O \textsc{iii}]/H$\beta$ vs [N \textsc{ii}]/H$\alpha$ diagram for NGC 1365, without grids for clarity. Green dashed lines indicate 20\%, 40\%, 60\%, and 80\% AGN.}
%\label{fig:1365BPTagnfrac_nogrids}
%\end{figure*}%\hspace{\columnwidth}

AGN activity south-east of the nucleus matches the biconical outflow seen in \citet{Veilleux2003}, supporting the claim made by \citet{Veilleux2003} that the outflowing material is the result of AGN activity. Very little AGN activity, and in general structure, is seen on the north-west side of the nucleus. However, the north-west component of the [O \textsc{iii}] biconical outflow is fainter than its south-east counterpart \citep[Figure \ref{fig:oIIImaps} and][]{Veilleux2003}. 

As a source of ionisation and excitation, the AGN of NGC 1365 is relatively weak. This can be seen when considering the fraction of strong collisionally-excited emission line fluxes as a result of AGN activity, as the flux of collisionally-excited emission lines such as [O \textsc{iii}] and [N \textsc{ii}] are greatly enhanced by the extremely hard radiation field from the accretion disk of AGN \citep{Kewley2006,Kewley2013a}. We measure the fraction of [O \textsc{iii}] luminosity from AGN activity in NGC 1365 within the TYPHOON field-of-view to be 10.02 $\pm$ 0.71\%. Conversely, this implies that ${\sim} 90$\% of the total [O \textsc{iii}] luminosity in NGC 1365 within the TYPHOON field-of-view is from star formation, showing that star formation is the dominant process in NGC 1365. This is supported by the BPT diagram in Figure~\ref{fig:1365_normal_radius}, which not only shows an extensive star-forming sequence, but also shows star formation occurring in the nucleus of NGC 1365 along with AGN activity. Further, we calculate the radius at which star formation and AGN activity dominate equally (AGN fraction = 50\%) for NGC 1365 to be 1.350 $\pm$ 0.362 kpc. This calculation shows the majority of the AGN activity to be contained tightly within the nucleus with a very small radius of influence. 

We calculate the fraction of the total H$\alpha$ luminosity attributable to AGN activity in NGC 1365 within the TYPHOON field-of-view to be 5.12 $\pm$ 0.07\%. As noted in \citet{Davies2014a}, the AGN fractions in galaxies which display both processes indicate the degree of error imposed when using H$\alpha$ as a star formation rate (SFR) indicator. In the case of NGC 1365, SFRs calculated for NGC 1365 using H$\alpha$ can be considered largely accurate, as we observe ${\sim} 95$\% of the total H$\alpha$ luminosity to be from star formation. At most, an overestimate of ${\sim} 5$\% is expected when calculating SFRs in NGC 1365.

% Biconical outflow

%AGN activity on the south-east side of the nucleus matches the biconical outflow seen in \citet{Veilleux2003}. However, very little AGN activity, and in general structure, is seen on the north-west side of the nucleus.

%Seen in Figure~\ref{fig:1365BPTagnfrac}, our new method for calculating the starburst-AGN fraction gives more realistic results in the star-forming regime than previously, when compared with Figure~\ref{fig:1365agnmap_old}. The map of the starburst-AGN fraction distribution across the galaxy can be seen in Fig.~\ref{fig:1365BPTagnmap}. AGN activity on the south-east side of the nucleus matches the biconical outflow seen in \citet{Veilleux2003}. However, very little AGN activity, and in general structure, is seen on the north-west side of the nucleus. %The AGN activity present in the nucleus appears overall to be roughly 70\% of the maximum AGN activity measured in this galaxy, with the spaxel containing 100\% of the maximum AGN activity located slightly off the nucleus towards the west of the galaxy. 

\subsubsection{NGC 1068}
\label{sec:1068}
% Can talk about specifics of both galaxies; mention strangeness of 1068 (ionisation echo, Hanny's voorwerp)

\begin{figure*}
\centering
\includegraphics[width=\textwidth]{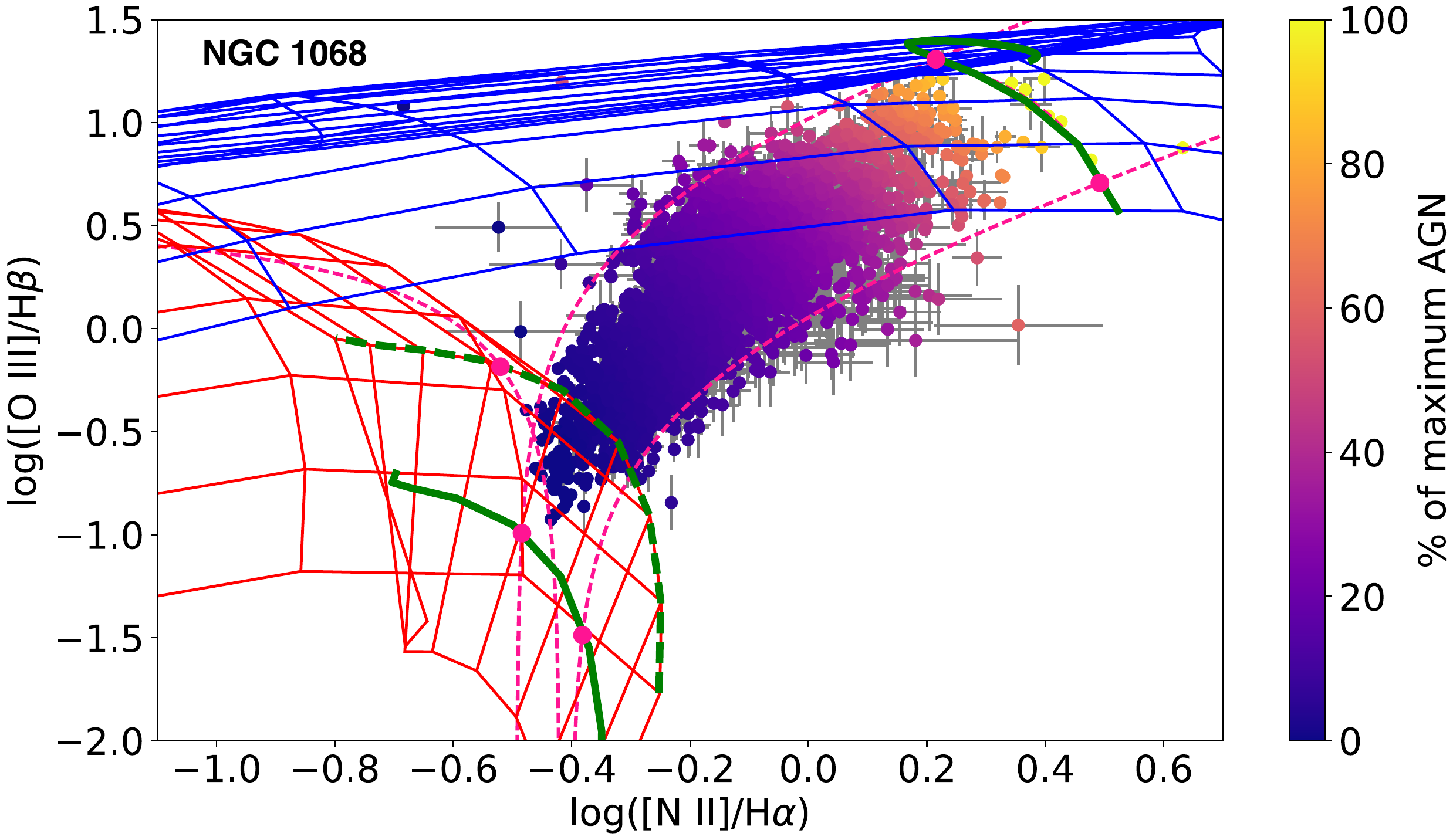}
\caption{AGN fraction [O \textsc{iii}]/H$\beta$ vs [N \textsc{ii}]/H$\alpha$ diagram for NGC 1068.}
\label{fig:1068BPTagnfrac}
\end{figure*}

\begin{figure*}%{\textwidth}
\includegraphics[width=\textwidth]{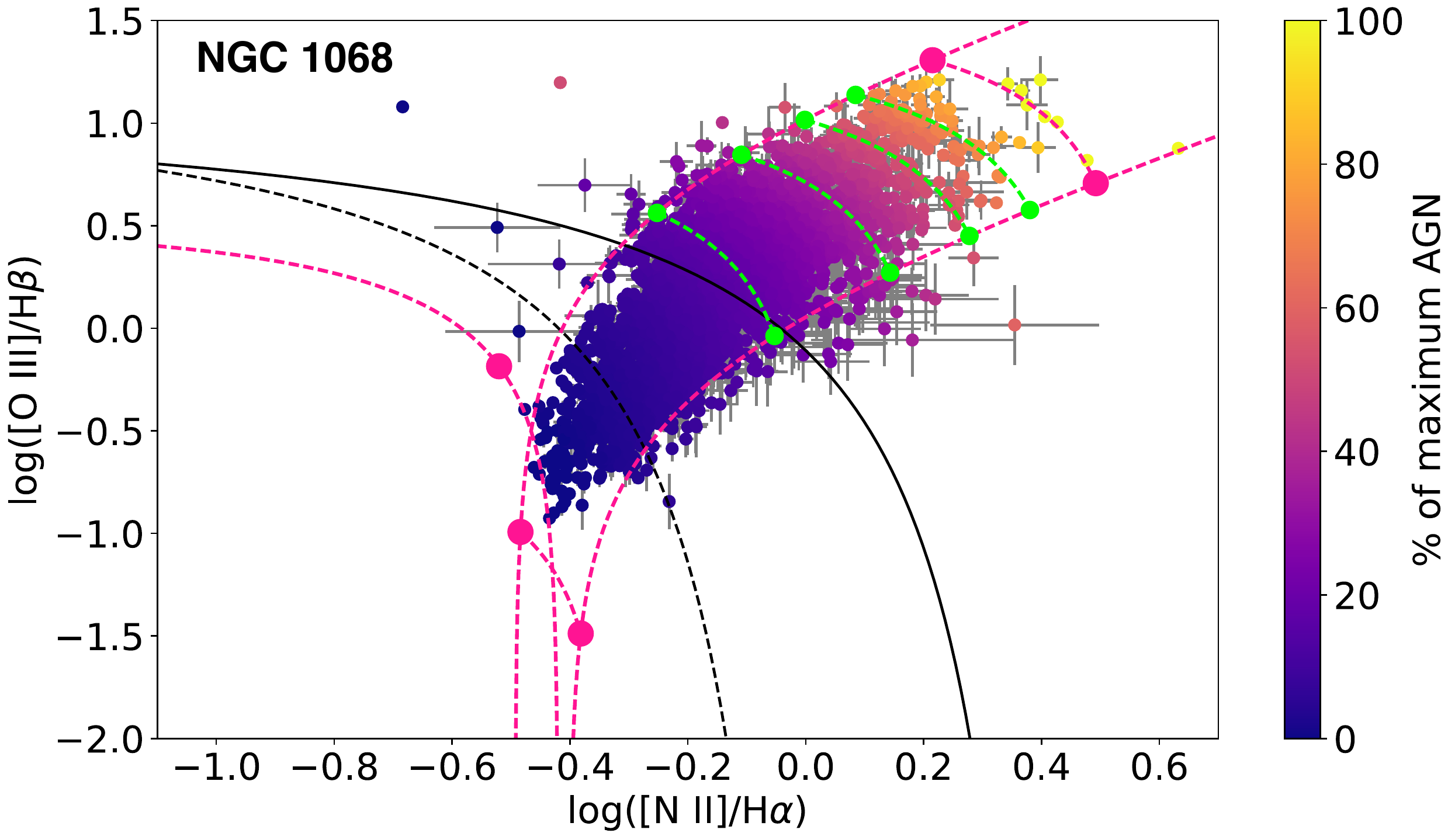}
\caption{AGN fraction [O \textsc{iii}]/H$\beta$ vs [N \textsc{ii}]/H$\alpha$ diagram, without grids for clarity. Green dashed lines indicate 20\%, 40\%, 60\%, and 80\% AGN.}
\label{fig:1068BPTagnfrac_nogrids}
\end{figure*}

\begin{figure*}
\centering
\begin{subfigure}{\columnwidth}
\includegraphics[width=\textwidth]{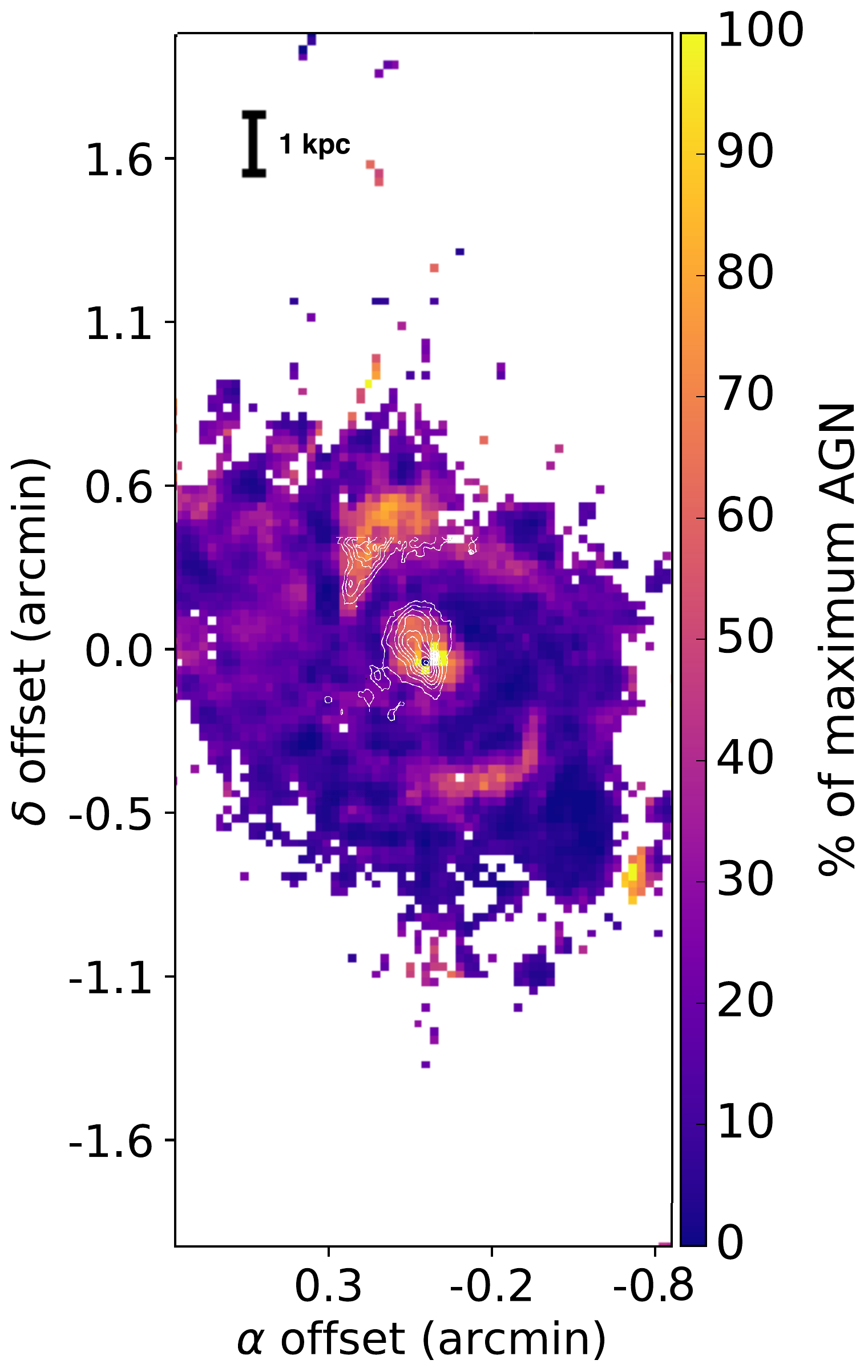}
\caption{}%[O \textsc{iii}]/H$\beta$ vs [N \textsc{ii}]/H$\alpha$ starburst-AGN fraction map, showing the [O \textsc{iii}]/(H$\alpha$ + [N \textsc{ii}]) contours from \citet{Pogge1988} in white.}
\label{fig:1068BPTagnmap}
\end{subfigure}
\begin{subfigure}{\columnwidth}
\includegraphics[scale=0.55]{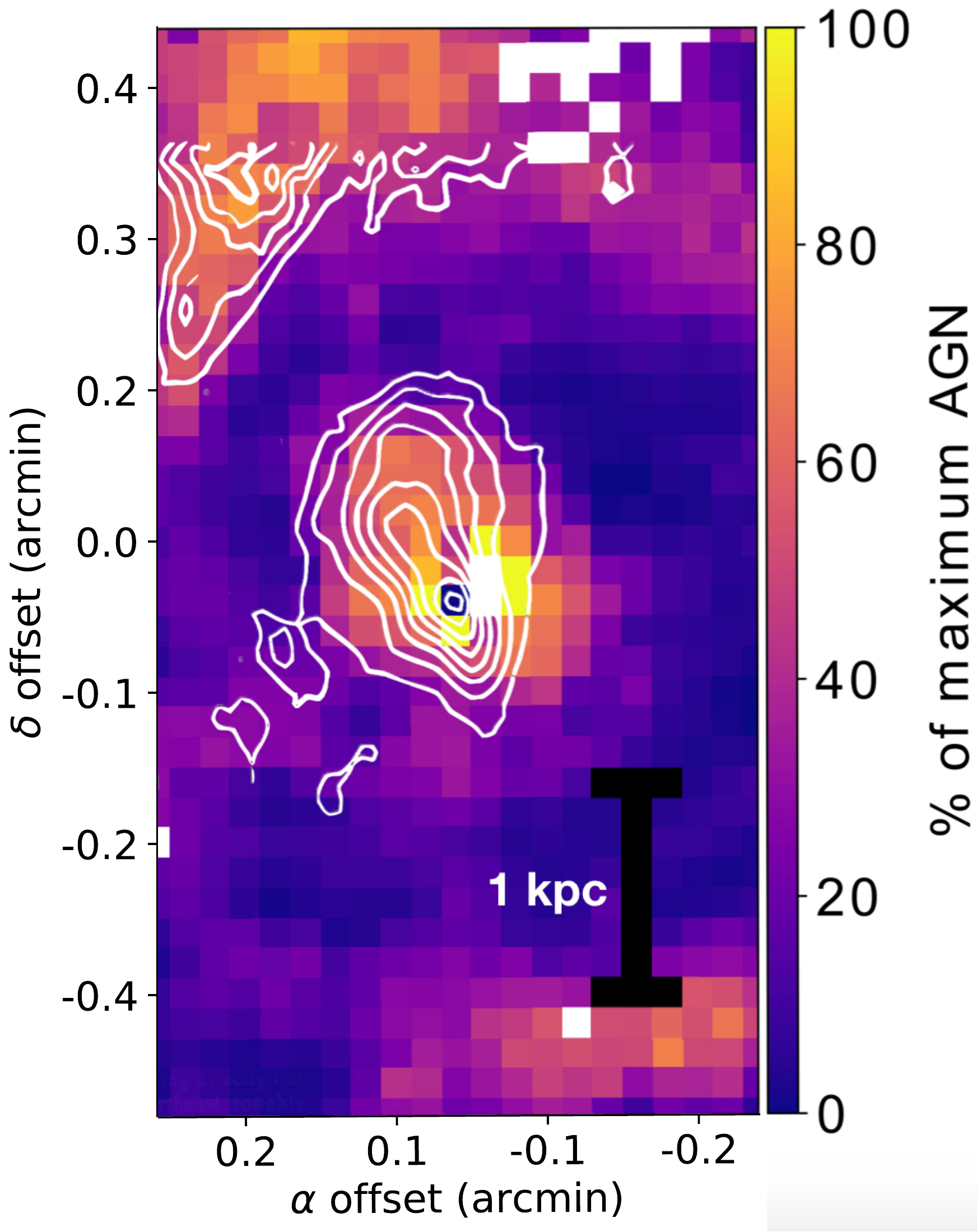}
\caption{}%[O \textsc{iii}]/H$\beta$ vs [N \textsc{ii}]/H$\alpha$ starburst-AGN fraction map of the centre of NGC 1068.}
\label{fig:1068BPTagnmap_cz}
\end{subfigure}
%\begin{subfigure}{\columnwidth}
%\includegraphics[width=\columnwidth]{1365_SII_agnmap_normal.pdf}
%\caption{[O \textsc{iii}]/H$\beta$ vs [S \textsc{ii}]/H$\alpha$ starburst-AGN fraction map}
%\label{fig:1365SIIagnmap}
%\end{subfigure}
\caption{%The starburst-AGN fraction BPT diagram (a) and map (b) for NGC 1365 using the new fraction method. The basis points at the centre of the galaxy are constrained by the grids in both ionisation parameter and metallicity (green lines), whilst the basis point for the spaxels located at the edge of the galaxy is constrained solely by metallicity (orange line), and is located at the point where it fits the data.
Map of NGC 1068 (a), with spaxels coloured to the O \textsc{iii}]/H$\beta$ vs [N \textsc{ii}]/H$\alpha$ AGN fraction. The nucleus of NGC 1068 is shown in (b). The [O \textsc{iii}]/(H$\alpha$ + [N \textsc{ii}]) contours from \citet{Pogge1988} are shown on NGC 1068 in (a) and (b) in white.}
\label{fig:1068}
\end{figure*}

Figures~\ref{fig:1068BPTagnfrac} and~\ref{fig:1068BPTagnfrac_nogrids} show the BPT diagram and subsequent AGN fraction calculation for NGC 1068. The map of the AGN fraction for NGC 1068 is shown in Figure~\ref{fig:1068}. Our findings support the claims made by \citet{Cecil2002} and \citet{Dopita2002b} regarding the driving force behind the large-scale outflows seen in NGC 1068. Material with a very high AGN fraction is seen in Figure~\ref{fig:1068BPTagnmap} to both the north-east of the galaxy, and to the south-west through openings in the galactic disk, coinciding with the biconical structure shown in \citet{Das2006}. This suggests that the large-scale outflowing plume of [O \textsc{iii}] seen in \citet{Pogge1988} is the result of AGN-driven activity (shown in Figure~\ref{fig:1068BPTagnmap}). %\citet{Das2006} eliminate the radio jet seen emanating from NGC 1068 as a possible source of the large-scale outflow.

%\hspace{\textwidth}

%\hspace{\columnwidth}

AGN-driven outflow from NGC 1068 has also been concluded from results in radio wavelengths. Using data from ALMA, \citet{GB2014} map emission from a set of dense molecular gas tracers, namely CO(3-2), CO(6-5), HCN(4-3), HCO$^+$
(4-3), and CS(7-6). Within the inner 50 -- 400 pc, they report a massive outflow in all molecular gas tracers of $M_{\mathrm{mol}} = 2.7^{+0.9}_{-1.2} \times 10^7 M_\odot$. They also show a tight correlation between the ionised gas outflow, the radio jet, and outward motions within the circumnuclear disk of NGC 1068, suggesting the outflow is AGN-driven. They strengthen this suggestion by noting the outflow rate in the circumnuclear disk is $63^{+21}_{-37} M_\odot$ yr$^{-1}$, much larger than the star formation rate calculated for NGC 1068 at these radii. 

Star formation within the nucleus of NGC 1068 has been seen previously, with \citet{SB2012} locating young stars within the central 100 pc. The nuclear star formation rate in NGC 1068 is quite modest however, with estimates of $SFR_{\mathrm{nuclear}} {\sim} 0.4-0.7 M_\odot$ yr$^{-1}$ out to a radius of 35 pc \citep{Davies2007}, $SFR_{\mathrm{nuclear}} {\sim} 0.4 M_\odot$ yr$^{-1}$ out to a radius of 12 pc \citep{Esquej2014}, and $SFR_{\mathrm{nuclear}} {\sim} 1 M_\odot$ yr$^{-1}$ out to a radius of 140 pc.

Further out from the nucleus, up to radii of a few kpc, the star formation rate in NGC 1068 has been shown to be enormous. \citet{Thronson1989} calculate a SFR of ${\sim} 100 M_\odot$ yr$^{-1}$ in the ring surrounding the circumnuclear disk. The starburst ring in NGC 1068 can be seen in our results in Figure~\ref{fig:1068BPTagnmap_cz} as a ring of spaxels with low AGN fraction surrounding the central region containing high AGN fractions. This large SFR is likely a consequence of the very large molecular mass in the circumnuclear region (${\sim} 2-6 \times 10^9 M_\odot$), which appears to have been confined to the circumnuclear region by the bar at the centre of NGC 1068. As a result, star formation in NGC 1068 is seen to primarily take place in the star-forming ring surrounding the galaxy's central region. Hence, observations showing extensive star formation towards the outermost regions of NGC 1068 have been rare. This finding is reflected in the BPT diagram of NGC 1068 shown in Figure~\ref{fig:1068BPTagnfrac} which lacks a large star-forming sequence like that seen in Figure~\ref{fig:1365BPTagnfrac} for NGC 1365. Our results suggest that star formation as a source of ionisation and excitation weakens at radii $\leq 2.246 \pm 1.277$ kpc, as these radii contain spaxels with AGN fractions greater than 50\%.

We show star formation nevertheless is influential in the total emission from NGC 1068, despite the lack of a star-forming sequence and a relatively low nuclear SFR. Presumably a result of the enormous SFR surrounding the circumnuclear disk, we calculate roughly ${\sim} 59$\% of the total [O \textsc{iii}] luminosity to be from star formation within the TYPHOON field-of-view (shown as an AGN fraction of 41.68 $\pm$ 0.28\% in Table~\ref{tab:agnfracs}). Our calculations in Table~\ref{tab:agnfracs} also show star formation to be the dominant process in the emission of all collisionally-excited lines considered within the TYPHOON field-of-view, with between ${\sim} 60-70$\% of the total luminosity in the lines [O \textsc{iii}]$\lambda$5007 (${\sim} 58$\%), [O \textsc{ii}]$\lambda \lambda$3726,3729 (${\sim} 68$\%), [S \textsc{ii}]$\lambda \lambda$6716,6731(${\sim} 67$\%) and [N \textsc{ii}]$\lambda$6854 (${\sim} 59$\%) attributable to star formation rather than AGN emission.

We claim the use of the total H$\alpha$ luminosity as a SFR indicator in NGC 1068 will overestimate the SFR by ${\sim} 25$\%. Shown in Table~\ref{tab:agnfracs}, the luminosity of H$\alpha$ from AGN activity is calculated to be 24.38 $\pm$ 0.12\%, implying only ${\sim} 75$\% of the luminosity from H$\alpha$ in NGC 1068 is from star formation.

\section{Investigating the effect of spatial resolution}
\label{sec:spatres}

We investigate the AGN fraction and other properties as a function of spatial resolution. We apply the same calculations and methods described in Section~\ref{sec:agnfrac} to datacubes rebinned to 330, 500, and 1000 pc/pixel; resolutions found in other IFU surveys such as MaNGA, SAMI and CALIFA. 

%Since the basis points are intended to represent pure star formation and pure AGN, the position of the points remains consistent throughout the rebinning process. 

%Figures~\ref{fig:1365spatres_frac} and \ref{fig:1068spatres_frac} show the BPT diagram colour-coded by radius and the corresponding starburst-AGN fraction for both NGC 1365 and NGC 1068 respectively. The starburst-AGN fraction galaxy maps are shown for both galaxies in Figures~\ref{fig:1365spatres_map} and \ref{fig:1068spatres_map}, and nucleus maps are shown in Figures~\ref{fig:1365spatres_map_cz} and \ref{fig:1068spatres_map_cz} for NGC 1365 and NGC 1068 respectively.

%The basis point locations remain consistent amongst the different datacube resolutions seen in Figures~\ref{fig:1365BPTagnfrac} and~\ref{fig:1068BPTagnfrac}, despite the changing metallicity gradient. %In addition to the metallicity and ionisation parameter spread, the true level of AGN dominance may also be miscalculated. Shown in Fig.~\ref{fig:agnfracpc}, a higher percentage of spaxels appear to be AGN dominated the further the resolution is decreased. 

\subsection{Effect of resolution on the AGN fraction}
\label{sec:agnfracres}
% points to include:
% Might not be shocks, maybe DIG
% AGN position may be dependent on resolution (e.g. SIIHa 500pc)
% Index of 1.2 rather than 1.4 because better fit to data?

The AGN fraction calculations for NGC 1365 and NGC 1068 at differing resolutions are shown in Figures~\ref{fig:1365spatres_frac} and~\ref{fig:1068spatres_frac} as the BPT diagrams and in Figures~\ref{fig:1365spatres_map} and~\ref{fig:1068spatres_map} as maps of the galaxies. Figures~\ref{fig:1365spatres_map_cz} and~\ref{fig:1068spatres_map_cz} show maps of the AGN fraction in the nuclear regions of NGC 1365 and NGC 1068 respectively. 

The basis point positions remain the same across all rebinned data for consistency, set to the values shown in Table~\ref{tab:derived}. However, we do note following the work of \citet{Poetrodjojo}, the metallicity gradient of a galaxy flattens with coarser resolution. Hence, if the metallicity gradient for both galaxies was recalculated using the rebinned data, the estimate for the central and outer metallicities would differ for each datacube, resulting in different positions for the basis points. We avoid this issue by noting the basis points are intended to represent line ratios associated with 100\% star formation and 100\% AGN emission. Binning together spaxels of AGN and star formation combines the processes in these spaxels, raising the overall contribution from AGN activity in the star-forming spaxels, and vice-versa. The result is a diminishing spread of spaxels on the BPT diagram with coarser resolution. Hence, as the resolution coarsens, the likelihood of a spaxel containing 100\% emission from AGN activity or 100\% emission from star-formation continues to decrease. This spread of spaxels on the BPT diagram is highest with the native resolution of both galaxies, suggesting the best indication of 100\% AGN emission and 100\% star-forming emission is provided by the highest-resolution datacubes available.

\begin{figure*}
\centering
\includegraphics[scale=0.5]{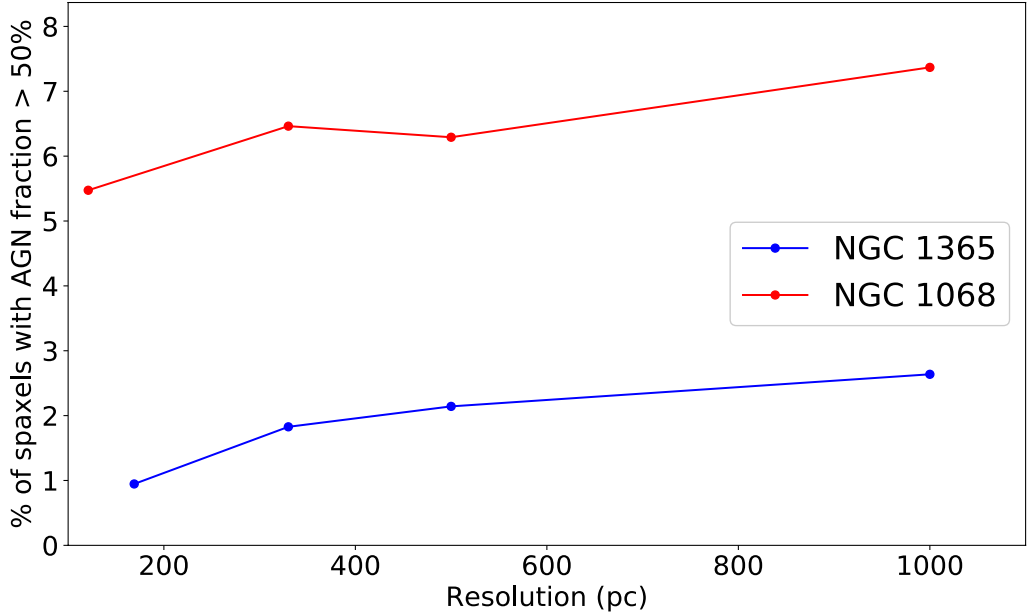}
\caption{The percentage of spaxels in the BPTs of both NGC 1365 and NGC 1068 with AGN fractions of greater than 50\% as a function of spatial resolution. The four markers indicate (in order): native resolution (169pc and 121pc for NGC 1365 and NGC 1068 respectively), 330pc, 500pc, and 1000pc (1kpc).}
\label{fig:agnfracpc}
\end{figure*}

\begin{figure*}
\centering
\begin{subfigure}{\columnwidth}
\centering
\includegraphics[width=\columnwidth]{1365_ms_normal_new_name.pdf}
\caption{BPT, Native resolution (169 pc/pix)}
\label{fig:1365bptnormal}
\end{subfigure}%\hspace{\columnwidth}
\begin{subfigure}{\columnwidth}
\centering
\includegraphics[width=\columnwidth]{1365_agnfrac_normal_new_name.pdf}
\caption{AGN fraction, Native resolution (169 pc/pix)}
\label{fig:1365fracnormal}
\end{subfigure}\hspace{\columnwidth}
\begin{subfigure}{\columnwidth}
\centering
\includegraphics[width=\columnwidth]{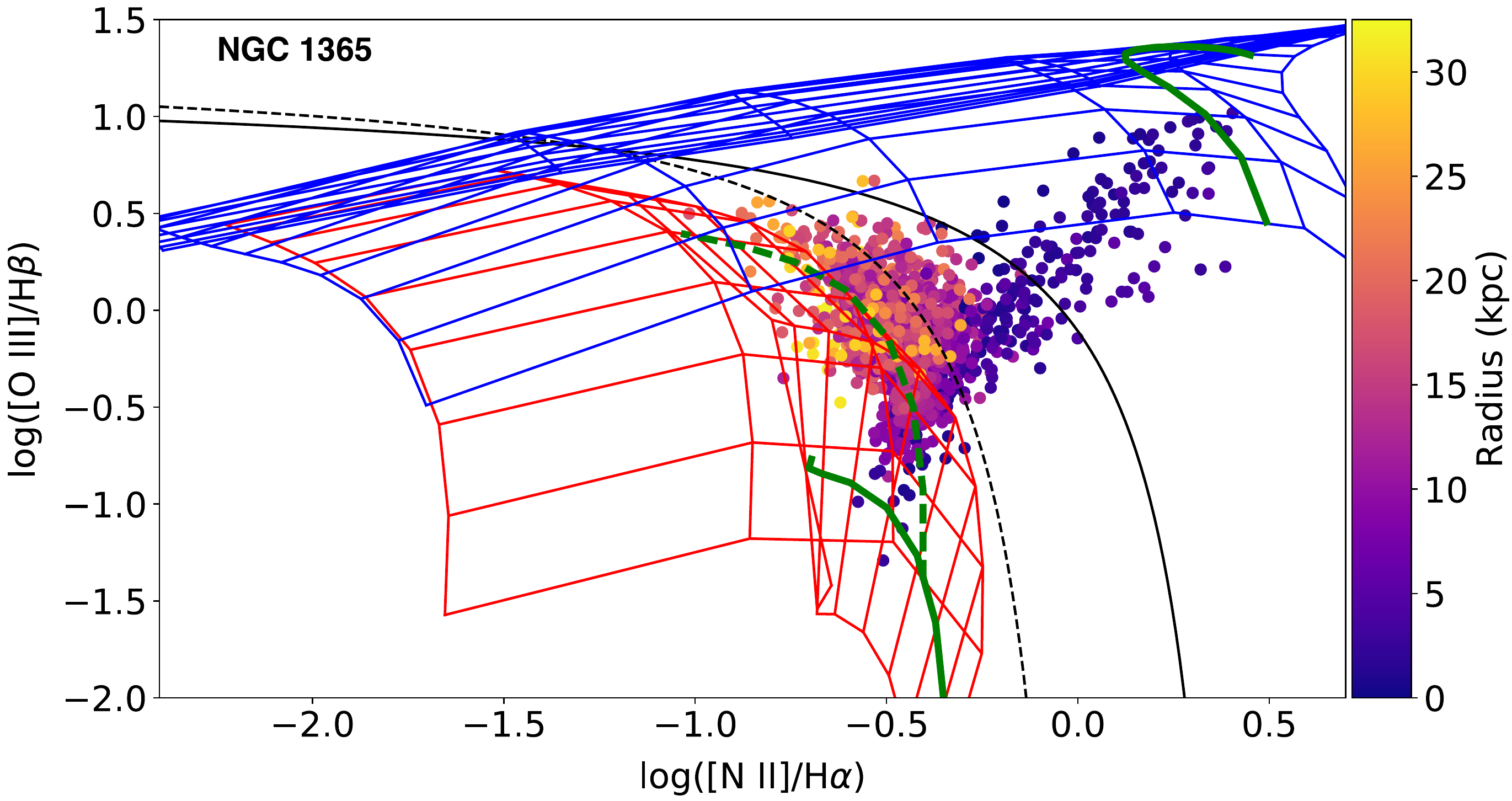}
\caption{BPT, 330 pc/pix}
\label{fig:1365bpt330}
\end{subfigure}
\begin{subfigure}{\columnwidth}
\centering
\includegraphics[width=\columnwidth]{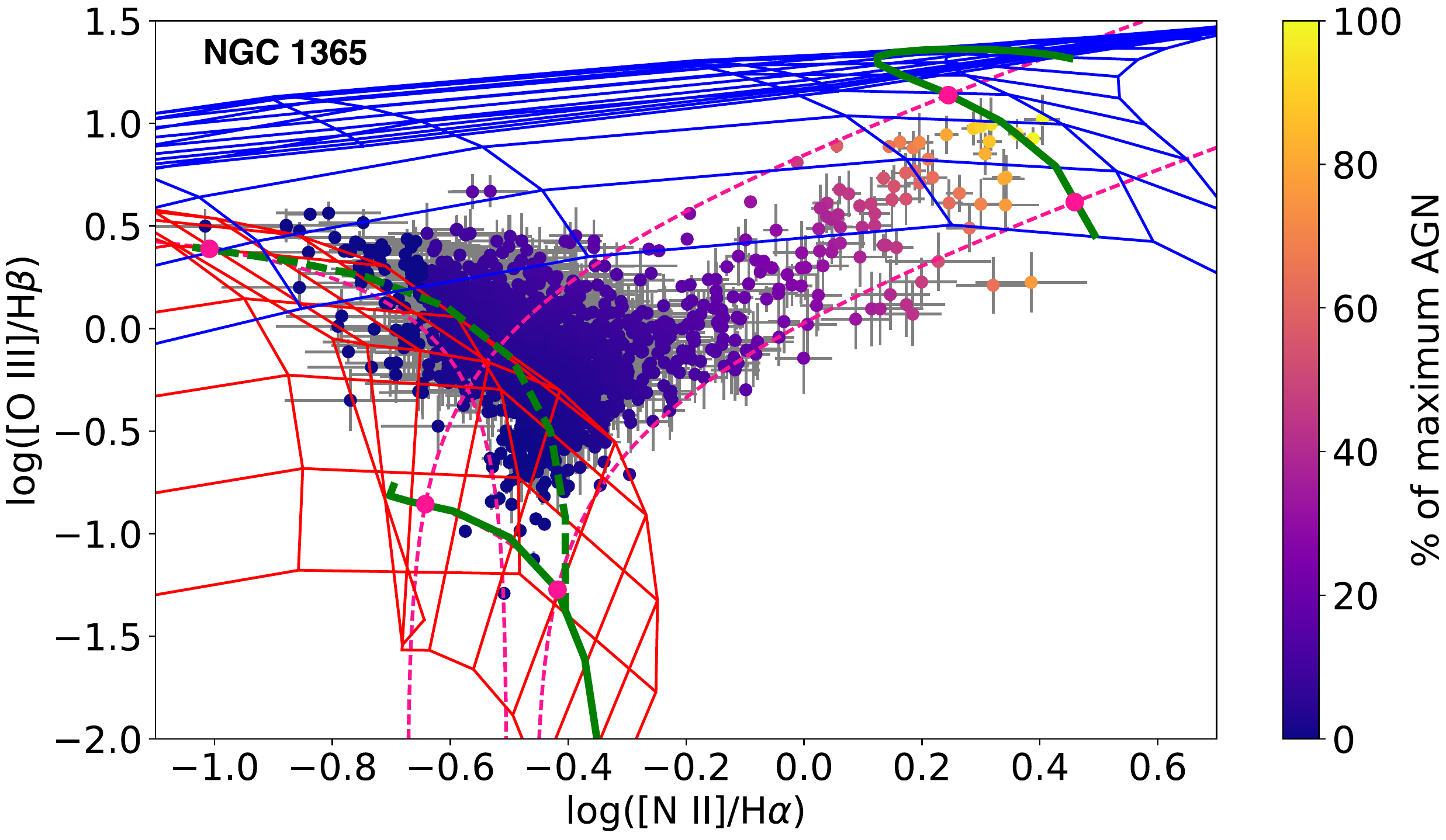}
\caption{AGN fraction, 330 pc/pix}
\label{fig:1365frac330}
\end{subfigure}
\begin{subfigure}{\columnwidth}
\centering
\includegraphics[width=\columnwidth]{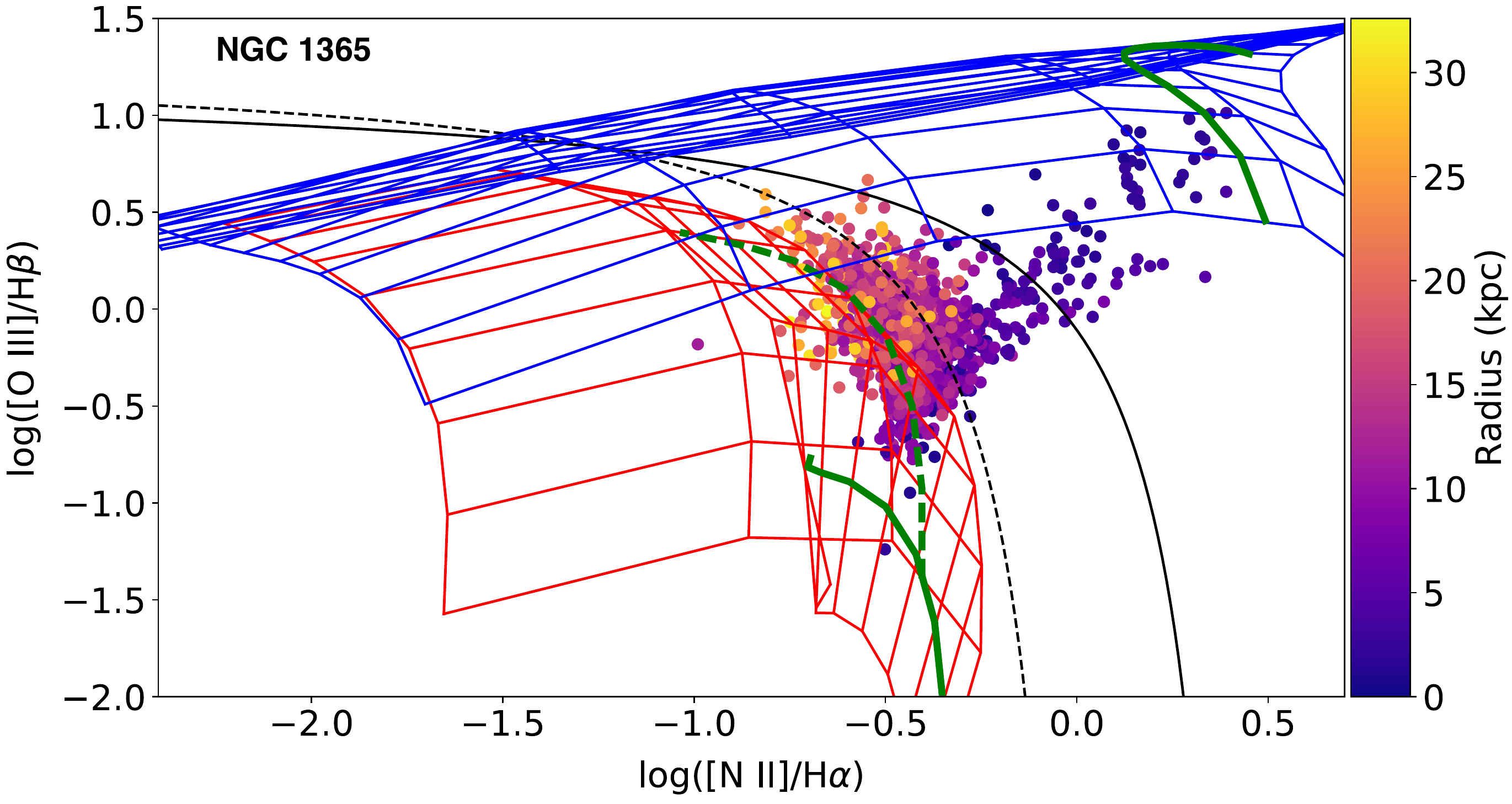}
\caption{BPT, 500 pc/pix}
\label{fig:1365bpt500}
\end{subfigure}%\hspace{\columnwidth}
\begin{subfigure}{\columnwidth}
\centering
\includegraphics[width=\columnwidth]{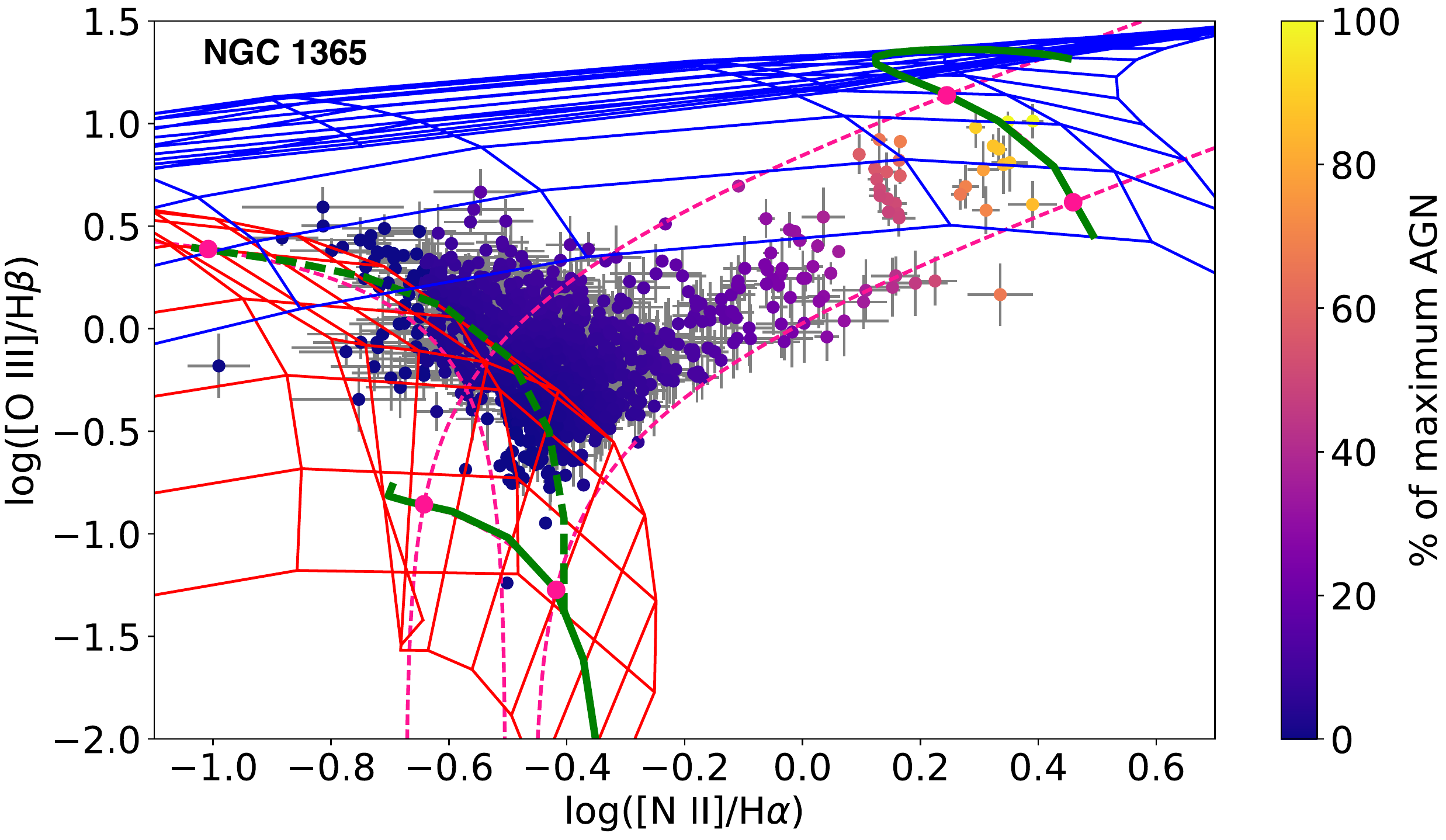}
\caption{AGN fraction, 500 pc/pix}
\label{fig:1365frac500}
\end{subfigure}\hspace{\columnwidth}
\begin{subfigure}{\columnwidth}
\centering
\includegraphics[width=\columnwidth]{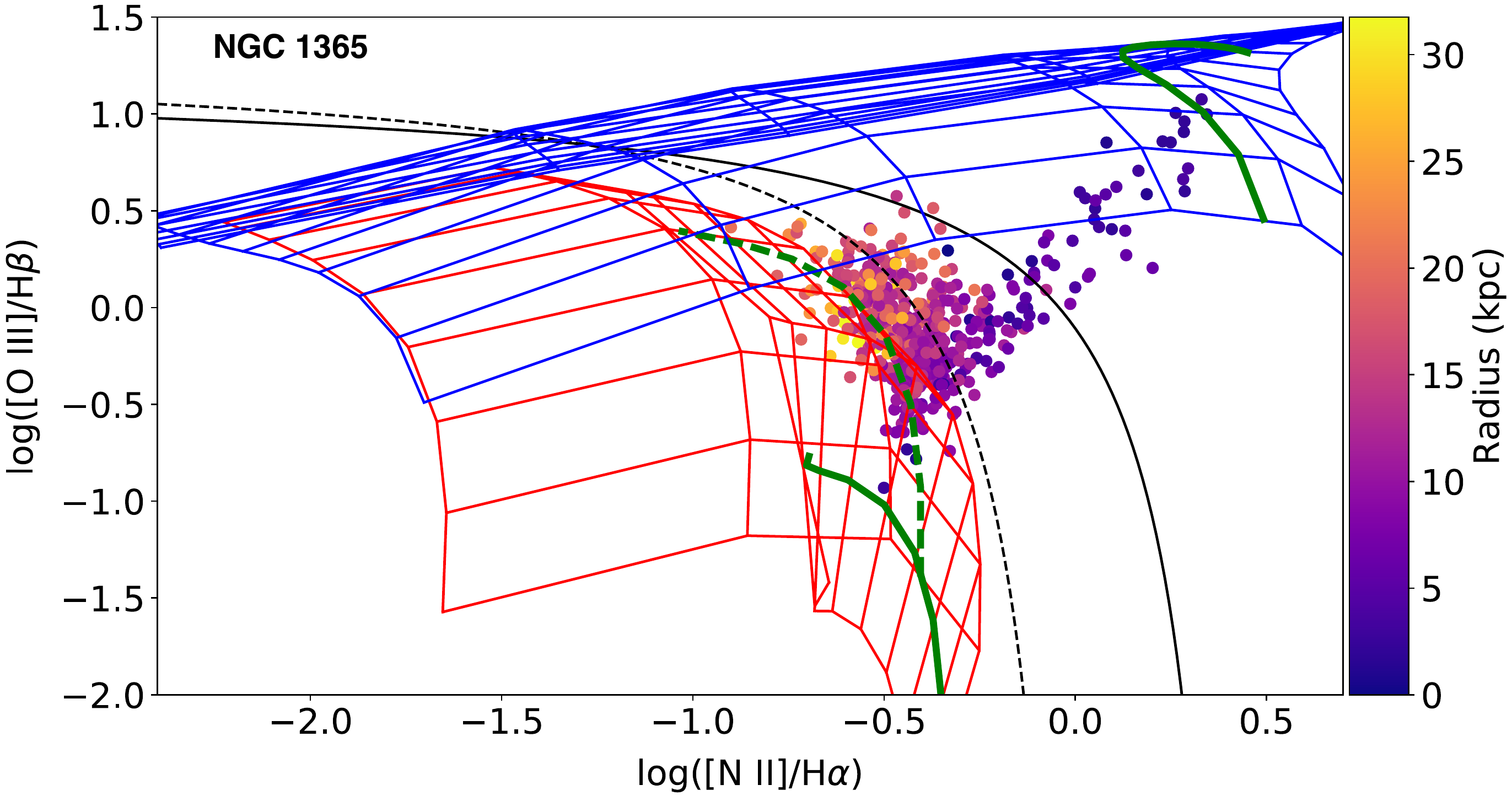}
\caption{BPT, 1 kpc/pix}
\label{fig:1365bpt1k}
\end{subfigure}
%\caption{The BPT diagram for NGC 1365, after rebinning to lower spatial resolutions.}
\begin{subfigure}{\columnwidth}
\centering
\includegraphics[width=\columnwidth]{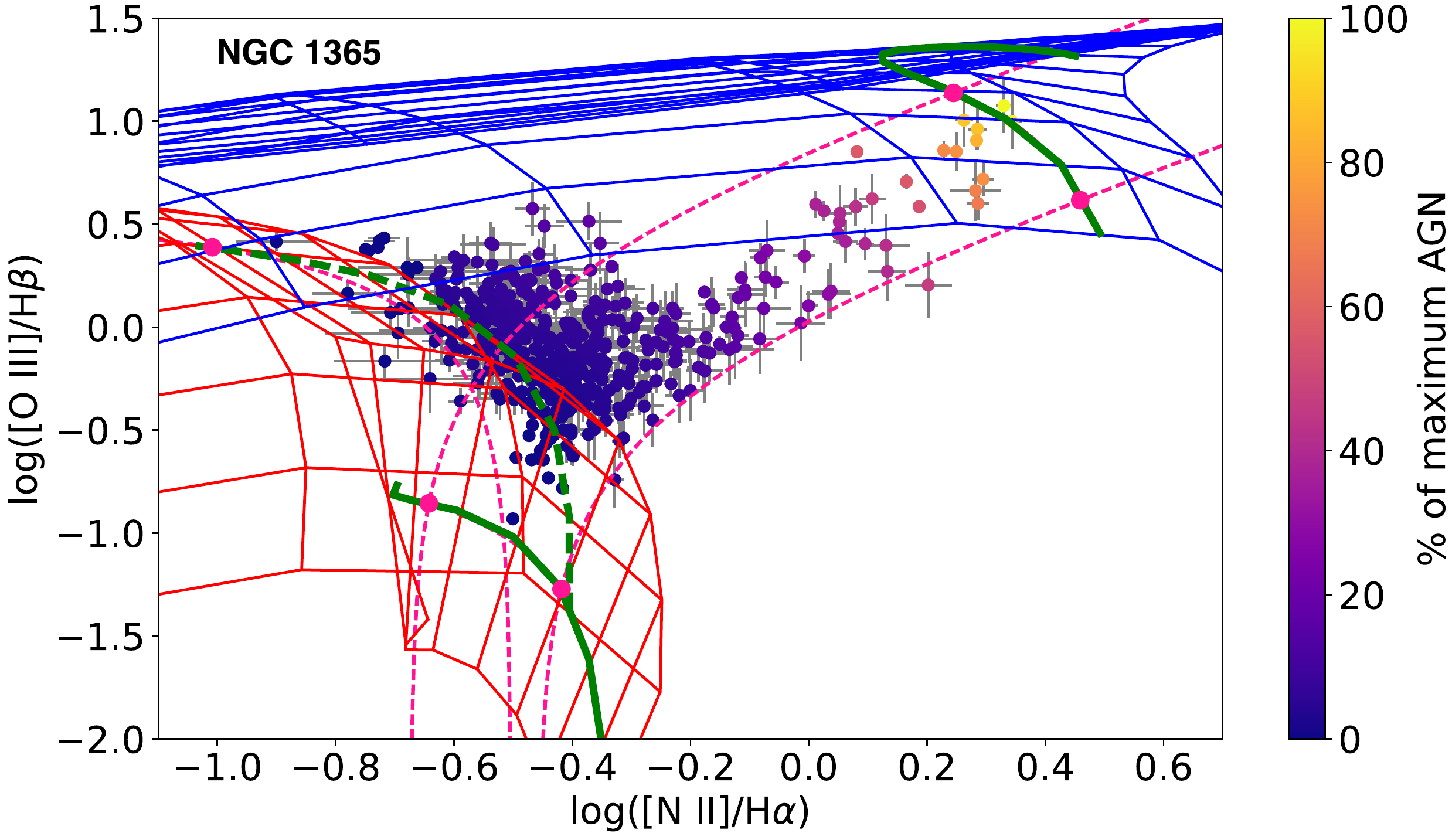}
\caption{AGN fraction, 1 kpc/pix}
\label{fig:1365frac1k}
\end{subfigure}
\caption{The AGN fractions and BPT diagrams coloured by radius for NGC 1365, after rebinning to lower spatial resolutions. Errors for the spaxels on the BPT diagrams are the same as those on the AGN fraction plots.}
\label{fig:1365spatres_frac}
\end{figure*}

\begin{figure*}
\centering
\begin{subfigure}{\columnwidth}
\centering
\includegraphics[scale=0.32]{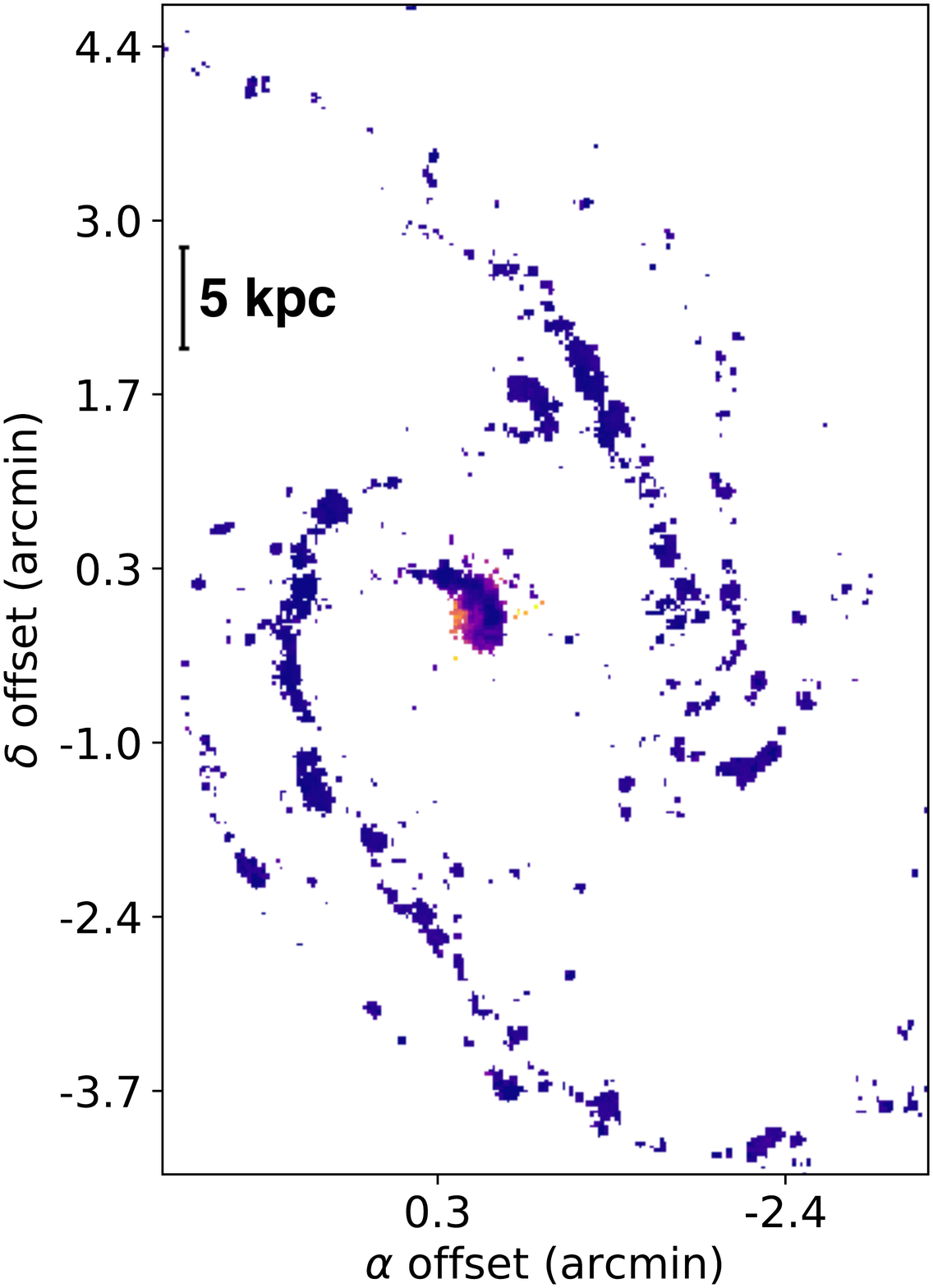}
\caption{Native resolution (169 pc/pix)}
\label{fig:1365mapnormal}
\end{subfigure}%\hspace{\columnwidth}
\begin{subfigure}{\columnwidth}
\includegraphics[width=\columnwidth]{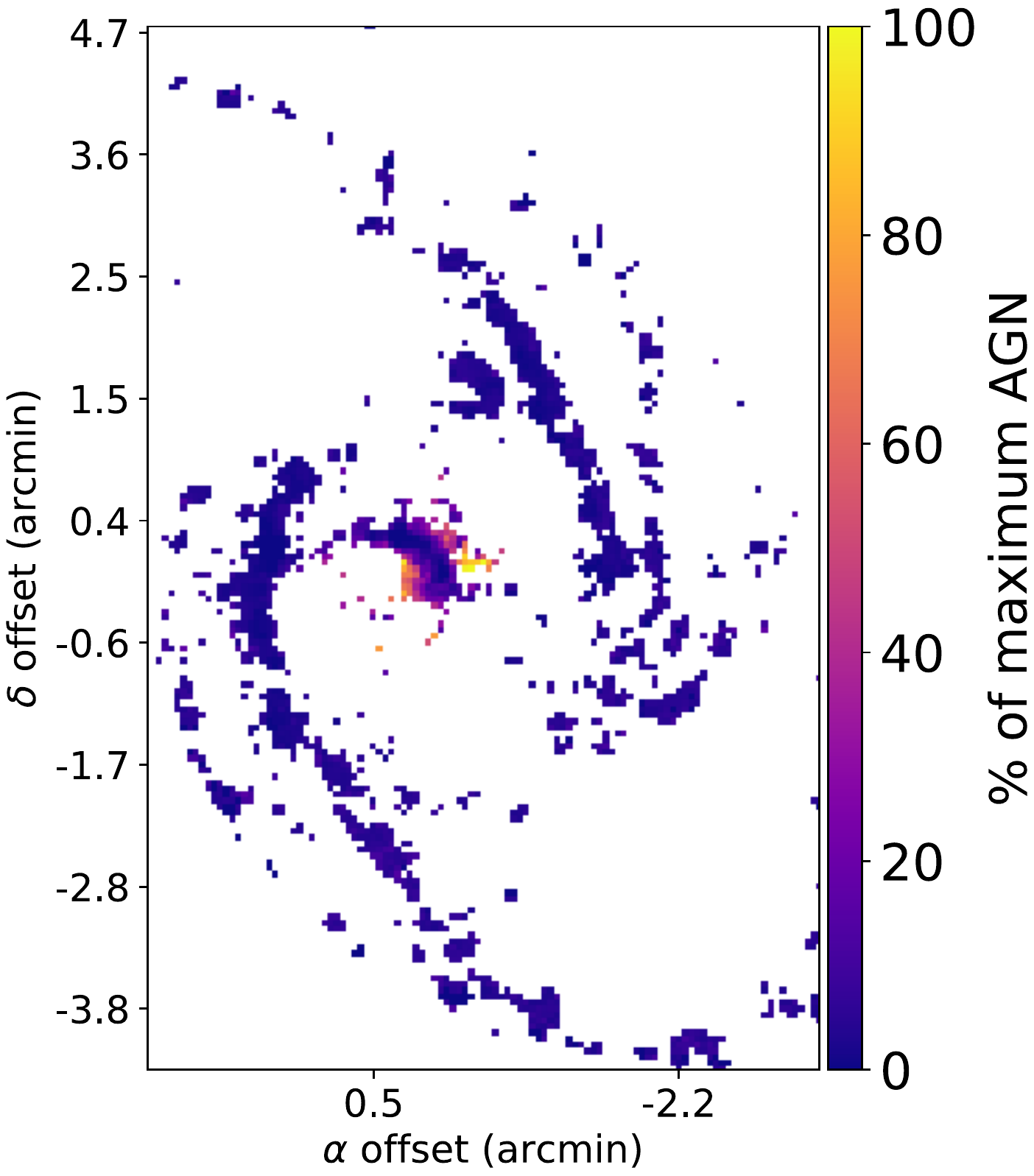}
\caption{330 pc/pix}
\label{fig:1365map330}
\end{subfigure}
\begin{subfigure}{\columnwidth}
\centering
\includegraphics[scale=0.59]{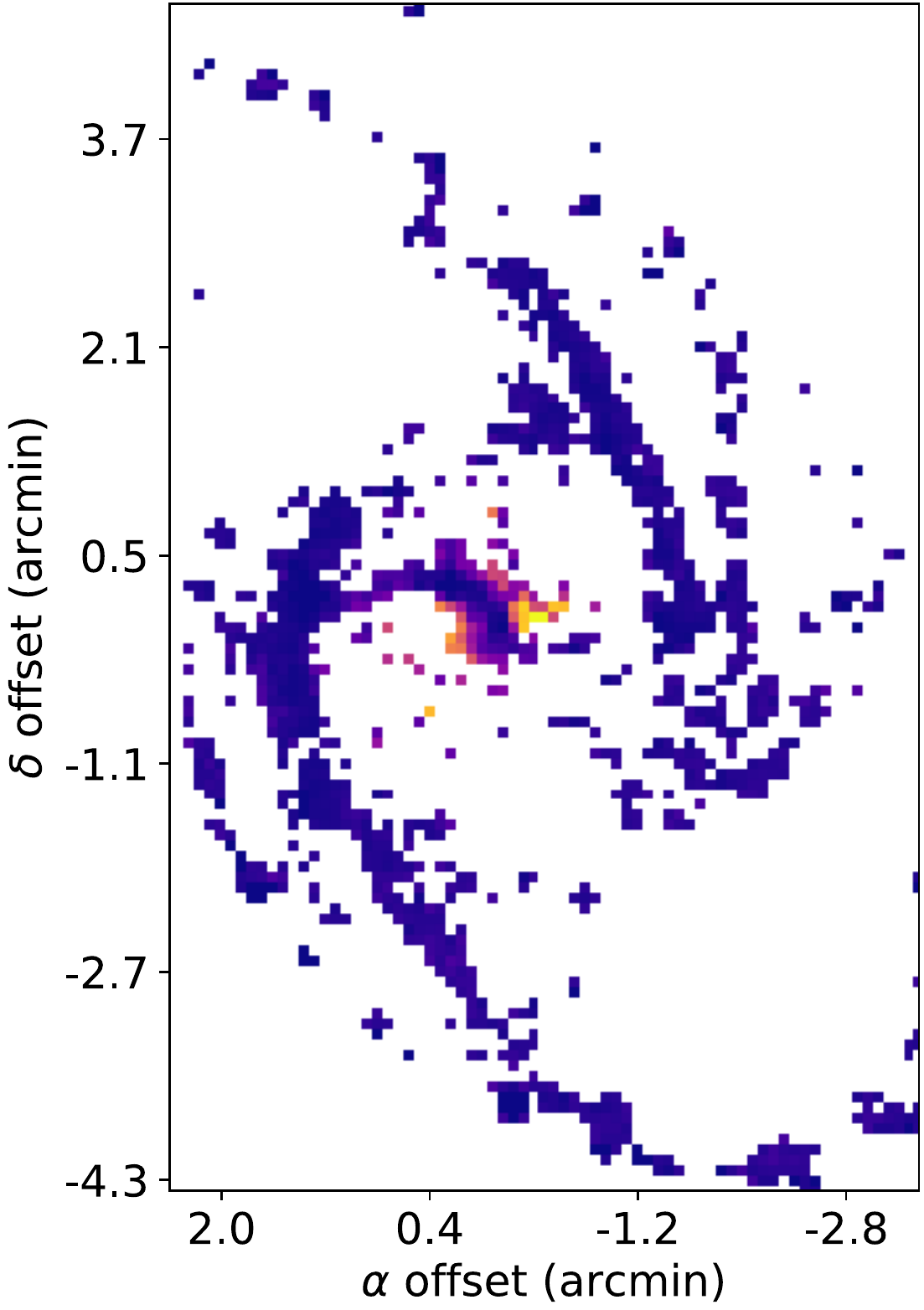}
\caption{500 pc/pix}
\label{fig:1365map500}
\end{subfigure}%\hspace{\columnwidth}
\begin{subfigure}{\columnwidth}
\includegraphics[width=\columnwidth]{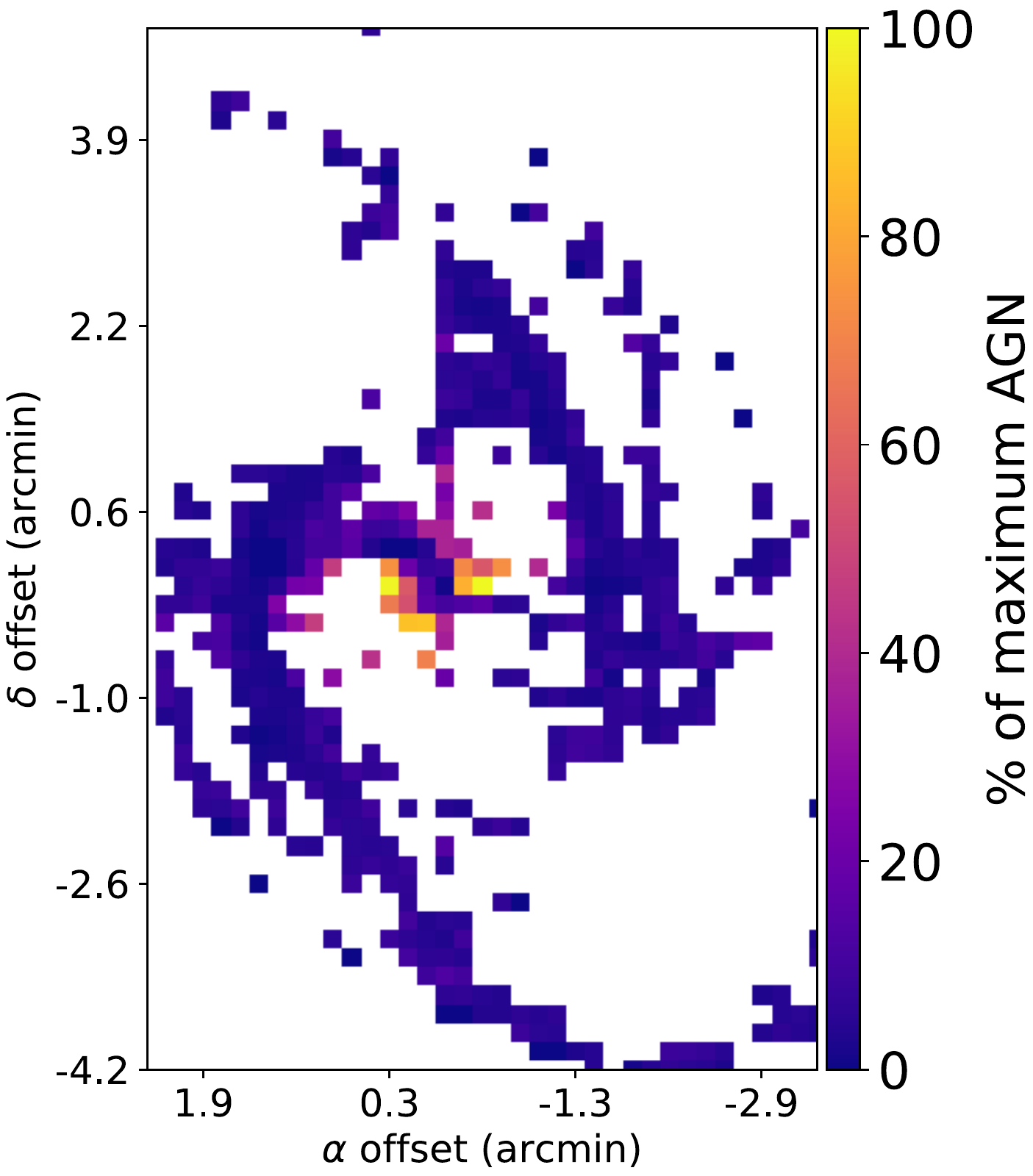}
\caption{1 kpc/pix}
\label{fig:1365map1k}
\end{subfigure}
\caption{The [O \textsc{iii}]/H$\beta$ vs [N \textsc{ii}]/H$\alpha$ AGN fraction map for NGC 1365, after rebinning to lower spatial resolutions.}
\label{fig:1365spatres_map}
\end{figure*}

\begin{figure*}
\centering
\begin{subfigure}{\columnwidth}
\centering
\includegraphics[scale=0.33]{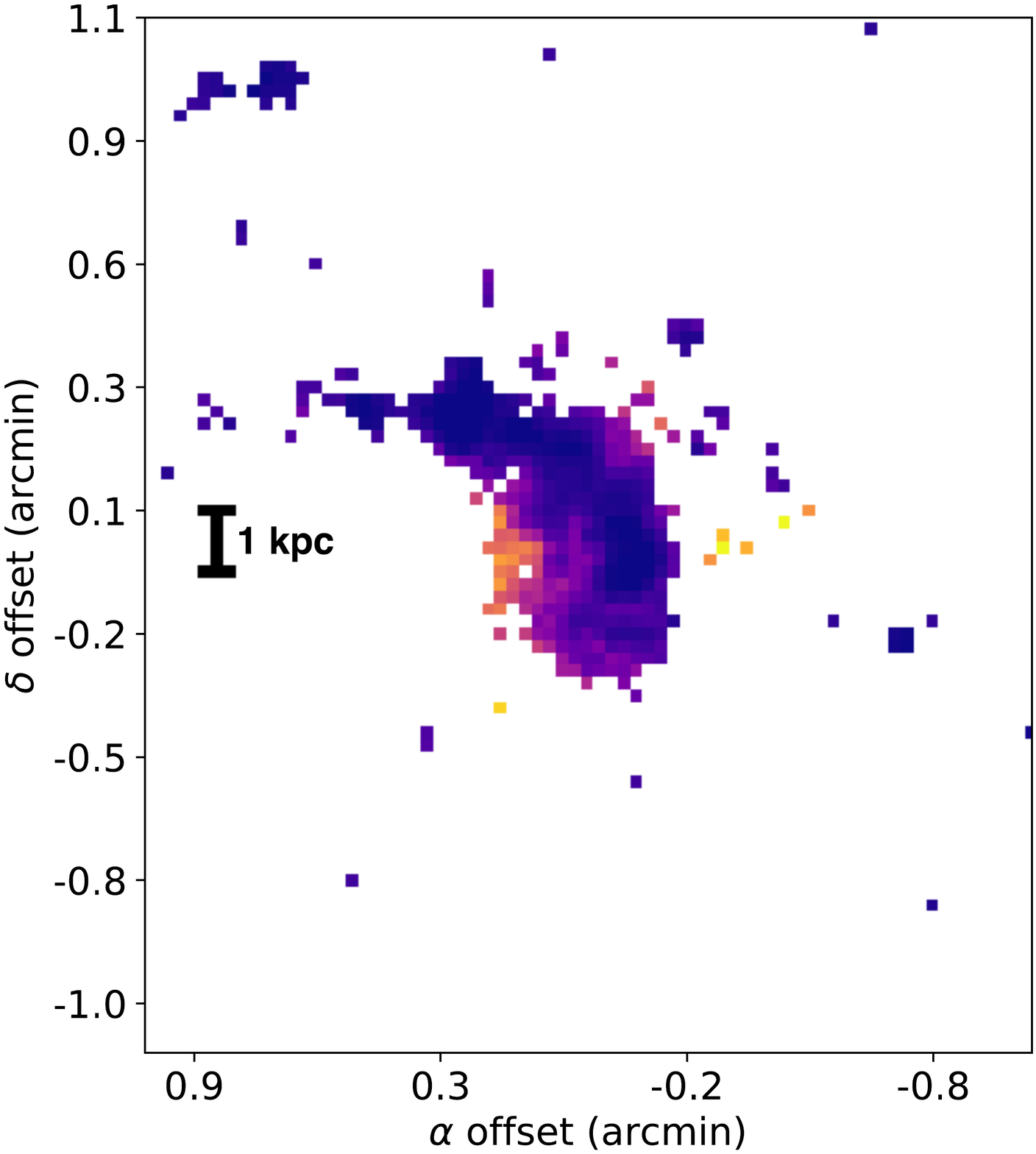}
\caption{Native resolution (169 pc/pix)}
\label{fig:1365mapnormal_cz}
\end{subfigure}%\hspace{\columnwidth}
\begin{subfigure}{\columnwidth}
\includegraphics[width=\columnwidth]{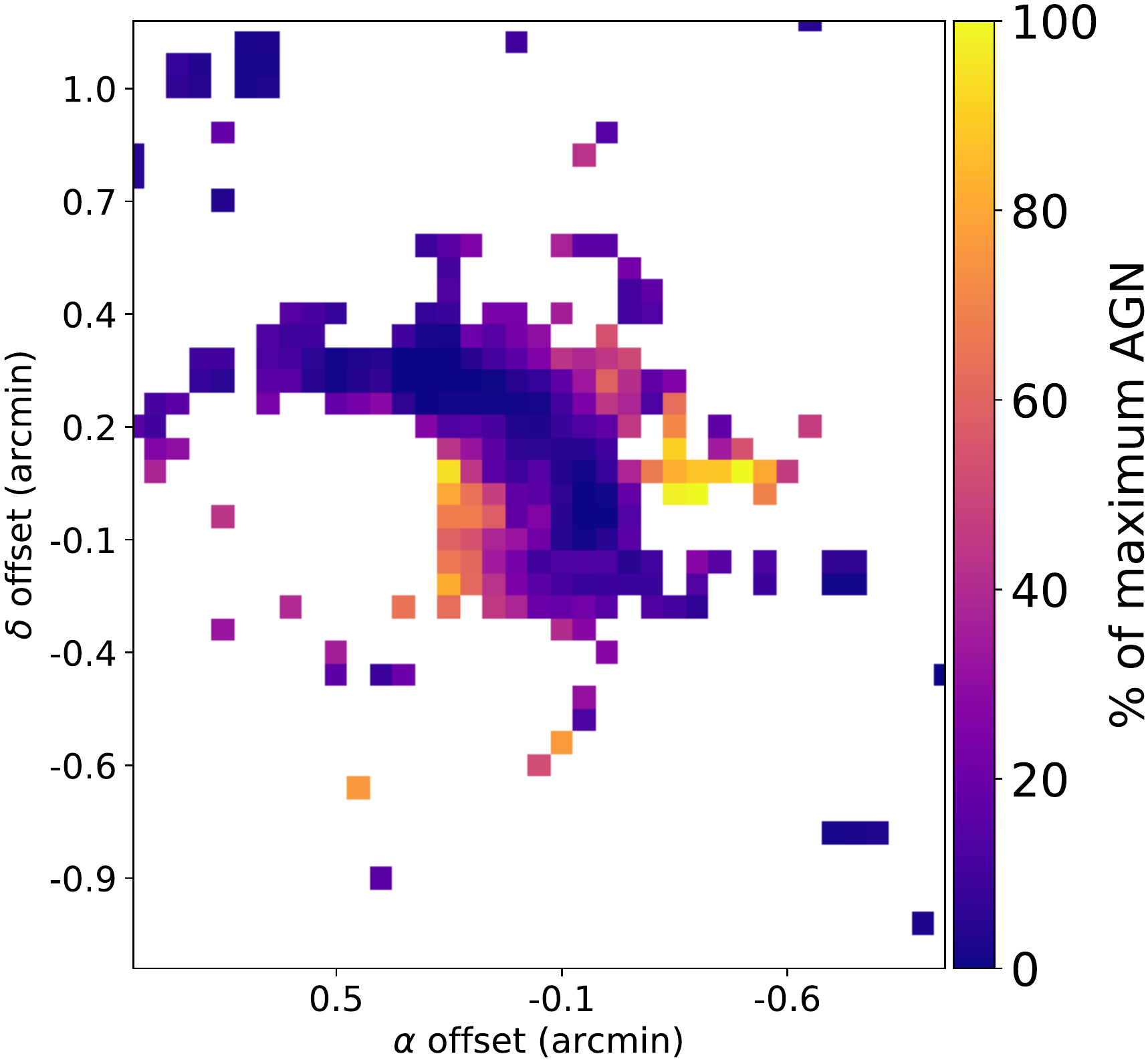}
\caption{330 pc/pix}
\label{fig:1365map330_cz}
\end{subfigure}
\begin{subfigure}{\columnwidth}
\centering
\includegraphics[scale=0.5]{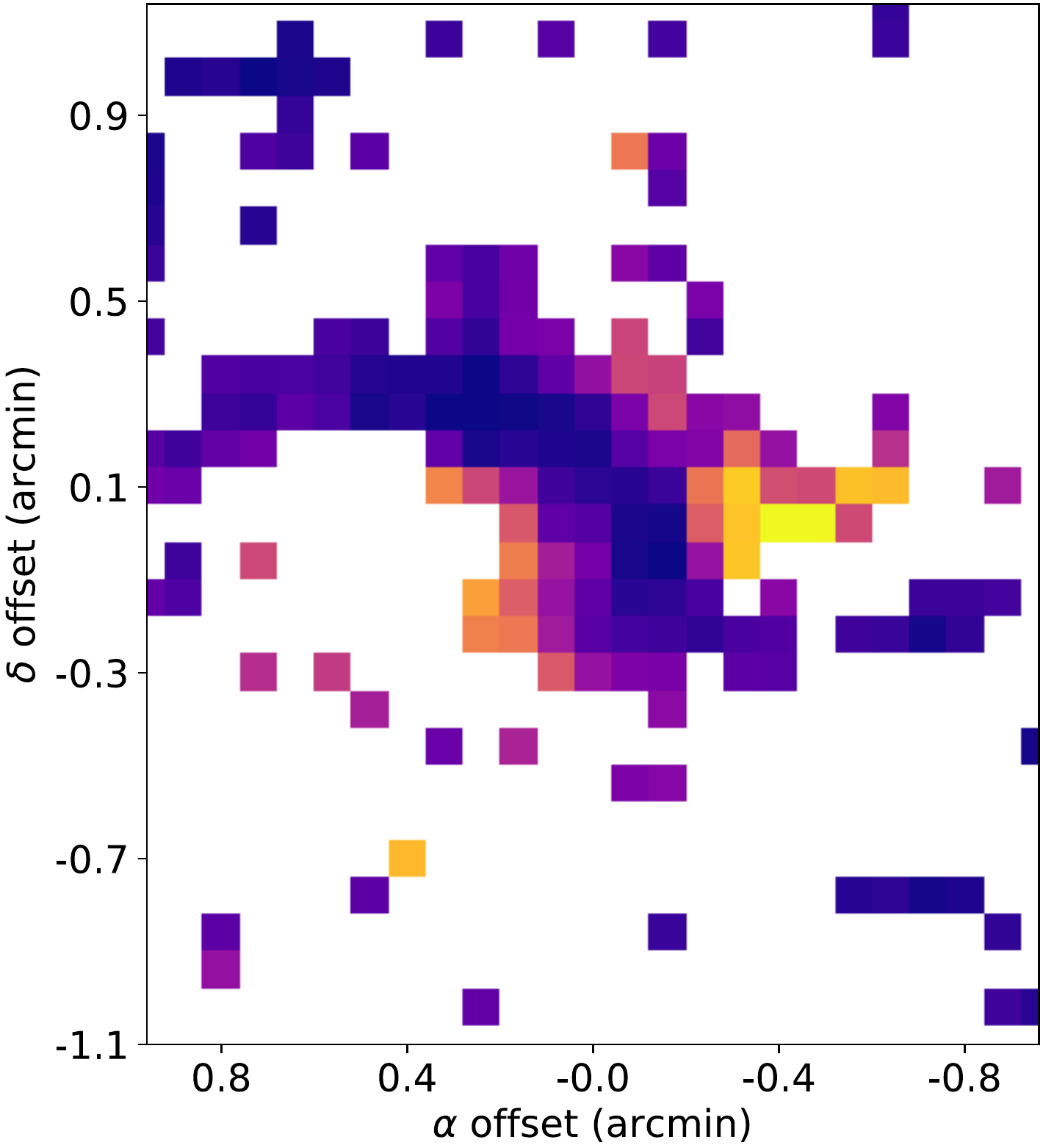}
\caption{500 pc/pix}
\label{fig:1365map500_cz}
\end{subfigure}%\hspace{\columnwidth}
\begin{subfigure}{\columnwidth}
\includegraphics[scale=0.5]{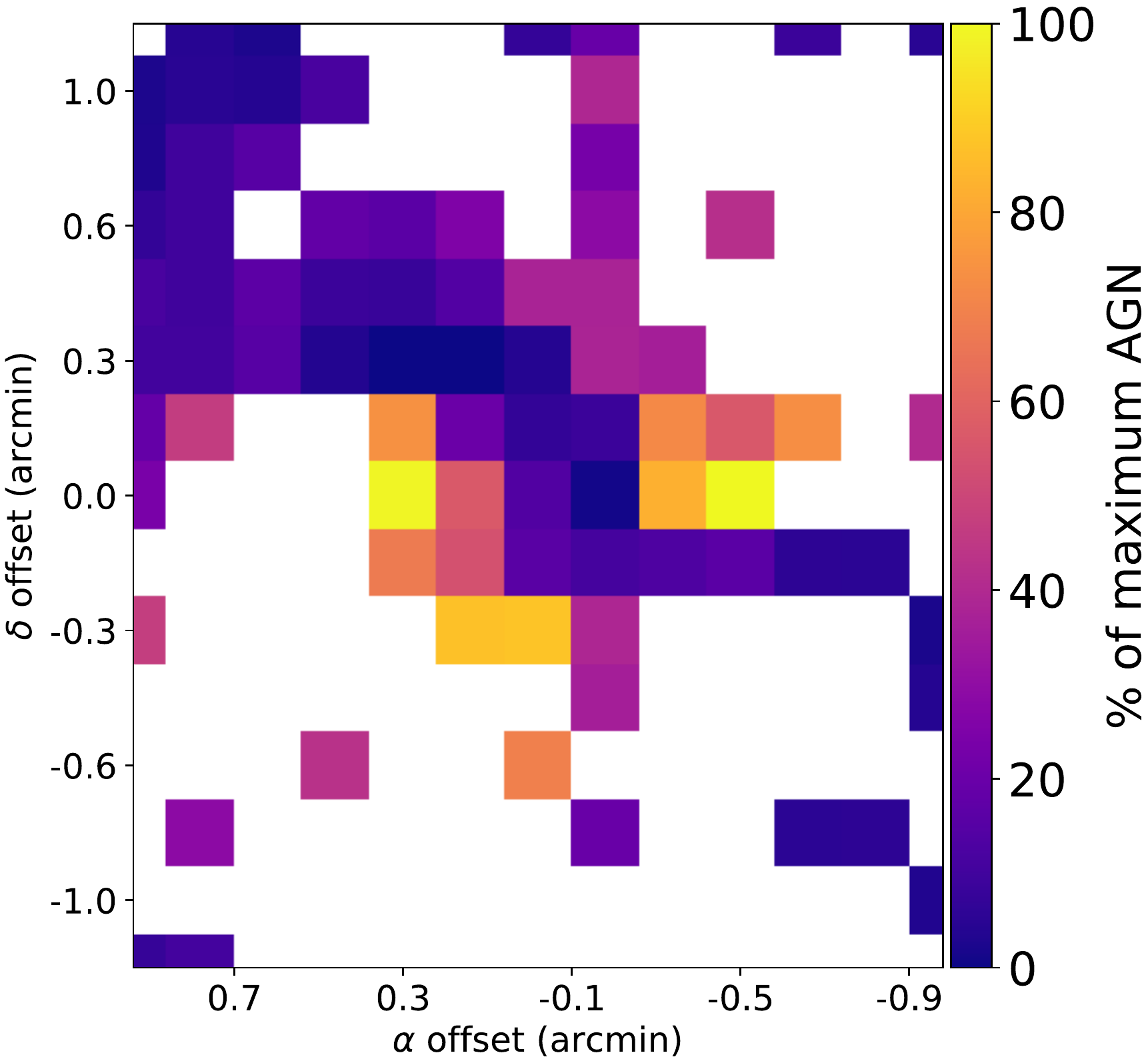}
\caption{1 kpc/pix}
\label{fig:1365map1k_cz}
\end{subfigure}
\caption{Images from Figure~\ref{fig:1365spatres_map} with a zoom on the nuclear region of NGC 1365}
\label{fig:1365spatres_map_cz}
\end{figure*}

The increase in AGN contribution in spaxels as the data is rebinned is shown in Table~\ref{tab:more} which gives a constant increase in the radius of equal starburst-AGN domination for both galaxies (the radius where the AGN fraction = 50\%). It is also shown in Figure~\ref{fig:agnfracpc}, showing an average increase in the percentage of spaxels with AGN fractions > 50\% as resolution decreases. The AGN radiation field produces large [O \textsc{iii}] and [N \textsc{ii}] line strengths, which dominate over the weaker line strengths from star formation, even at low AGN fractions.

In general, the AGN fractions for several strong emission lines (H$\alpha$, H$\beta$, [O \textsc{ii}]$\lambda \lambda$3726,3729, [O \textsc{iii}]$\lambda$5007, [S \textsc{ii}]$\lambda \lambda$6716,6731, and [N \textsc{ii}]$\lambda$6584) may increase with coarser resolution (Table~\ref{tab:agnfracs}). Considering NGC 1365, we find no significant change in the AGN fractions within the error bounds of the aforementioned strong emission lines between native resolution and a resolution of 1 kpc/pix, averaged over all lines shown in Table~\ref{tab:agnfracs}. NGC 1068 however, shows an average increase of 17.67 $\pm$ 7.09\% in the AGN fractions for the same strong emission lines between native resolution and a resolution of 1 kpc/pix. This has large ramifications for calculating the SFR in a galaxy. NGC 1068 shows as increase in the H$\alpha$ AGN fraction between native (121 pc/pix) and 1 kpc/pix resolutions of $14.59 \pm 0.14$\%. As a result, in general, the SFR from a galaxy with AGN activity may be continually and further overestimated with coarser spatial resolution.

We find that there is no limiting spatial resolution which allows the AGN fraction to be reliably estimated. The radius of equal starburst-AGN domination continues to decrease as the resolution increases (seen in Table~\ref{tab:more}), as does the percentage of spaxels with AGN fraction greater than 50\% (seen in Figure~\ref{fig:agnfracpc}). As such, one can expect the measurable emission from AGN to continue to decrease once the resolution has increased beyond the resolutions of the TYPHOON galaxies (169 pc/pixel for NGC 1365; 121 pc/pixel for NGC 1068). Hence, when attempting to accurately measure and quantify the influence and emission from an AGN, one should obtain data at a resolution as high as possible. 

%Rebinning the data reduces the amount of star-forming spaxels seen in the centre of both galaxies (as seen in the radius BPT diagrams in Figs.~\ref{fig:1365spatres_frac} and \ref{fig:1068spatres_frac}) due to the binning with spaxels which are dominated by the AGN. As a result, the centre of both galaxies appear more dominated by AGN activity as seen in the starburst-AGN fraction maps. Shown in Figure~\ref{fig:agnfracpc}, a higher percentage of spaxels appear to be AGN dominated the further the resolution is decreased. In the case of NCG 1365, rebinning the datacubes to lower resolutions recovers features in the galaxy such as the large-scale biconical outflow to the south-east and north-west of the nucleus as reported in \citet{Veilleux2003}, seen in Figure~\ref{fig:1365spatres_map}. This biconical outflow appears at lower resolutions with a large starburst-AGN fraction, supporting claims that the outflow is the result of AGN photoionisation \citep{Veilleux2003}.
% NGC 1068

\subsection{Effect of resolution on low surface brightness features}
\label{sec:lsb}

As the resolution decreases, low surface brightness (LSB) features become apparent. Seen in the BPT diagrams at lower resolutions are spaxels in the regions commonly associated with features such as shocks \citep[for the position of shock features on diagnostic diagrams, see][]{Rich2010,RKD2011,Kewley2013a}. These LSB features typically have low signal-to-noise ratio values. The rebinning process produces an increase in the S/N of the LSB spaxels. Hence, LSB features become observable in the data. Table~\ref{tab:more} shows the relative difference in spaxels with emission attributable to LSB features. In total, we see an increase in the percentage of spaxels which we attribute to low surface brightness features between the native resolution data and the rebinned data for both NGC 1365 and NGC 1068. The increase in the relative amount of spaxels attributed to LSB features differs for both galaxies, with up to a factor of ${\sim} 12.5$ increase in NGC 1365 with coarser resolution, and up to a factor of ${\sim} 1.6$ increase in NGC 1068. These spaxels are defined as being outside the nuclear column with moderate [O \textsc{iii}]/H$\beta$ and moderate-high [N \textsc{ii}]/H$\alpha$ ratios, in the region of the BPT where LSB features are typically found. 

\subsubsection{Shocks}
\label{sec:shocks}

Shocks (or shock waves) are compression waves which propagate at a speed greater than the sound speed in the medium. Due to the supersonic nature of the shock wave, instantaneous changes in properties of the medium (such as density, pressure and temperature) occur at the shock front, causing discontinuities in the values of these properties before and after the boundary of the shock. The source of supersonic compression may originate from astrophysical phenomena such as supernova explosions, stellar winds, or outflows from young stellar clusters or AGN \citep[][and references therein]{ADU}.

Shocks are typically diagnosed and categorised by their high velocity dispersions \citep[typically ${\sim} 100 - 500$ kms${-1}$;][]{Rich2010,RKD2011,RKD2014,Ho2014}. Figure~\ref{fig:1365vdisp_map} shows the velocity dispersion maps across all resolutions considered for NGC 1365. Regions of high velocity dispersion ($\sigma {\sim} 160$ kms$^{-1}$) can be seen around the nucleus of NGC 1365 across all resolutions. However, the high velocity dispersion regions surrounding the nucleus grow in size as resolution decreases. The high-$\sigma$ areas surrounding the nucleus are aligned with the biconical outflow in [N \textsc{ii}]/H$\alpha$ and [O \textsc{iii}]/H$\beta$ shown in NGC 1365 from \citet{Veilleux2003} (shown in Figure~\ref{fig:Veilleux2003outflow}), suggesting the shocks in NGC 1365 are the result of outflow from the central AGN.

The number of regions of high velocity dispersion along the spiral arms of NGC 1365 increases as the resolution decreases, with velocity dispersions up to $\sigma {\sim} 300 \mathrm{kms}^{-1}$ at a resolution of 1kpc/pix. The high levels of star formation in the spiral arms of NGC 1365 shown in Figures~\ref{fig:1365BPTagnfrac} and~\ref{fig:1365BPTagnmap} suggest that regions of high velocity dispersion within the spiral arms are a result of shocks from stellar sources, such as supernova explosions and stellar winds. It appears that shocks from stellar sources are intensified and enhanced with coarser resolution.

\begin{figure*}
\centering
\begin{subfigure}{\columnwidth}
\centering
\includegraphics[width=\columnwidth]{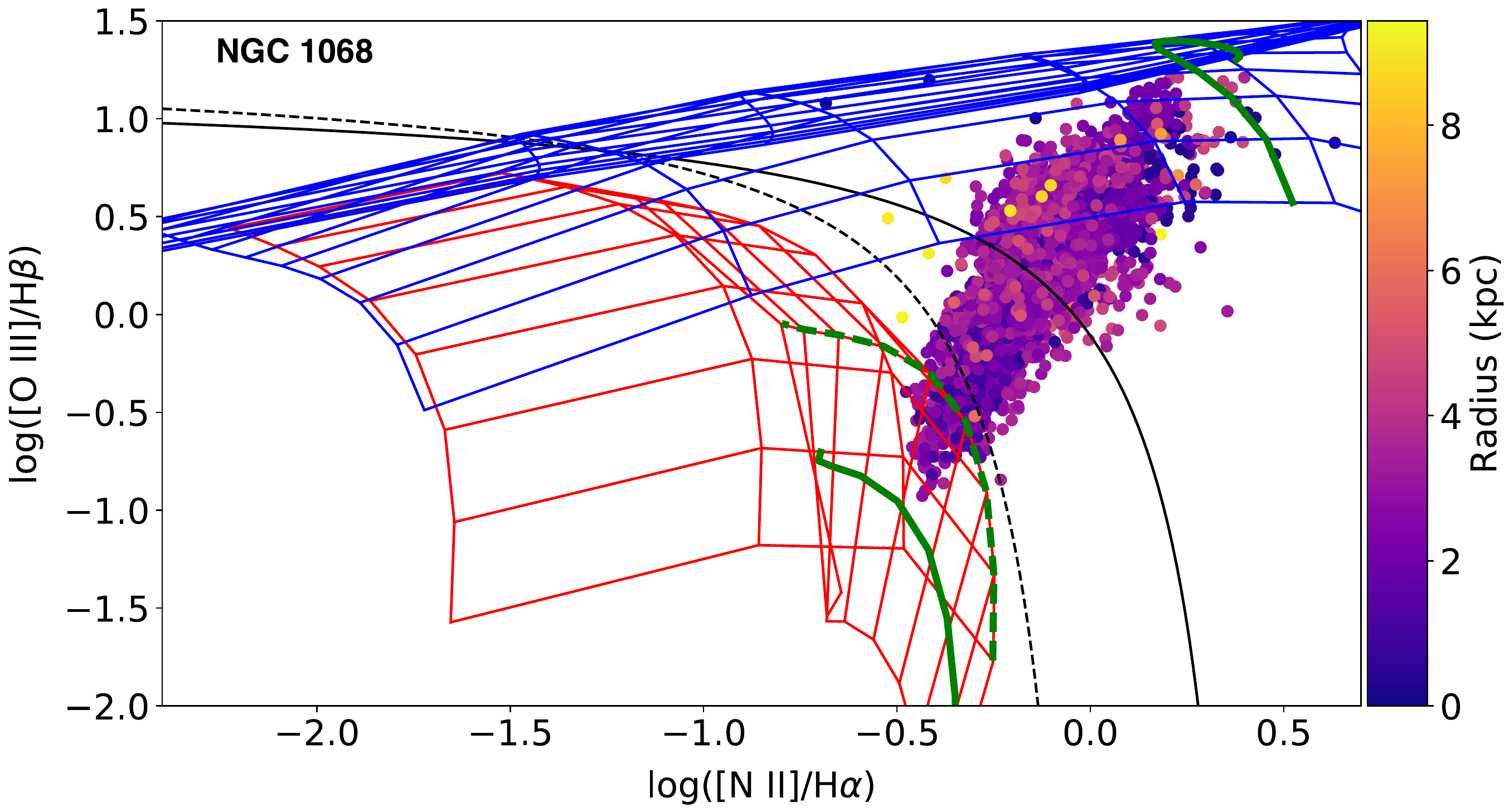}
\caption{BPT, Native resolution (121 pc/pix)}
\label{fig:1068bptnormal}
\end{subfigure}%\hspace{\columnwidth}
\begin{subfigure}{\columnwidth}
\centering
\includegraphics[width=\columnwidth]{1068_agnfrac_normal_new_name.pdf}
\caption{AGN fraction, Native resolution (121 pc/pix)}
\label{fig:1068fracnormal}
\end{subfigure}\hspace{\columnwidth}
\begin{subfigure}{\columnwidth}
\centering
\includegraphics[width=\columnwidth]{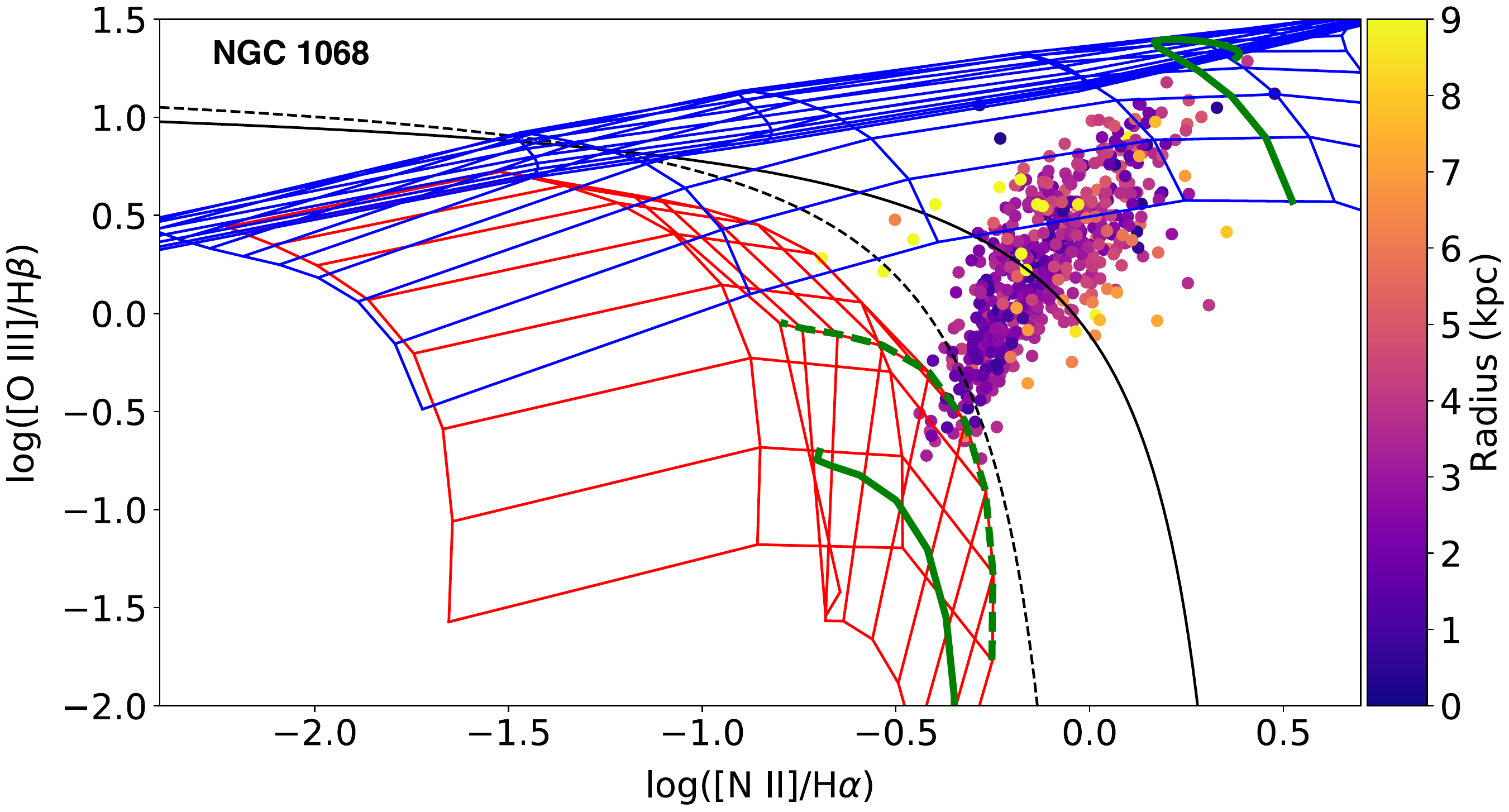}
\caption{BPT, 330 pc/pix}
\label{fig:1068bpt330}
\end{subfigure}
\begin{subfigure}{\columnwidth}
\centering
\includegraphics[width=\columnwidth]{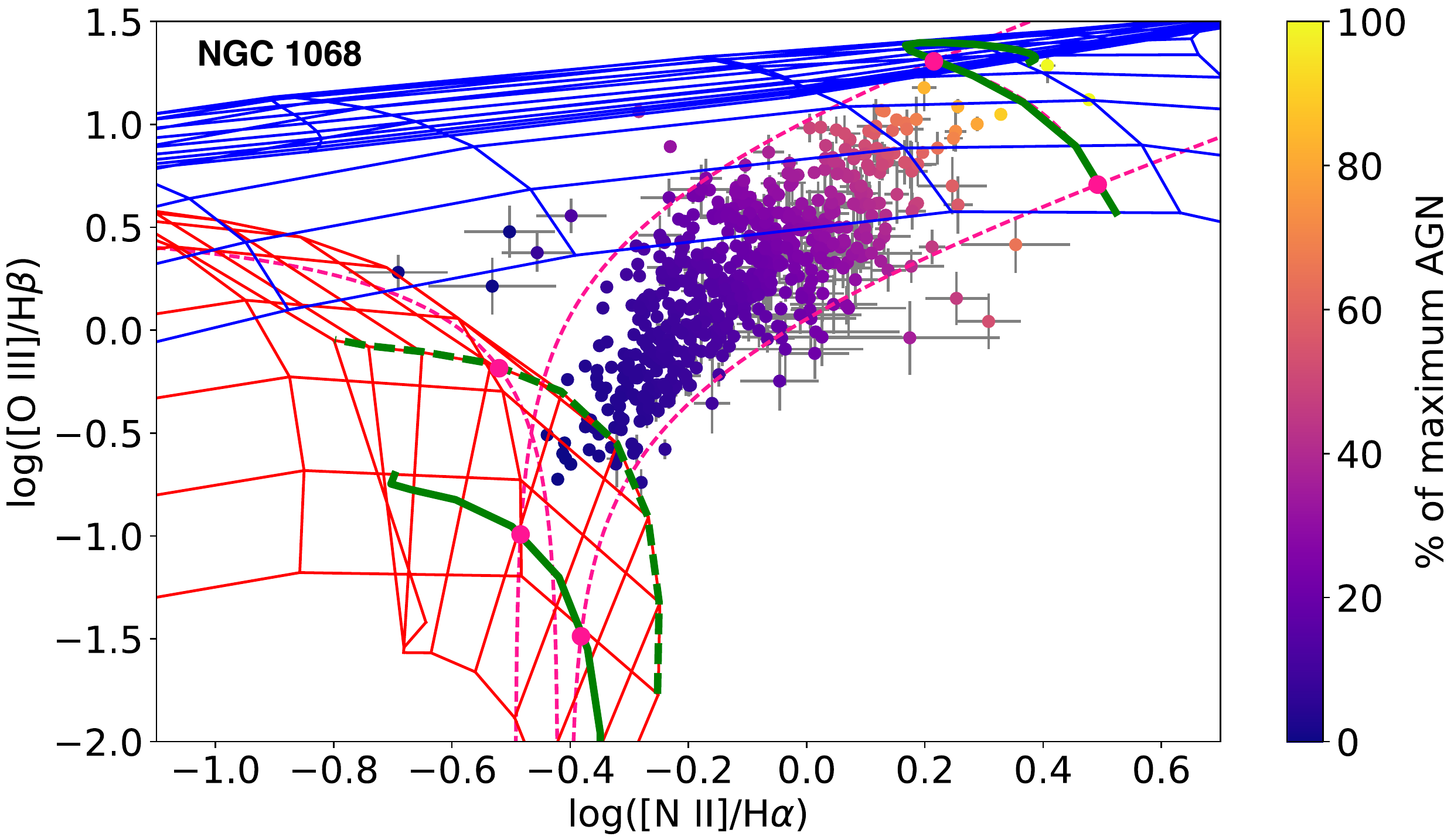}
\caption{AGN fraction, 330 pc/pix}
\label{fig:1068frac330}
\end{subfigure}
\begin{subfigure}{\columnwidth}
\centering
\includegraphics[width=\columnwidth]{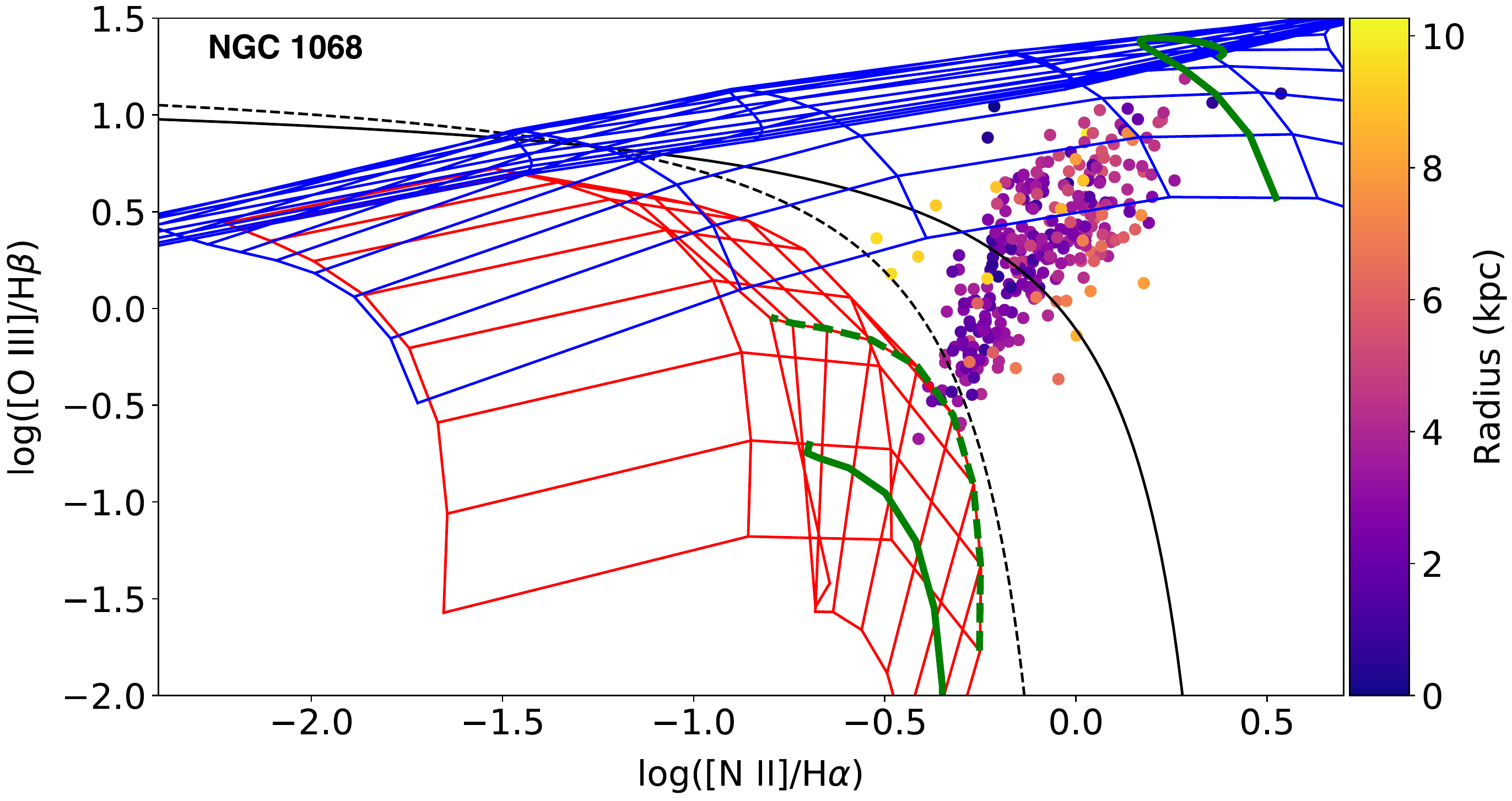}
\caption{BPT, 500 pc/pix}
\label{fig:1068bpt500}
\end{subfigure}%\hspace{\columnwidth}
\begin{subfigure}{\columnwidth}
\centering
\includegraphics[width=\columnwidth]{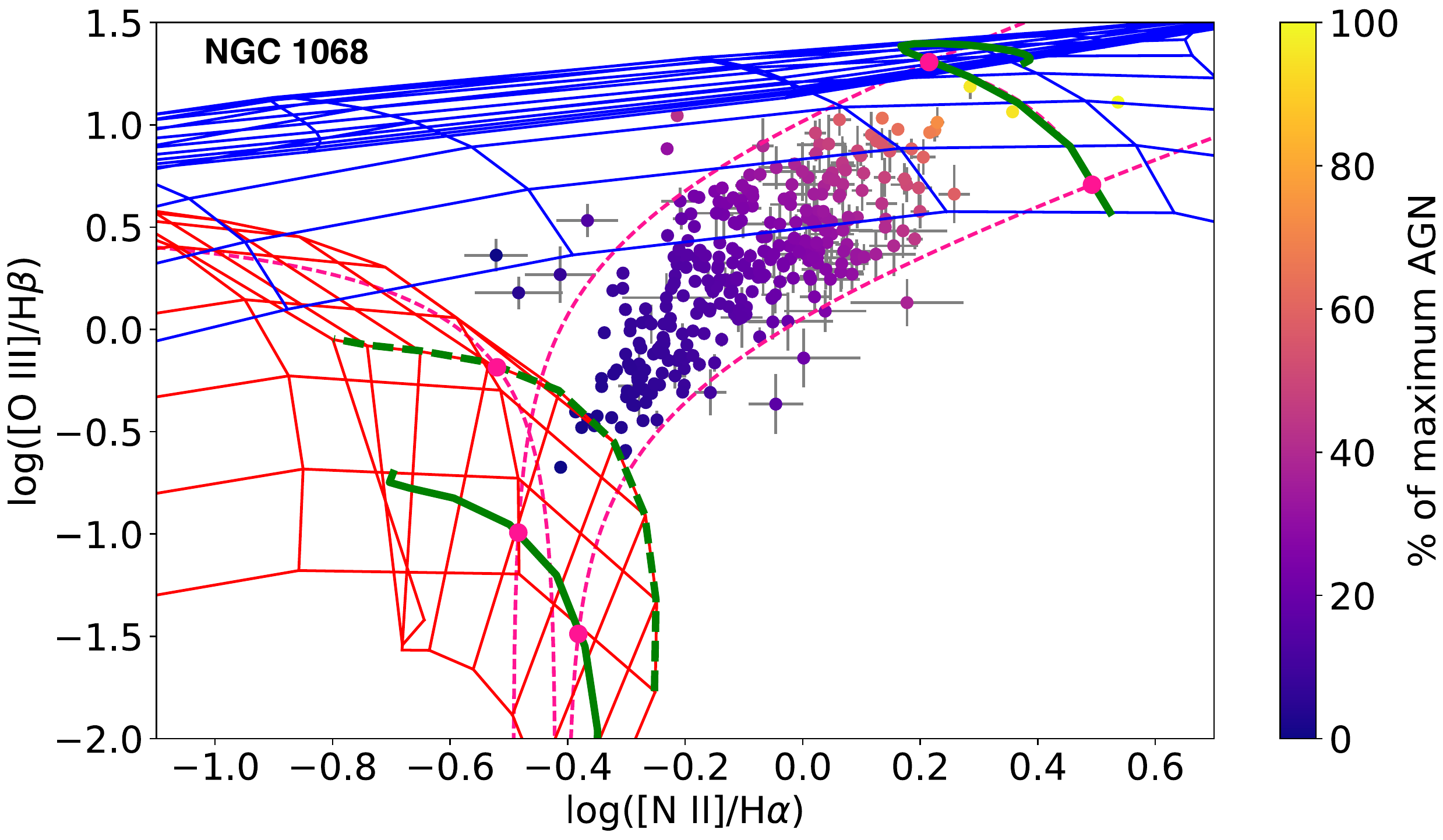}
\caption{AGN fraction, 500 pc/pix}
\label{fig:1068frac500}
\end{subfigure}\hspace{\columnwidth}
\begin{subfigure}{\columnwidth}
\centering
\includegraphics[width=\columnwidth]{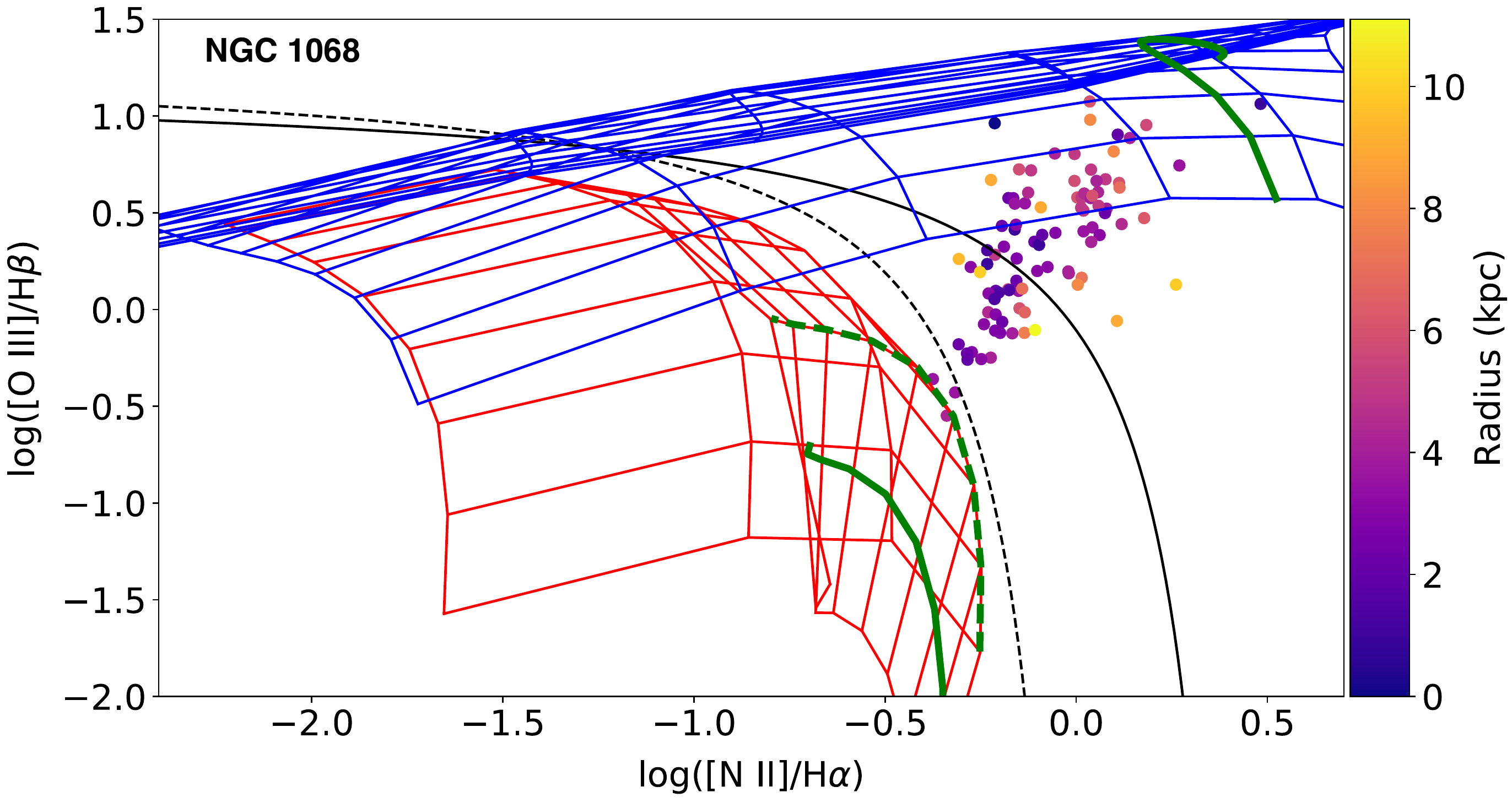}
\caption{BPT, 1 kpc/pix}
\label{fig:1068bpt1k}
\end{subfigure}
%\caption{The BPT diagram for NGC 1068, after rebinning to lower spatial resolutions.}
\begin{subfigure}{\columnwidth}
\centering
\includegraphics[width=\columnwidth]{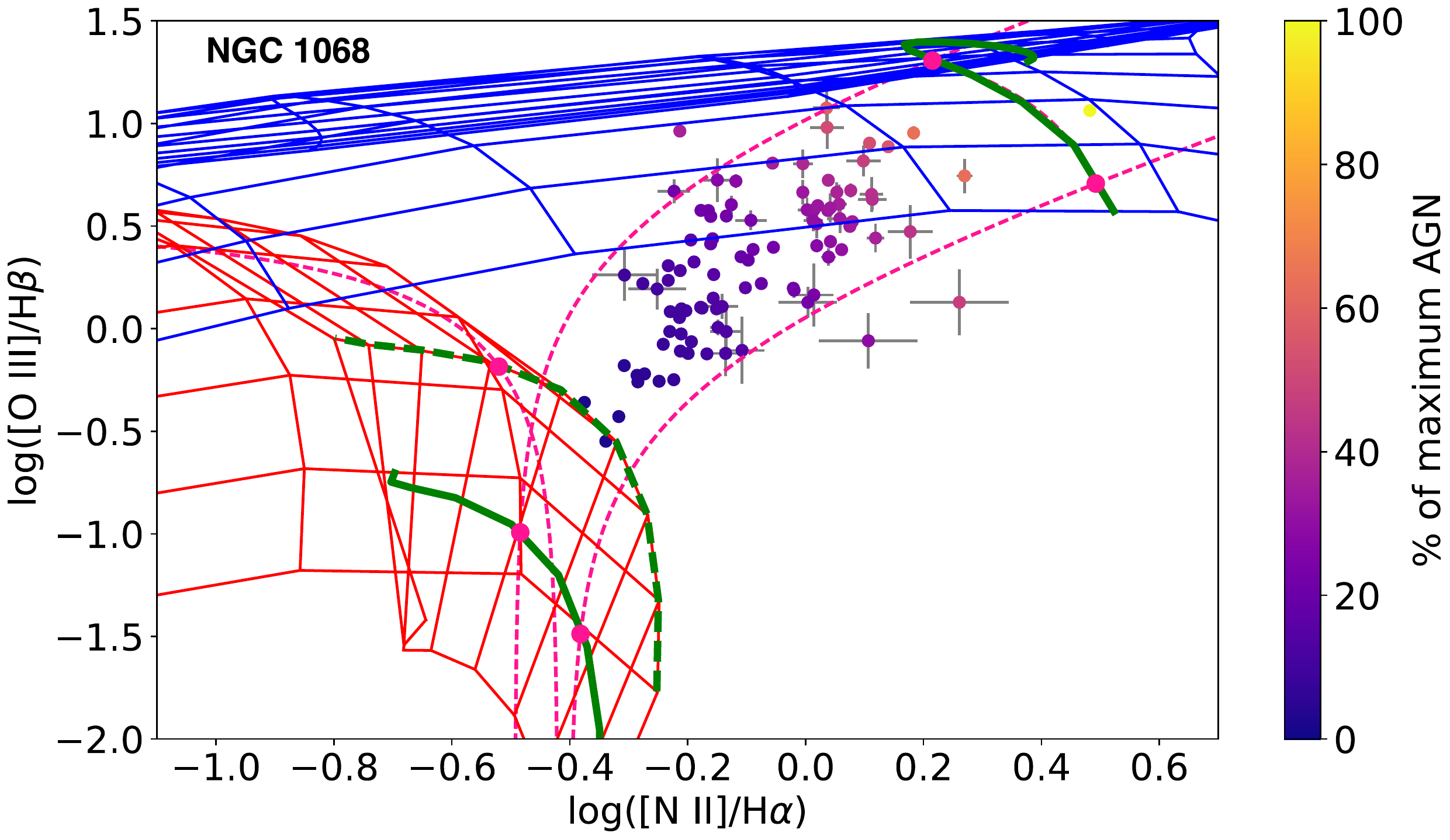}
\caption{AGN fraction, 1 kpc/pix}
\label{fig:1068frac1k}
\end{subfigure}
\caption{The AGN fractions and BPT diagrams coloured by radius for NGC 1068, after rebinning to lower spatial resolutions. Errors for the spaxels on the BPT diagrams are the same as those on the AGN fraction plots.}
\label{fig:1068spatres_frac}
\end{figure*}

\begin{figure*}
\centering
\begin{subfigure}{\columnwidth}
\centering
\includegraphics[scale=0.4]{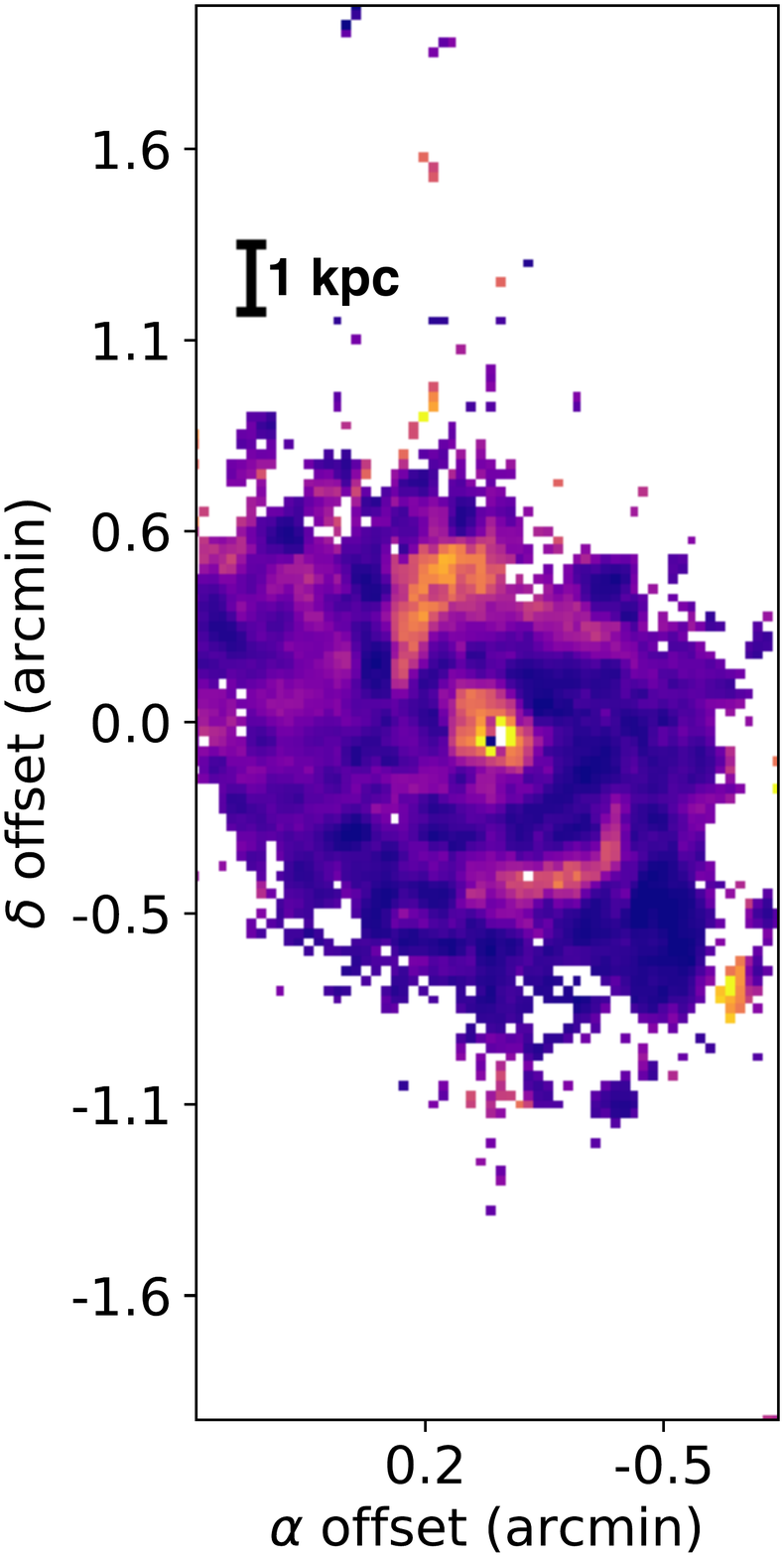}
\caption{Native resolution (121 pc/pix)}
\label{fig:1068mapnormal}
\end{subfigure}%\hspace{\columnwidth}
\begin{subfigure}{\columnwidth}
\centering
\includegraphics[scale=0.75]{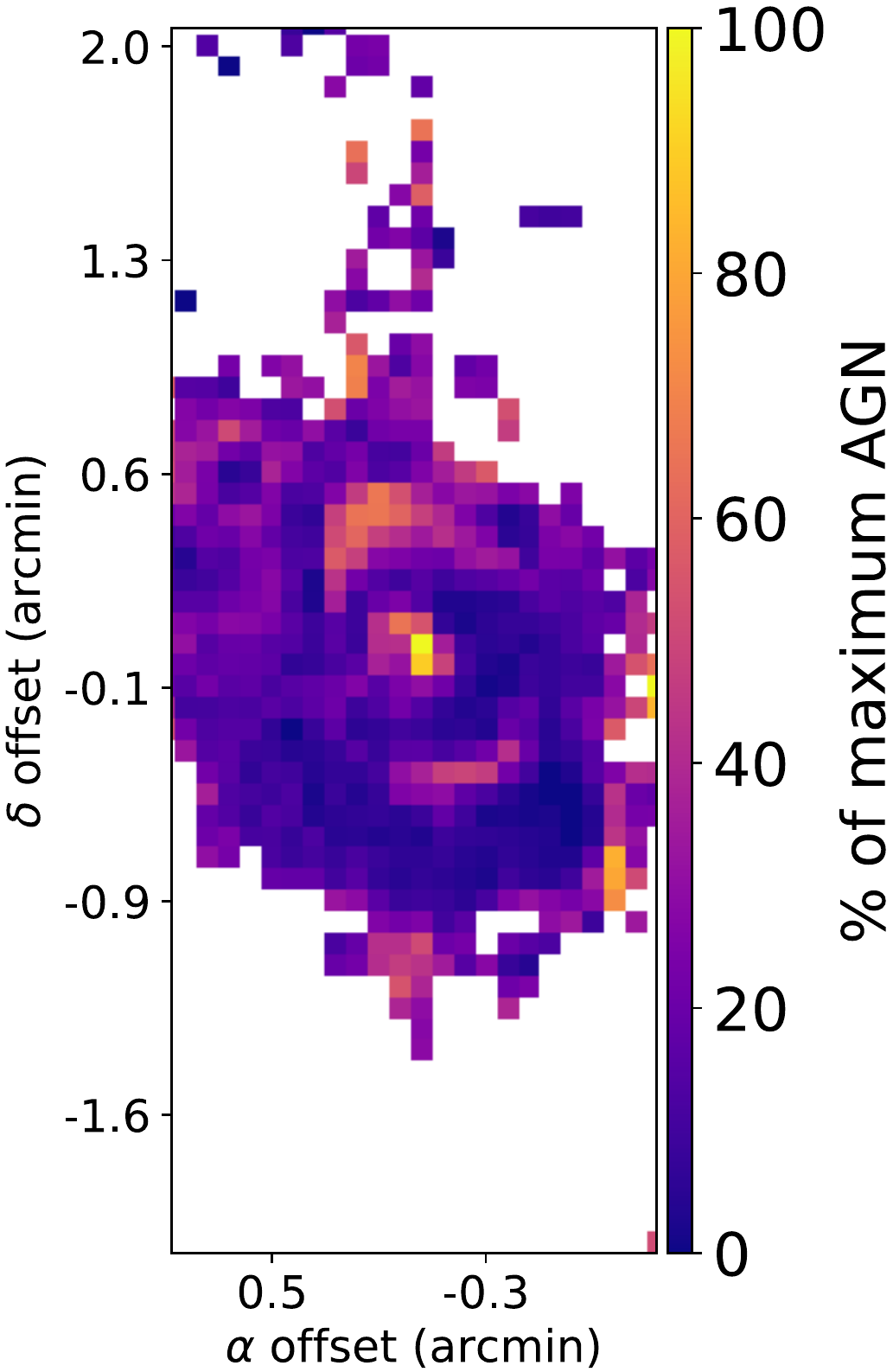}
\caption{330 pc/pix}
\label{fig:1068map330}
\end{subfigure}
\begin{subfigure}{\columnwidth}
\centering
\includegraphics[scale=0.75]{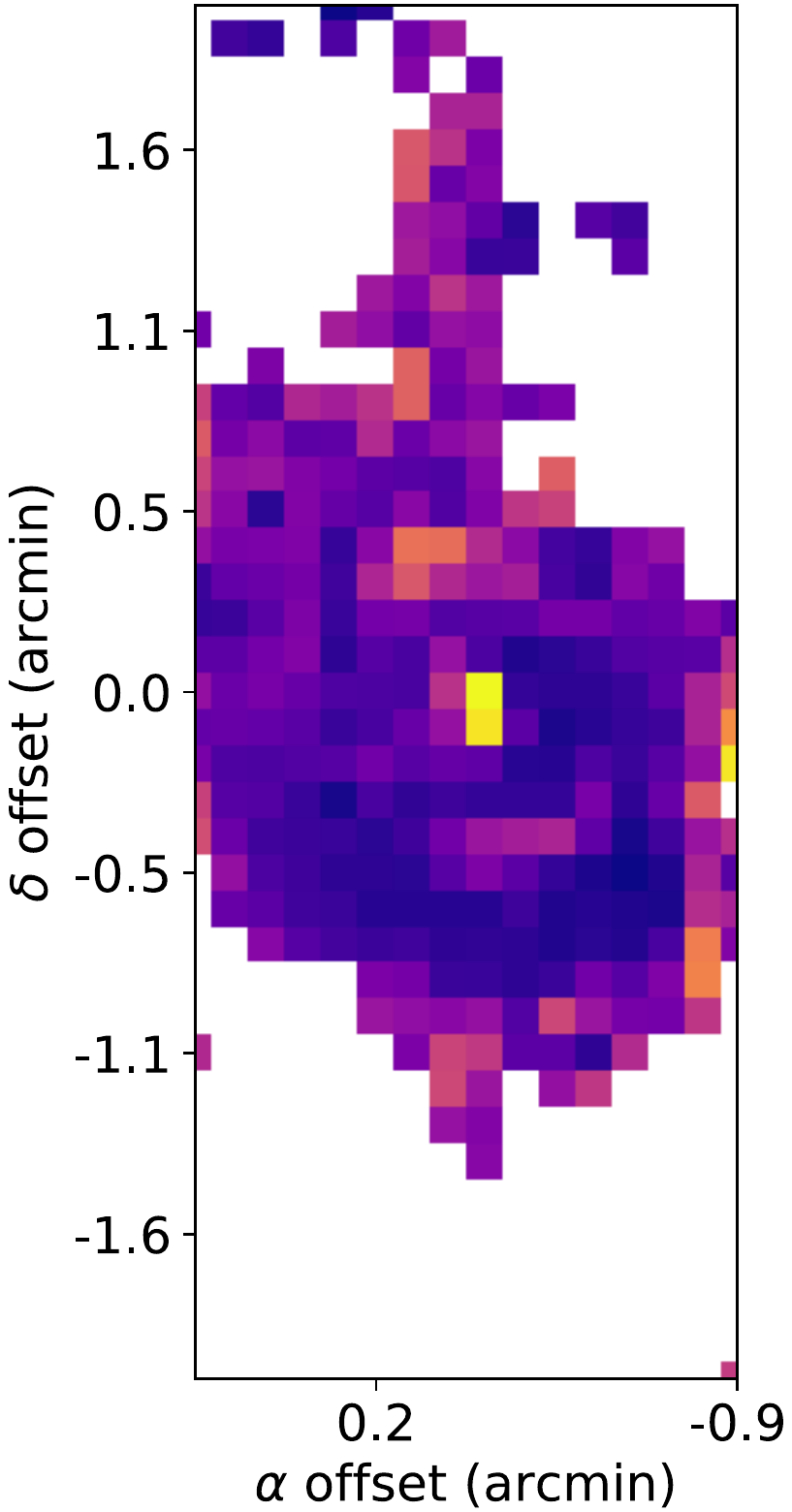}
\caption{500 pc/pix}
\label{fig:1068map500}
\end{subfigure}%\hspace{\columnwidth}
\begin{subfigure}{\columnwidth}
\centering
\includegraphics[scale=0.75]{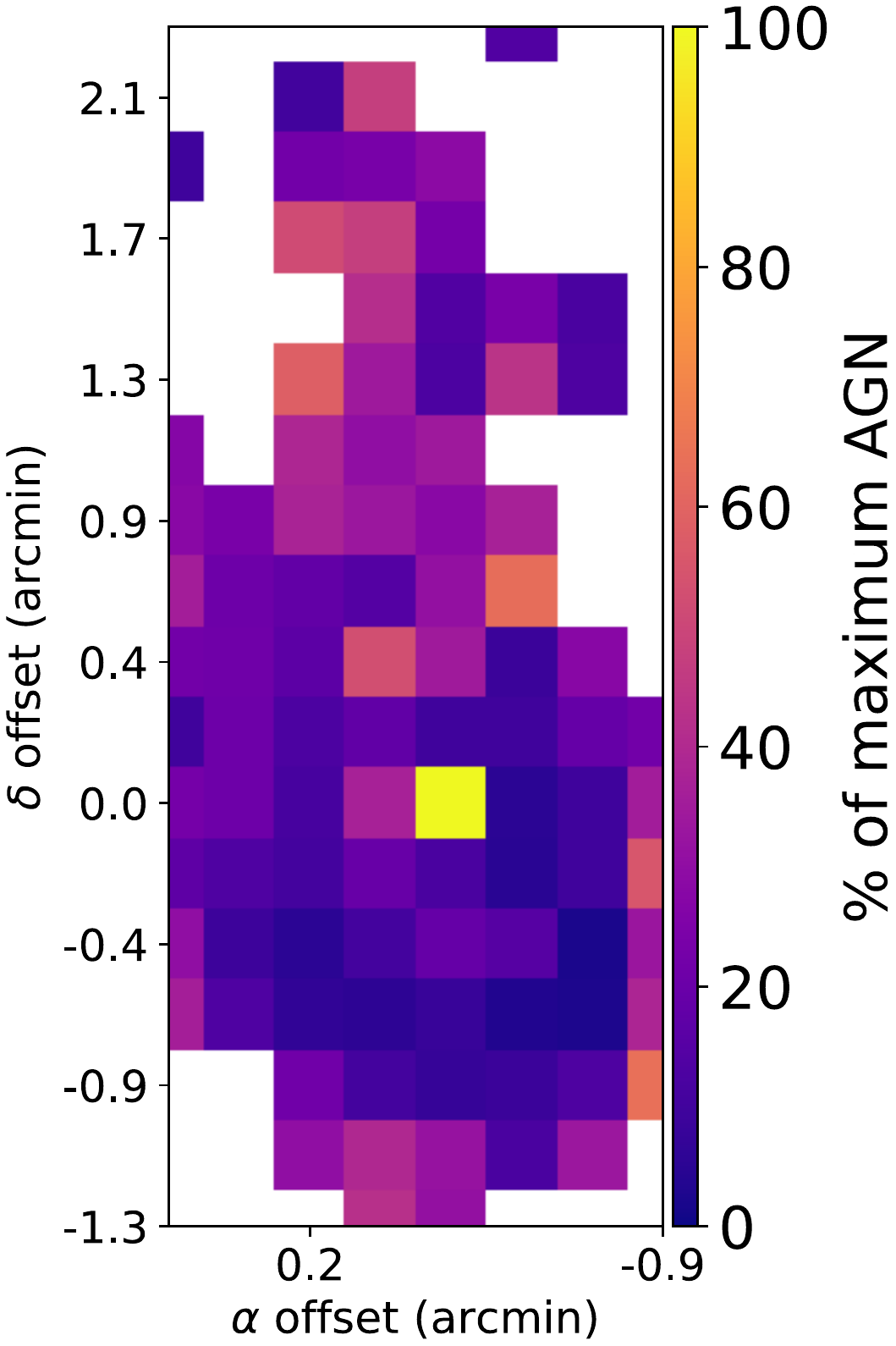}
\caption{1 kpc/pix}
\label{fig:1068map1k}
\end{subfigure}
\caption{The [O \textsc{iii}]/H$\beta$ vs [N \textsc{ii}]/H$\alpha$ AGN fraction map for NGC 1068, after rebinning to lower spatial resolutions.}
\label{fig:1068spatres_map}
\end{figure*}

\begin{figure*}
\centering
\begin{subfigure}{\columnwidth}
\centering
\includegraphics[scale=0.32]{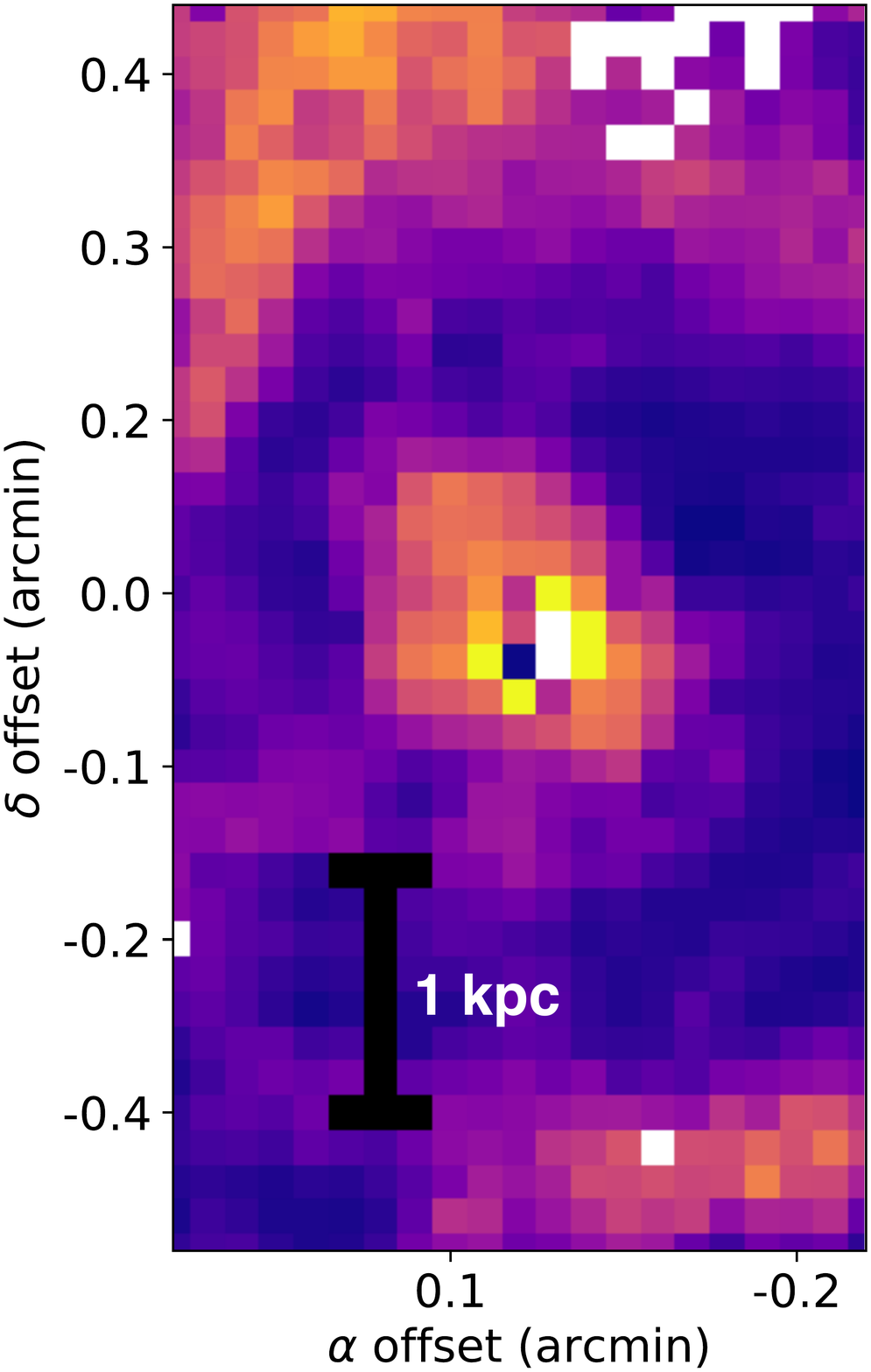}
\caption{Native resolution (169 pc/pix)}
\label{fig:1068mapnormal_cz}
\end{subfigure}%\hspace{\columnwidth}
\begin{subfigure}{\columnwidth}
\includegraphics[width=\columnwidth]{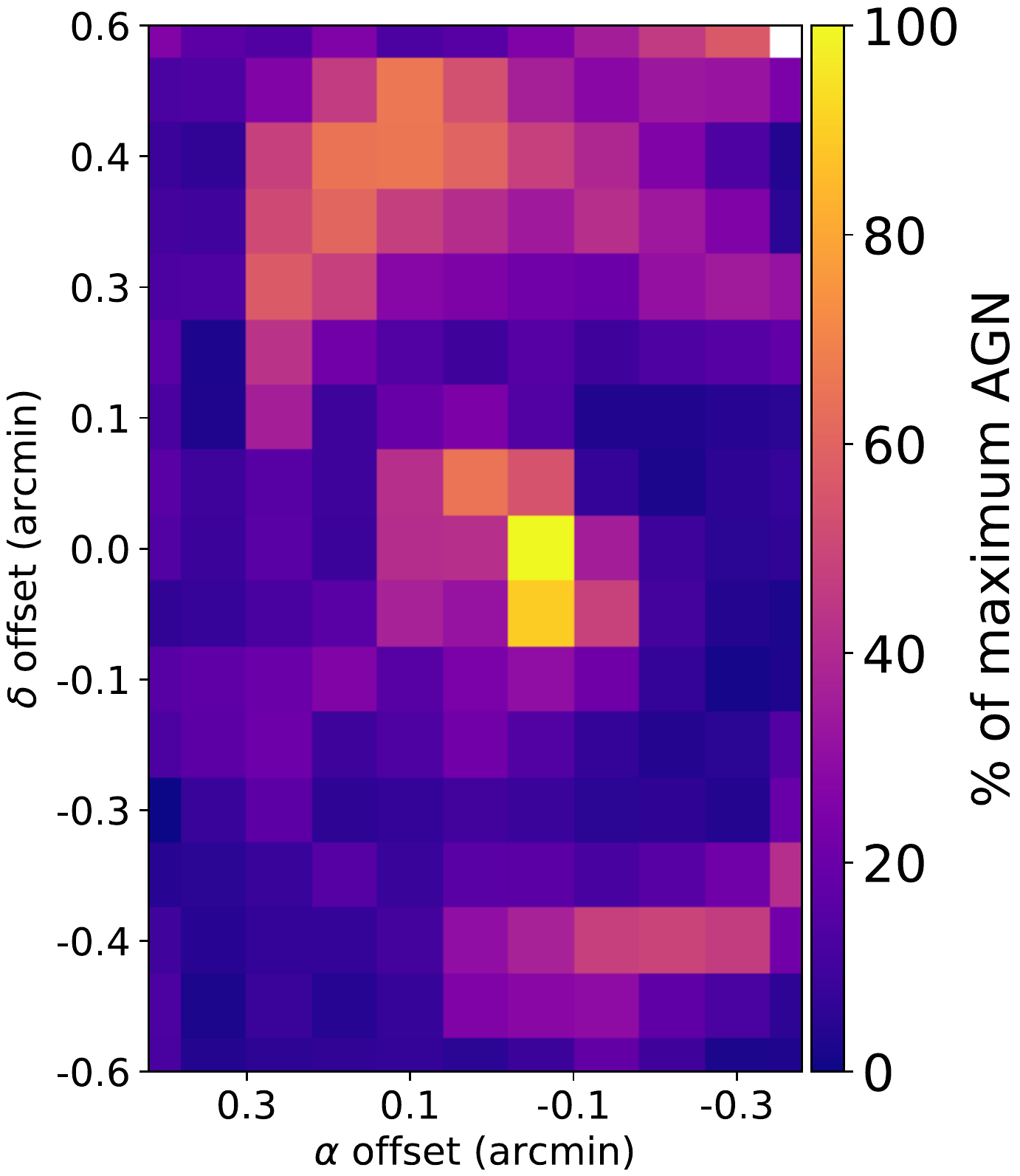}
\caption{330 pc/pix}
\label{fig:1068map330_cz}
\end{subfigure}
\begin{subfigure}{\columnwidth}
\centering
\includegraphics[scale=0.62]{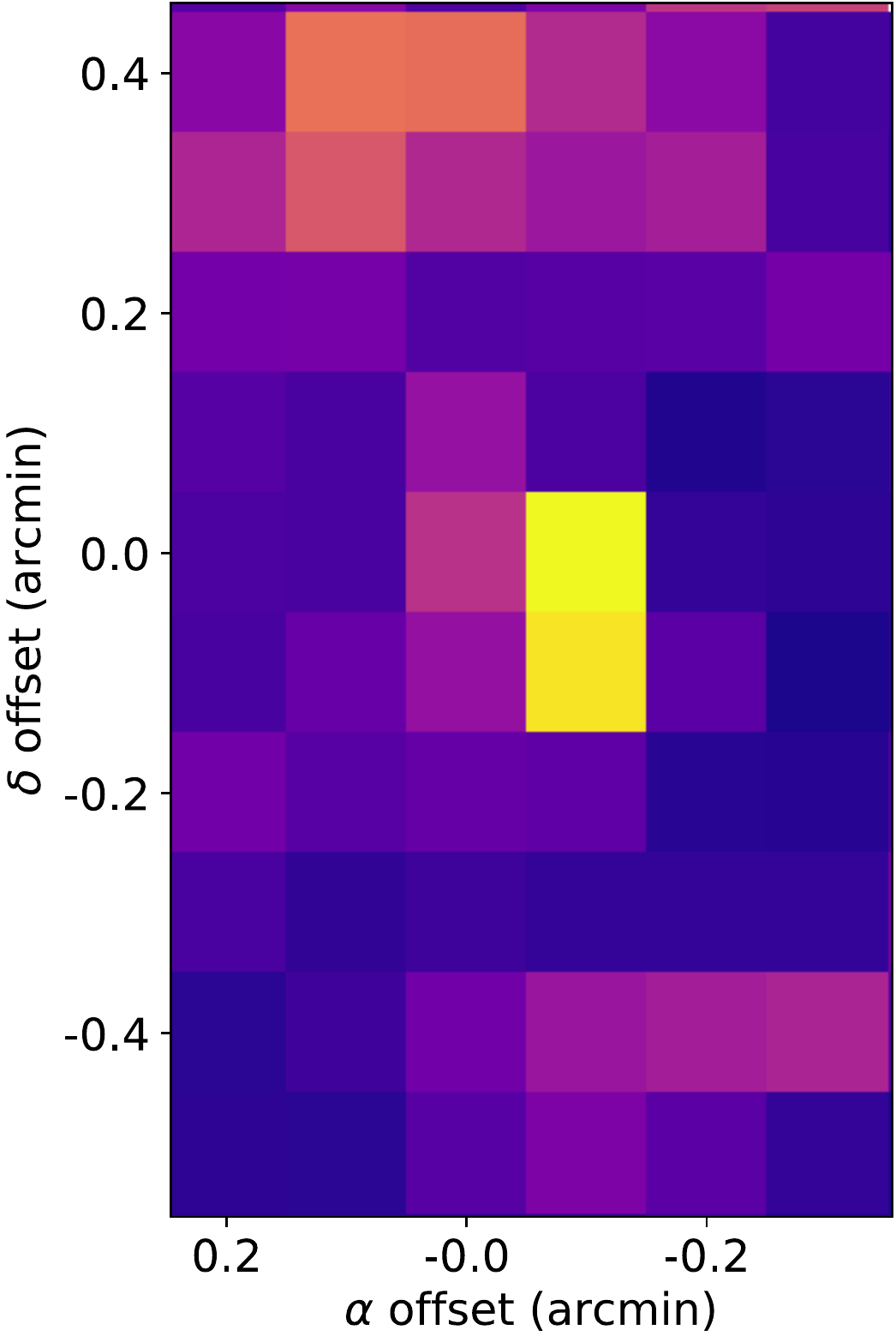}
\caption{500 pc/pix}
\label{fig:1068map500_cz}
\end{subfigure}%\hspace{\columnwidth}
\begin{subfigure}{\columnwidth}
\includegraphics[width=\columnwidth]{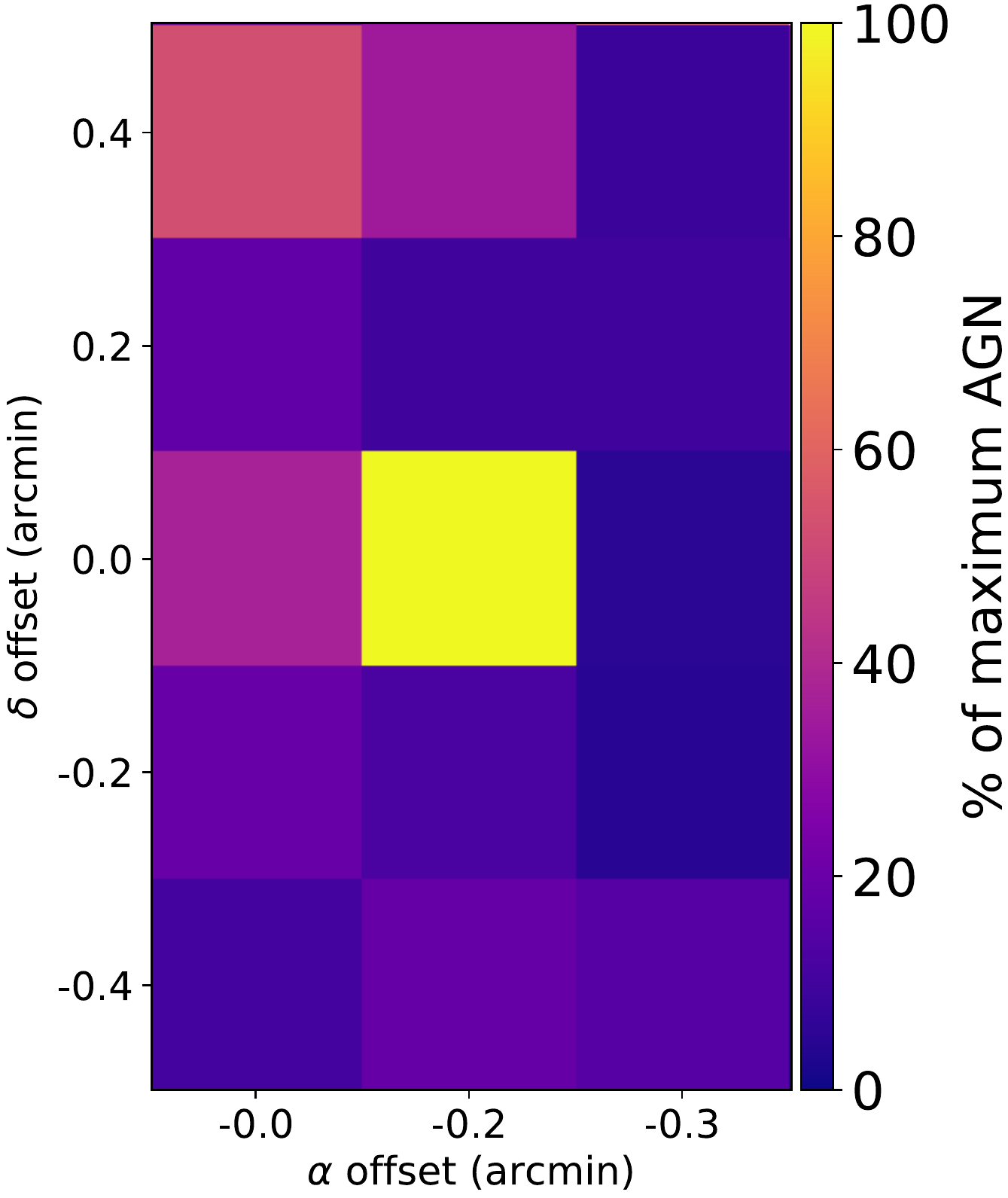}
\caption{1 kpc/pix}
\label{fig:1068map1k_cz}
\end{subfigure}
\caption{Images from Figure~\ref{fig:1068spatres_map} with a zoom on the nuclear region of NGC 1068}
\label{fig:1068spatres_map_cz}
\end{figure*}

\begin{figure*}
\centering
\begin{subfigure}{\columnwidth}
\centering
\includegraphics[scale=0.32]{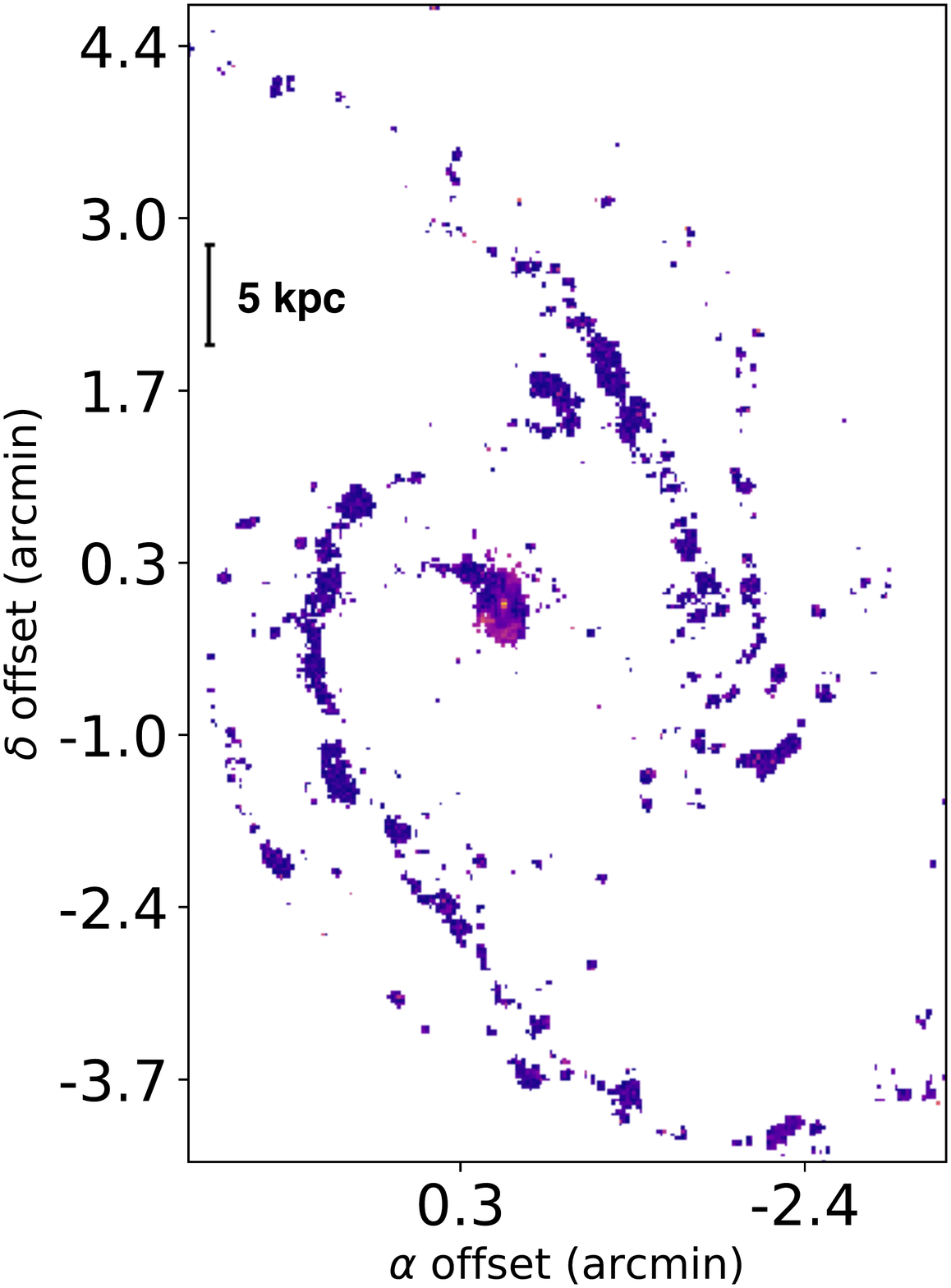}
\caption{Native resolution (169 pc/pix)}
\label{fig:1365vdispnormal}
\end{subfigure}%\hspace{\columnwidth}
\begin{subfigure}{\columnwidth}
%\centering
\includegraphics[scale=0.55]{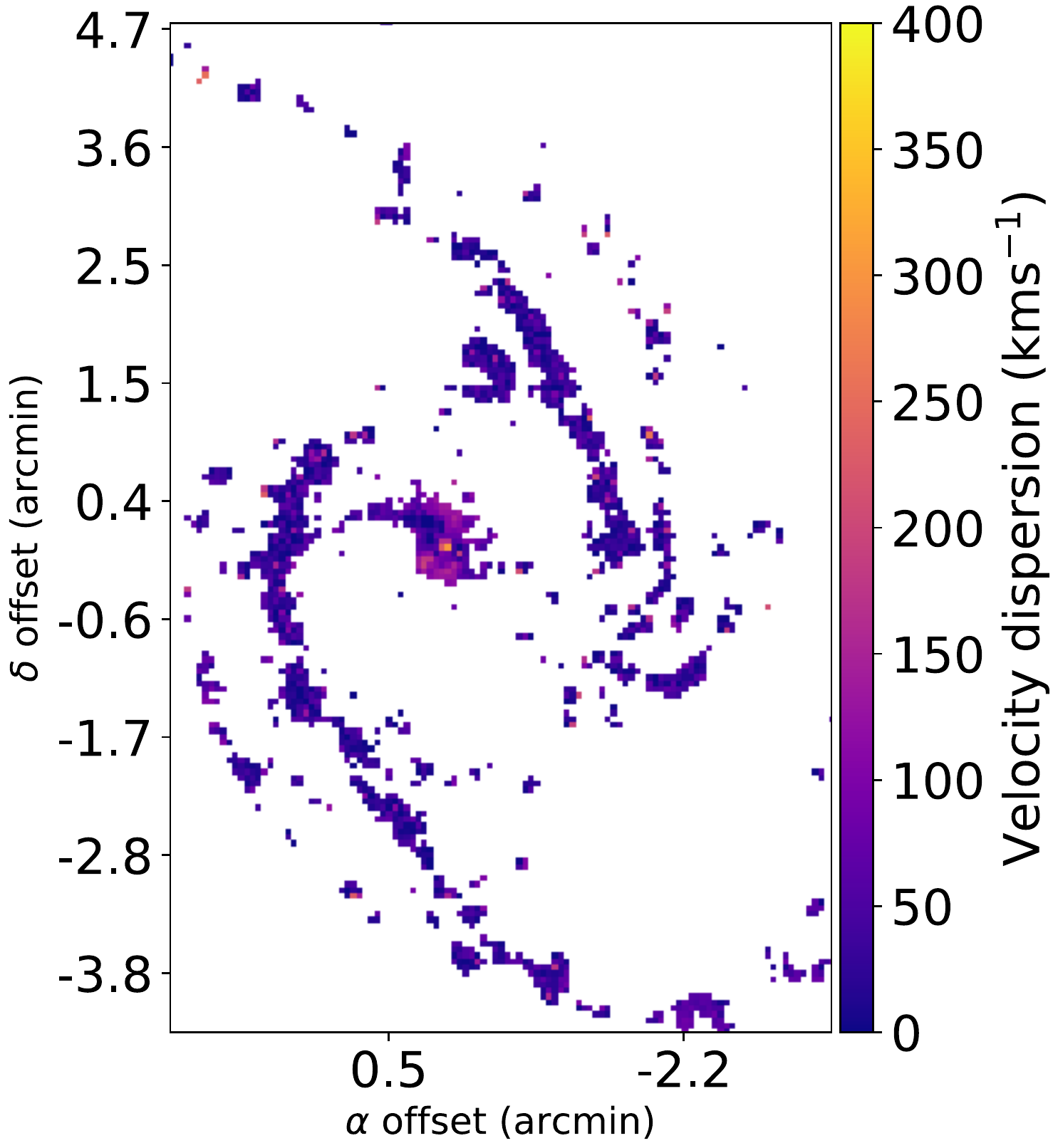}
\caption{330 pc/pix}
\label{fig:1365vdisp330}
\end{subfigure}
\begin{subfigure}{\columnwidth}
\centering
\includegraphics[scale=0.55]{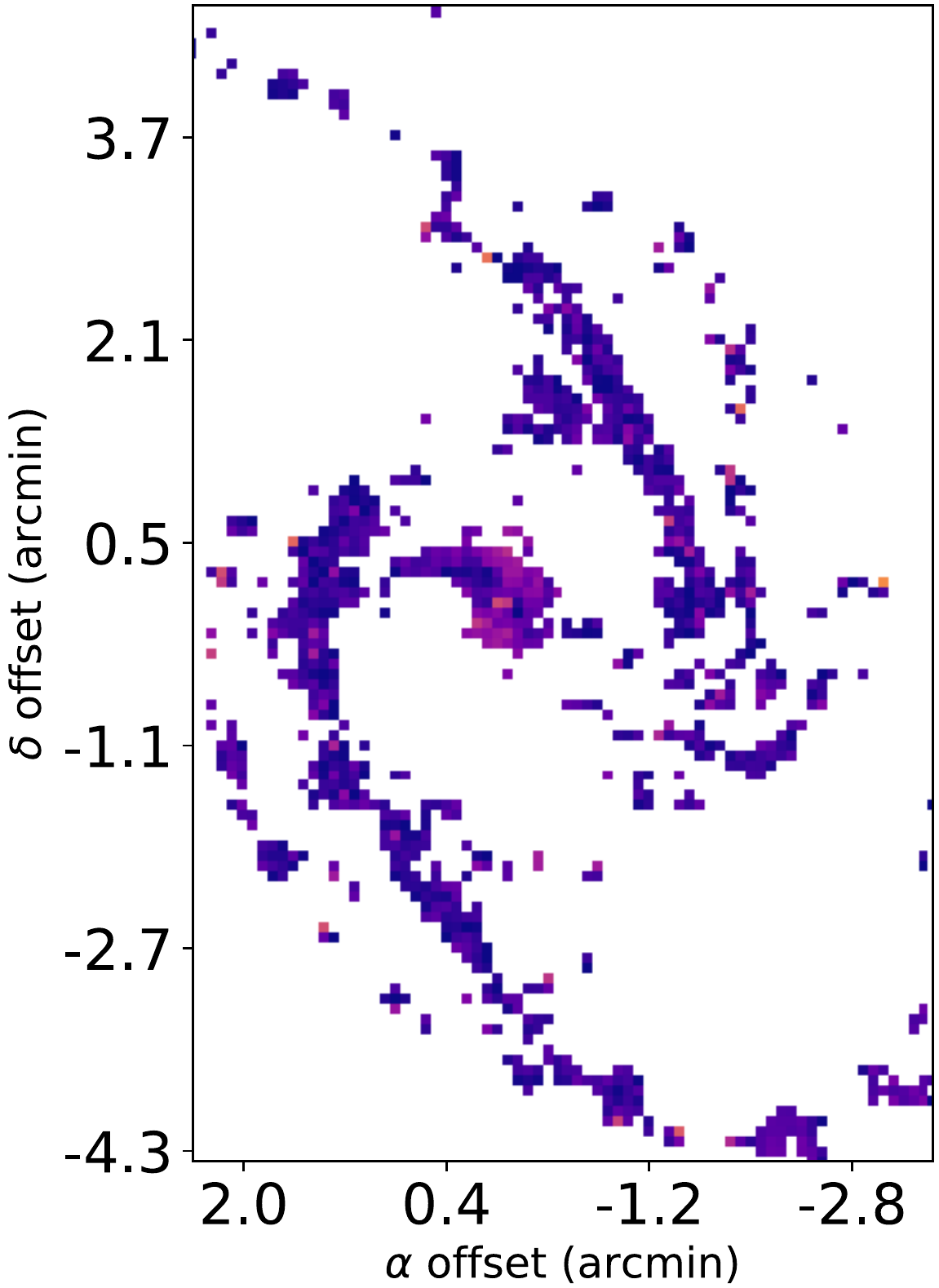}
\caption{500 pc/pix}
\label{fig:1365vdisp500}
\end{subfigure}%\hspace{\columnwidth}
\begin{subfigure}{\columnwidth}
%\centering
\includegraphics[scale=0.55]{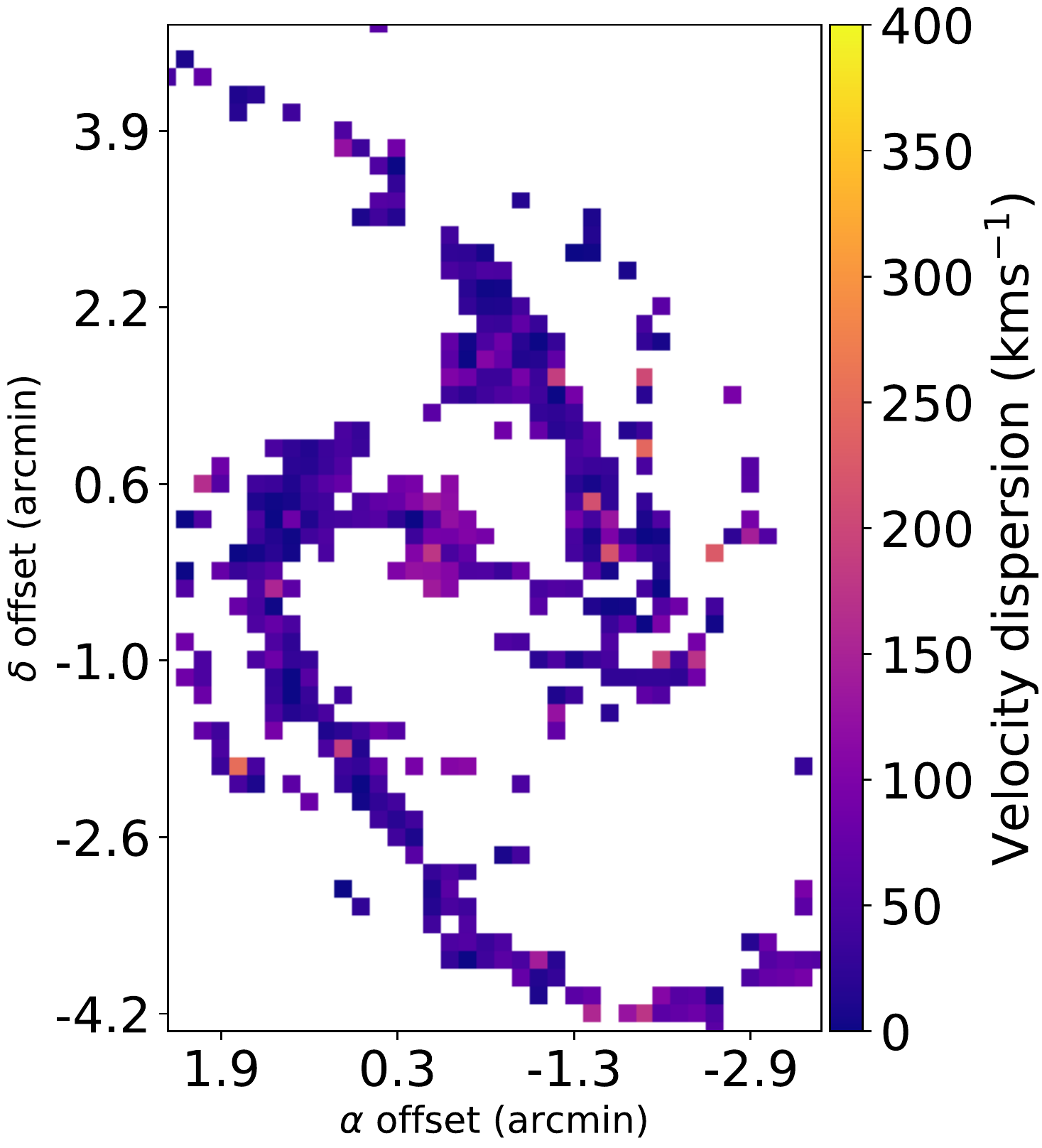}
\caption{1 kpc/pix}
\label{fig:1365vdisp1k}
\end{subfigure}
\caption{The velocity dispersion maps for NGC 1365, after rebinning to lower resolutions.}
\label{fig:1365vdisp_map}
\end{figure*}

\begin{figure*}
\centering
\begin{subfigure}{\columnwidth}
\centering
\includegraphics[width=\columnwidth]{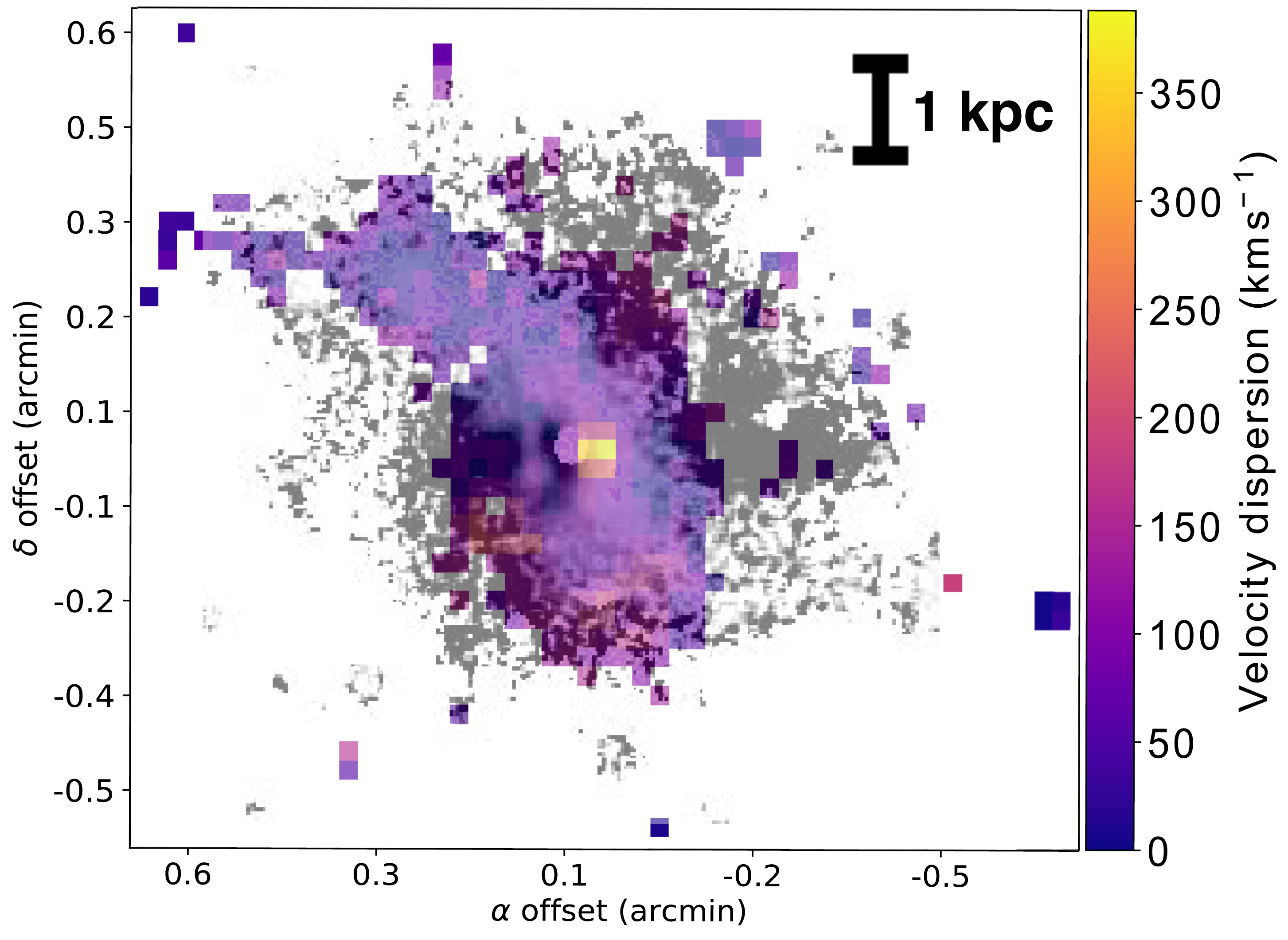}
\caption{[O \textsc{iii}]/H$\beta$}
\label{fig:1365vdispnormal}
\end{subfigure}%\hspace{\columnwidth}
\begin{subfigure}{\columnwidth}
%\centering
\includegraphics[scale=0.30]{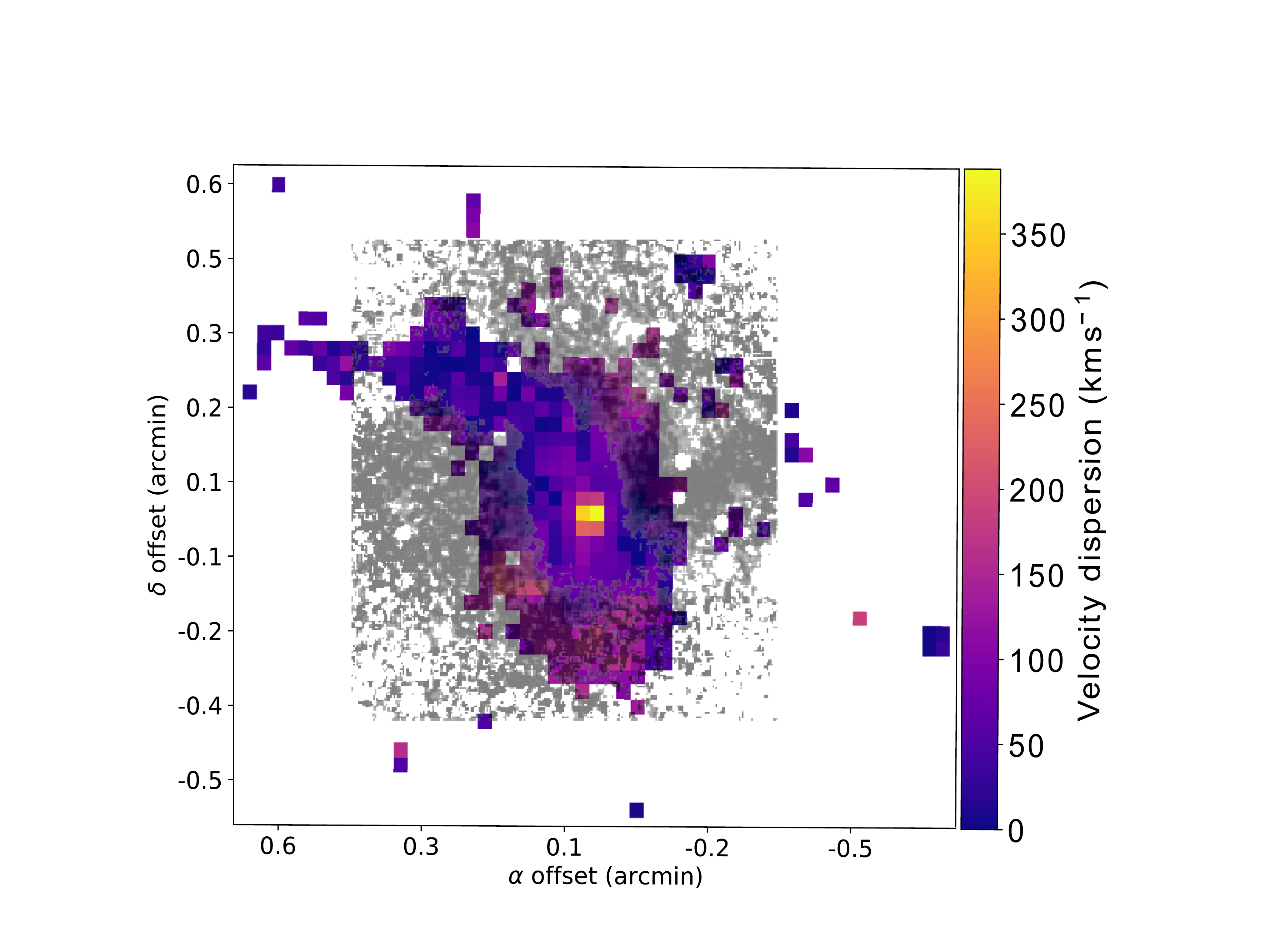}
\caption{[N \textsc{ii}]/H$\alpha$}
\label{fig:1365vdisp330}
\end{subfigure}
\caption{Outflows in [O \textsc{iii}]/H$\beta$ (a) and [N \textsc{ii}]/H$\alpha$ (b) seen in the nucleus of NGC 1365 from \citet{Veilleux2003}, coinciding with regions of high velocity dispersion. The [O \textsc{iii}]/H$\beta$ values in (a) range from 0 at the lightest to 3 at the darkest. The [N \textsc{ii}]/H$\alpha$ values in (b) range from 0 at the lightest to 1.5 at the darkest.}
\label{fig:Veilleux2003outflow}
\end{figure*}

\begin{figure*}
\centering
\begin{subfigure}{\columnwidth}
\centering
\includegraphics[scale=0.32]{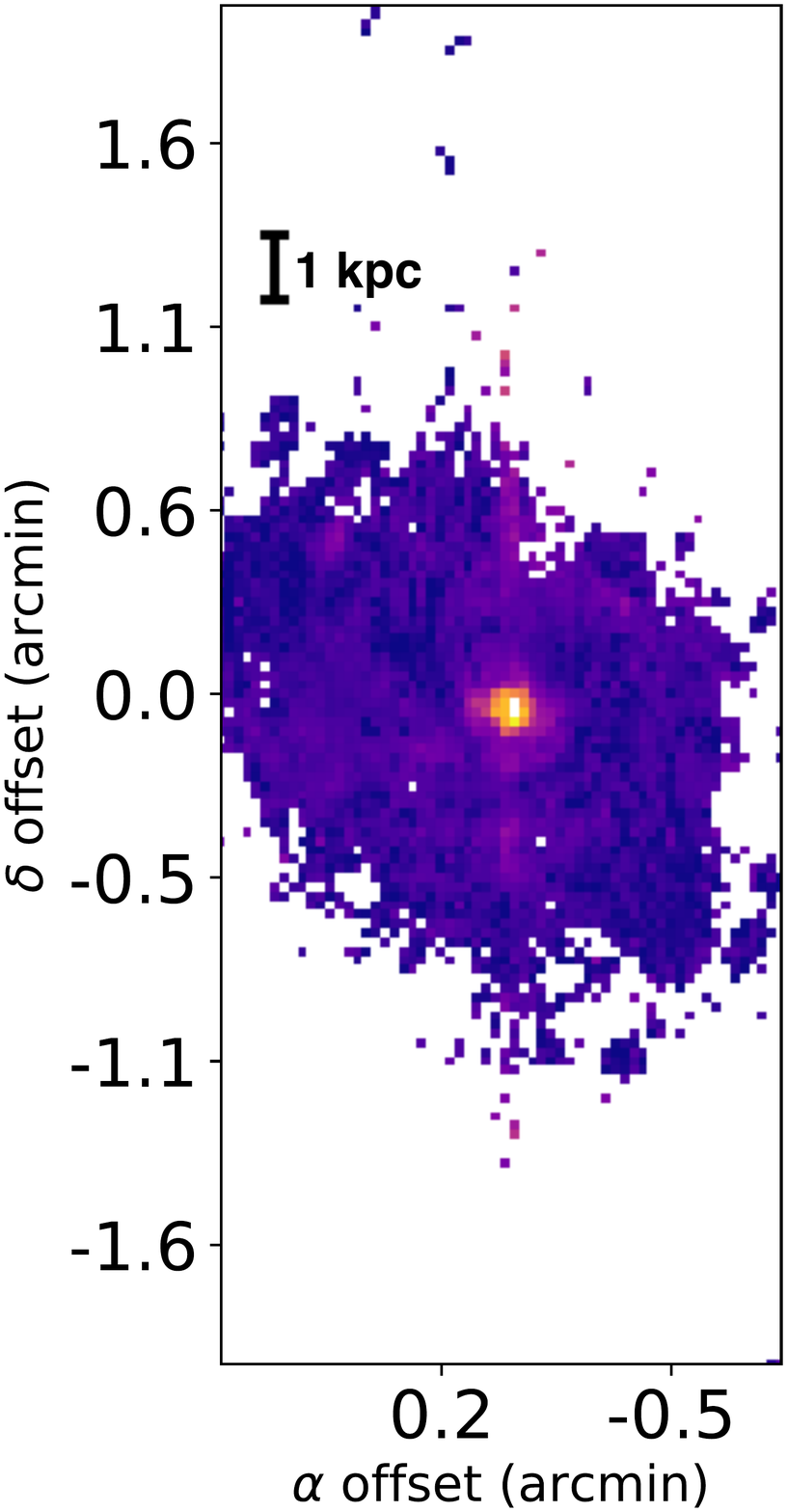}
\caption{Native resolution (121 pc/pix)}
\label{fig:1068vdispnormal}
\end{subfigure}%\hspace{\columnwidth}
\begin{subfigure}{\columnwidth}
%\centering
\includegraphics[scale=0.6]{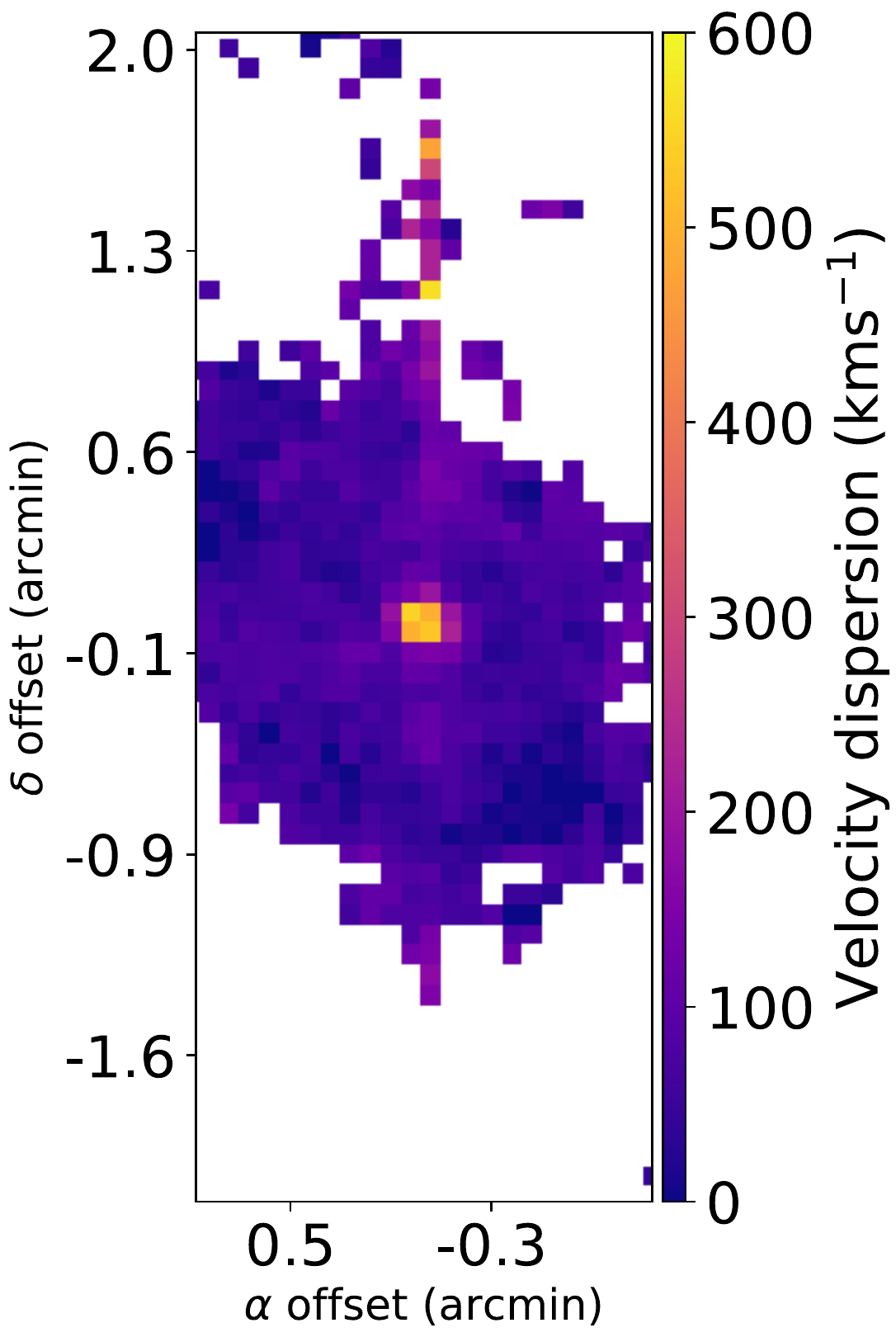}
\caption{330 pc/pix}
\label{fig:1068vdisp330}
\end{subfigure}
\begin{subfigure}{\columnwidth}
\centering
\includegraphics[scale=0.61]{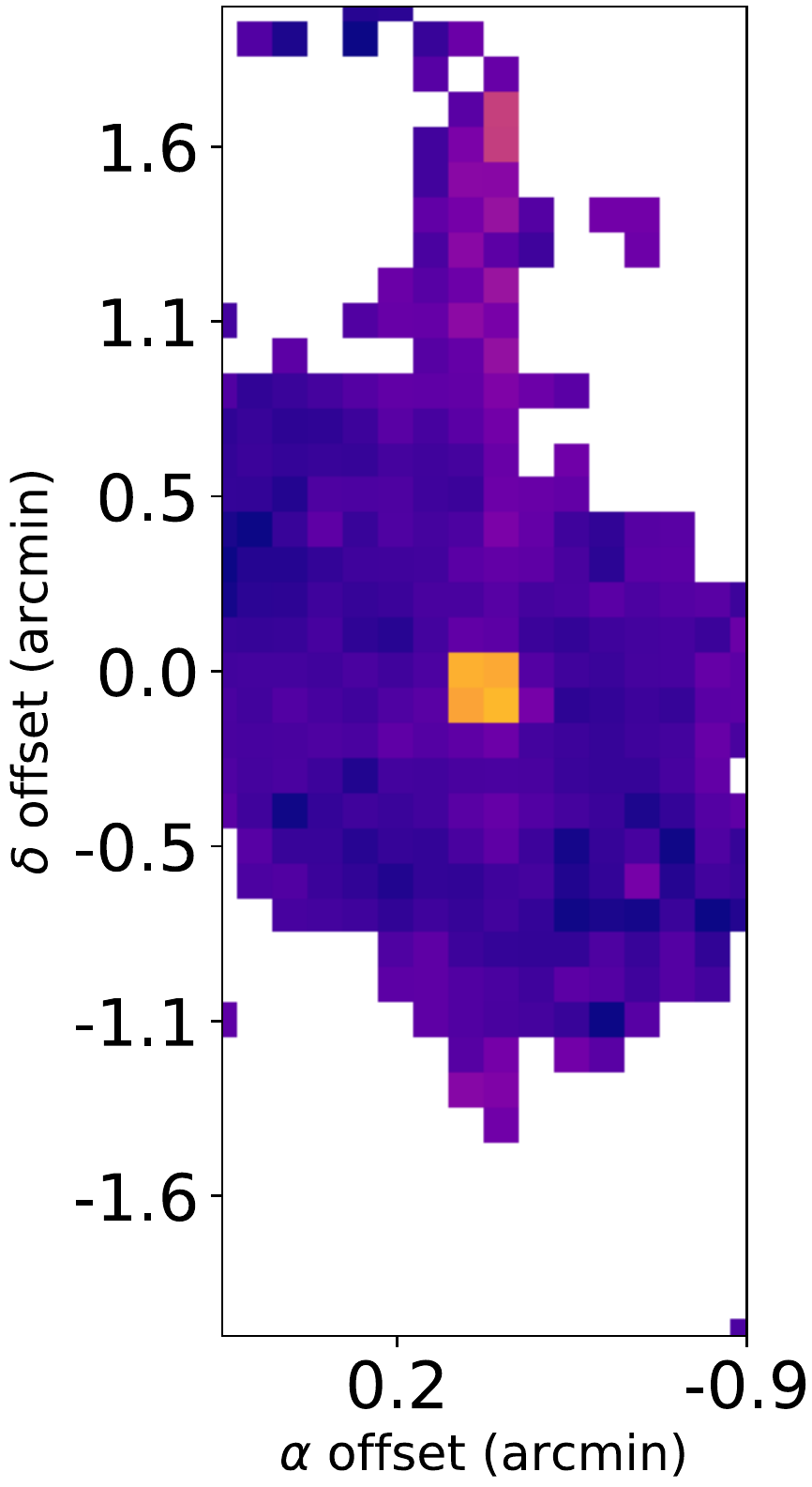}
\caption{500 pc/pix}
\label{fig:1068vdisp500}
\end{subfigure}%\hspace{\columnwidth}
\begin{subfigure}{\columnwidth}
%\centering
\includegraphics[scale=0.6]{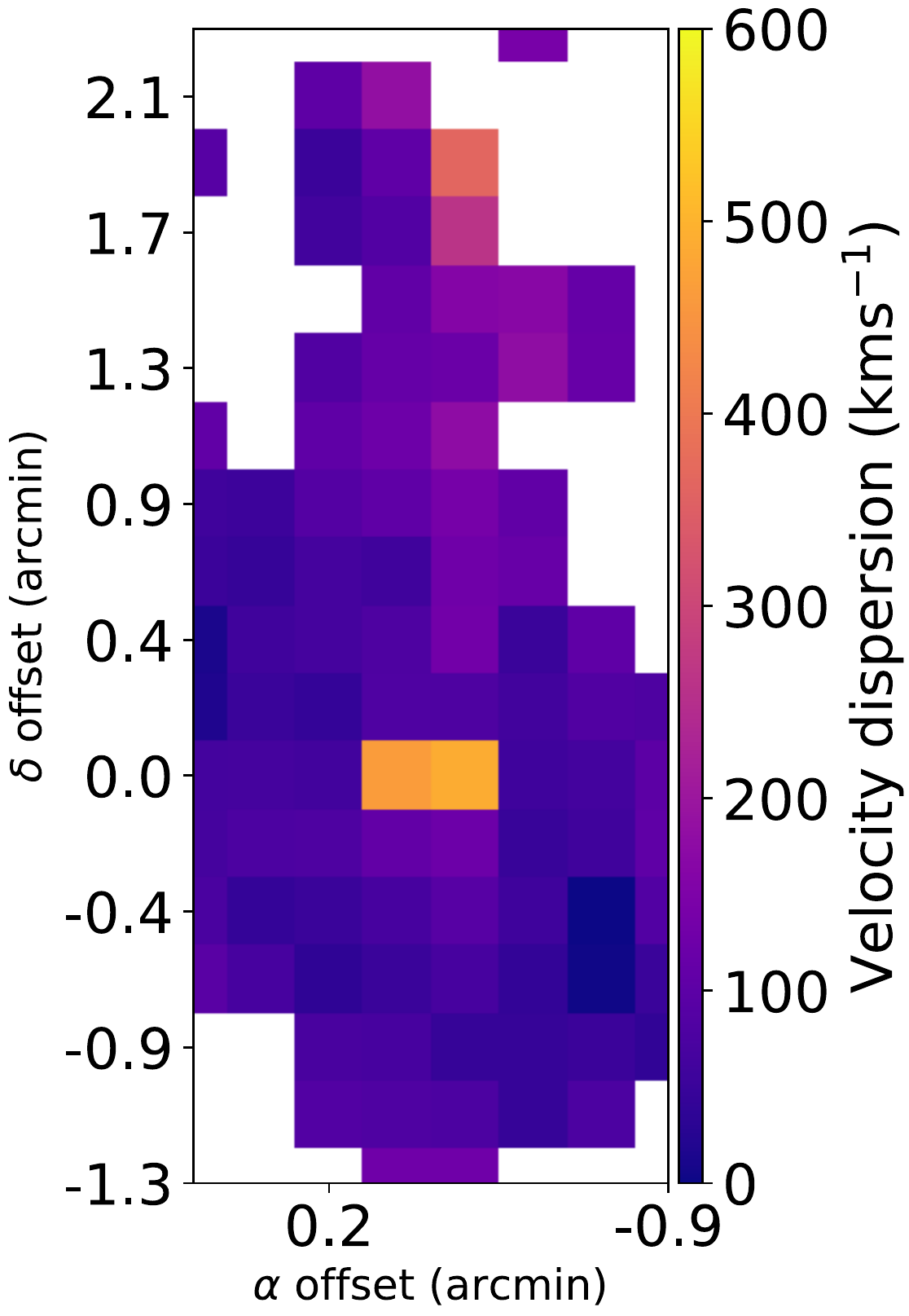}
\caption{1 kpc/pix}
\label{fig:1068vdisp1k}
\end{subfigure}
\caption{The velocity dispersion maps for NGC 1068, after rebinning to lower resolutions.}
\label{fig:1068vdisp_map}
\end{figure*}

NGC 1068 also shows evidence of shocks. Seen in the velocity dispersion maps for NGC 1068 in Figure~\ref{fig:1068vdisp_map}, regions of high velocity dispersion can be seen away from the central AGN, notably along the biconical outflowing material. An accreting supermassive black hole will produce thermal x-rays which provide a radiation pressure on nearby gas, causing expulsion of the gas in the form of a wind. The biconical outflow, believed to be the result of AGN activity \citep[and our Figure~\ref{fig:1068}]{Pogge1988,Cecil2002,Dopita2002b} is radiatively accelerated to high velocities \citep[${\sim} 3000$ kms$^{-1}$ in the case of NGC 1068;][]{Cecil2002} by the AGN wind. Once the wind enters the ISM, the high-velocity material interacts with the ISM material, which is travelling at a velocity on the order of the sound speed in the medium \citep[${\sim}$ a few tens of kms$^{-1}$;][]{ADU}. The high-velocity outflowing material, travelling at a velocity higher than the sound speed in the medium, compresses the ISM material upon interaction, causing shocks through the ISM \citep{ZK2012}. Hence, it is expected that regions of high velocity dispersion which correspond to shocks appear within the biconical outflowing material found in NGC 1068. The number and relative amount of the high velocity dispersion regions along the biconical outflow continues to increase as the resolution decreases. The spaxels in the area concerning shocks on the BPT diagrams for NGC 1068 shown in Figure~\ref{fig:1068spatres_frac} are assigned large radii ($r > 5$ kpc), constraining these spaxels to be outside the disk of NGC 1068.

\begin{table*}
\centering
\begin{tabular}{| c | c | c | c | c |}
\toprule
\multicolumn{5}{c}{\textbf{Percentage of spaxels from low surface brightness features}} \\
\midrule
\textbf{Resolution} & \textbf{Native} & \textbf{330 pc/pix} & \textbf{500 pc/pix} & \textbf{1 kpc/pix} \\
\midrule
\textbf{NGC 1365} & 0.03\%  & 0.61\%  & 0.77\%  & 0.20\%  \\
\textbf{NGC 1068} & 1.29\%  & 2.04\%  & 0.66\%  & 2.11\% \\
\midrule
\multicolumn{5}{c}{\textbf{Radii of equal starburst-AGN domination}} \\
\midrule
\textbf{Resolution} & \textbf{Native} & \textbf{330 pc/pix} & \textbf{500 pc/pix} & \textbf{1 kpc/pix} \\
\midrule
\textbf{NGC 1365} & 1.35 $\pm$ 0.36 kpc  &  2.49 $\pm$ 1.12 kpc & 2.79 $\pm$ 0.92 kpc  & 4.26 $\pm$ 2.08 kpc  \\
\textbf{NGC 1068} & 2.26 $\pm$ 1.27 kpc  & 3.55 $\pm$ 1.89 kpc  & 5.00 $\pm$ 1.81 kpc  & 6.97 $\pm$ 2.98 kpc \\
\bottomrule
\end{tabular}
\caption{(i) The percentage of spaxels showing emission from low surface brightness features. Spaxels showing emission from low surface brightness features are defined as being below the nuclear column. (ii) The radius of equal starburst and AGN domination (AGN fraction = 50\%) for varying resolutions for both NGC 1365 and NGC 1068. Values are calculated as averages of distances for spaxels with AGN fractions between 45\% and 55\%. The errors on these values represent the standard deviation of these sets of distances.}
\label{tab:more}
\end{table*}

\begin{table*}
\centering
\begin{tabular}{| c | c | c | c | c |}
%\begin{center}
\toprule
\multicolumn{5}{c}{\textbf{NGC 1365}} \\
\midrule
\textbf{Resolution} & \textbf{Native (169 pc/pix)} & \textbf{330 pc/pix} & \textbf{500 pc/pix} & \textbf{1 kpc/pix} \\
\midrule
\textbf{H$\alpha$} & 4.85 $\pm$ 0.07\% & 5.43 $\pm$ 0.06\% & 5.75 $\pm$ 0.06\% & 6.34 $\pm$ 0.05\% \\
\textbf{H$\beta$} & 4.41 $\pm$ 0.27\% & 4.86 $\pm$ 0.26\% & 5.09 $\pm$ 0.24\% & 5.54 $\pm$ 0.20\% \\
\textbf{[O \textsc{ii}]$\lambda \lambda$3726,3729} & 4.78 $\pm$ 1.76\% & 5.23 $\pm$ 1.13\% & 5.60 $\pm$ 1.02\% & 6.35 $\pm$ 0.81\% \\
\textbf{[O \textsc{iii}]$\lambda$5007} & 10.06 $\pm$ 0.71\% & 11.96 $\pm$ 0.72\% & 12.35 $\pm$ 0.66\% & 13.25 $\pm$ 0.52\% \\
\textbf{[S \textsc{ii}]$\lambda \lambda$6716,6731} & 5.65 $\pm$ 0.30\% & 6.47 $\pm$ 0.28\% & 6.97 $\pm$ 0.26\% &  7.93 $\pm$ 0.21\%\\
\textbf{[N \textsc{ii}]$\lambda$6584} & 6.95 $\pm$ 0.20\% & 8.49 $\pm$ 0.20\% & 9.17 $\pm$ 0.18\% & 10.32 $\pm$ 0.15\% \\
\midrule
\multicolumn{5}{c}{\textbf{NGC 1068}} \\
\midrule
\textbf{Resolution} & \textbf{Native (121 pc/pix)} & \textbf{330 pc/pix} & \textbf{500 pc/pix} & \textbf{1 kpc/pix} \\
\midrule
\textbf{H$\alpha$} & 25.23 $\pm$ 0.13\% & 39.40 $\pm$ 0.10\% & 39.07 $\pm$ 0.08\% & 39.82 $\pm$ 0.05\% \\
\textbf{H$\beta$} & 26.40 $\pm$ 0.81\% & 38.91 $\pm$ 0.51\% & 39.26 $\pm$ 0.38\% & 42.31 $\pm$ 0.25\% \\
\textbf{[O \textsc{ii}]$\lambda \lambda$3726,3729} & 32.23 $\pm$ 6.81\% & 25.27 $\pm$ 2.50\% & 31.56 $\pm$ 2.18\% &  42.68 $\pm$ 1.54\%\\
\textbf{[O \textsc{iii}]$\lambda$5007} & 42.69 $\pm$ 0.28\% & 68.19 $\pm$ 0.16\% & 68.75 $\pm$ 0.12\% &  68.88 $\pm$ 0.07\%\\
\textbf{[S \textsc{ii}]$\lambda \lambda$6716,6731} & 33.38 $\pm$ 0.79\% & 43.32 $\pm$ 0.43\% & 44.15 $\pm$ 0.38\% &  46.16 $\pm$ 0.21\%\\
\textbf{[N \textsc{ii}]$\lambda$6584} & 41.52 $\pm$ 0.31\% & 64.58 $\pm$ 0.15\% & 66.02 $\pm$ 0.11\% &  67.62 $\pm$ 0.07\%\\
\bottomrule
%\end{center}
\end{tabular}
\caption{AGN fractions for several strong emission lines at varying resolutions for both NGC 1365 and NGC 1068 within the TYPHOON field-of-view. Percentage indicates fraction of luminosity attributable to AGN activity for each line.}
\label{tab:agnfracs}
\end{table*}

%\begin{table*}
%\centering
%\begin{tabular}{| c | c | c | c | c |}
%\toprule
%\textbf{Resolution} & \textbf{Native} & \textbf{330 pc/pix} & \textbf{500 pc/pix} & \textbf{1 kpc/pix} \\
%\midrule
%\textbf{NGC 1365} & 1.350 $\pm$ 0.362 kpc  &  2.583 $\pm$ 1.146 kpc & 2.626 $\pm$ 1.019 kpc  & 4.262 $\pm$ 2.078 kpc  \\
%\textbf{NGC 1068} & 2.246 $\pm$ 1.277 kpc  & 3.551 $\pm$ 1.892 kpc  & 5.169 $\pm$ 1.820 kpc  & 6.392 $\pm$ 2.906 kpc \\
%\bottomrule
%\end{tabular}
%\caption{Table showing the radius of equal starburst and AGN domination (starburst-AGN fraction = 50\%) for varying resolutions for both NGC 1365 and NGC 1068. Values are calculated as averages of distances for spaxels with starburst-AGN fractions between 45\% and 55\%. The errors on these values represent the standard deviation of these sets of distances.}
%\label{tab:eqdist}
%\end{table*}

%As well as toward the centre, high velocity dispersion regions also seem to occur with significant low dispersion regions towards the outskirts of the galaxy -- most notably at the turning points of the spiral arms. These high velocity dispersion regions in NGC 1365 seem to suggest that the low surface brightness features seen in Figs.~\ref{fig:1365spatres_frac} and \ref{fig:1365spatres_map} are shocks which have been made evident through the rebinning process.

%\begin{figure*}
%\centering
%\begin{subfigure}{\columnwidth}
%\includegraphics[width=\columnwidth]{1365_SII_agnmap_normal.pdf}
%\caption{Native resolution (169 pc/pix)}
%%\label{fig:1365mapnormal}
%\end{subfigure}%\hspace{\columnwidth}
%\begin{subfigure}{\columnwidth}
%\includegraphics[width=\columnwidth]{1365_SII_agnmap_330pc.pdf}
%\caption{330 pc/pix}
%%\label{fig:1365map330}
%\end{subfigure}
%\begin{subfigure}{\columnwidth}
%\includegraphics[width=\columnwidth]{1365_SII_agnmap_500pc.pdf}
%\caption{500 pc/pix}
%%\label{fig:1365map500}
%\end{subfigure}%\hspace{\columnwidth}
%\begin{subfigure}{\columnwidth}
%\includegraphics[width=\columnwidth]{1365_SII_agnmap_1kpc.pdf}
%\caption{1 kpc/pix}
%%\label{fig:1365map1k}
%\end{subfigure}
%\caption{The [O \textsc{iii}]/H$\beta$ vs [S \textsc{ii}]/H$\alpha$ starburst-AGN fraction map for NGC 1365, after rebinning to lower spatial resolutions.}
%\label{fig:1365SIIspatres_map}
%\end{figure*}

% for NGC 1068, lower resolutions are using Mark's cube. Native res is I-Ting's cube

%\begin{figure*}
%\centering
%\begin{subfigure}{\columnwidth}
%\centering
%\includegraphics[width=\columnwidth]{1365_vdisphist_normal.eps}
%\caption{Native resolution (169 pc/pix)}
%\label{fig:1365vdisphistnormal}
%\end{subfigure}%\hspace{\columnwidth}
%\begin{subfigure}{\columnwidth}
%%\centering
%\includegraphics[width=\columnwidth]{1365_vdisphist_330pc.eps}
%\caption{330 pc/pix}
%\label{fig:1365vdisphist330}
%\end{subfigure}
%\begin{subfigure}{\columnwidth}
%\centering
%\includegraphics[width=\columnwidth]{1365_vdisphist_500pc.eps}
%\caption{500 pc/pix}
%\label{fig:1365vdisphist500}
%\end{subfigure}%\hspace{\columnwidth}
%\begin{subfigure}{\columnwidth}
%%\centering
%\includegraphics[width=\columnwidth]{1365_vdisphist_1kpc.eps}
%\caption{1 kpc/pix}
%\label{fig:1365vdisphist1k}
%\end{subfigure}
%\caption{The velocity dispersion distribution for NGC 1365, after rebinning to lower resolutions.}
%\label{fig:1365vdisphists}
%\end{figure*}
%

\section{Conclusions}

Using high spatial resolution data from the TYPHOON/PrISM survey, we analyse two AGN host galaxies NGC 1365 and NGC 1068. We extend the work of \citet{Davies2014a,Davies2014b} by using photoionisation grids to introduce a new method to calculate the AGN fraction of a galaxy from the BPT diagram. We also perform the same AGN fraction calculations on datacubes of NGC 1365 and NGC 1068 which have been rebinned to lower spatial resolutions. We report the following:

(i) IFU datacubes with resolutions as high as those used by TYPHOON (169 pc/pix for NGC 1365; 121 pc/pix for NGC 1068) show both star formation and AGN activity occurring in the nucleus of an AGN host galaxy.

(ii) The AGN in NGC 1365 is relatively weak, with only ${\sim} 10$\% of the total [O \textsc{iii}]$\lambda$5007 luminosity and ${\sim} 5$\% of the total H$\alpha$ luminosity from NGC 1365 being attributable to AGN activity within the TYPHOON field-of-view. This shows that the vast majority of the emission from NGC 1365 is from star formation. The AGN in NGC 1068 is stronger, accounting for ${\sim} 42$\% of the total [O \textsc{iii}] luminosity and ${\sim} 24$\% of the total H$\alpha$ luminosity within the TYPHOON field-of-view. This shows that whilst AGN activity in NGC 1068 is considerable, star formation does have alarge influence in the total emission from NGC 1068. The fraction of H$\alpha$ emission from AGN activity also has consequences for the calculation of SFRs in both of these galaxies. An overestimate of ${\sim} 5$\% and an overestimate of ${\sim} 24$\% on the SFR of NGC 1365 and NGC 1068 respectively will occur if H$\alpha$ for the calculation.

(iii) Low surface brightness features such as shocks in galaxies and certain outflows (such as that in NGC 1365) only appear at lower resolutions. Features with a low S/N ratio gain a S/N increase when binned with spaxels containing a higher S/N ratio, or when observed with a higher S/N feature in the same spaxel. Increases in the number of spaxels attributable to low surface brightness features up to factors of ${\sim} 12.5$ and ${\sim} 1.6$ are noticed with coarser resolution in NGC 1365 and NGC 1068 respectively. Hence, certain features within a galaxy are lost when the resolution of the observations increases.

(iv) The contribution of AGN to total galaxy emission increases with coarser resolution. An average 18.05\% increase from the emission from AGN activity in NGC 1068 in is seen in several strong lines as the resolution decreases from their respective native resolutions to a resolution of 1 kpc/pix. As the resolution lowers, the AGN fraction in individual spaxels and also individual emission lines increases, showing a larger effect from the AGN. This is also seen by the radius of equal starburst-AGN domination increasing with coarser resolution, with a roughly equal factor of 3 increase in the radius of equal domination for both NGC 1365 and NGC 1068 as the resolution decreases from their respective native resolutions to a resolution of 1 kpc/pix.

%(v) The maximum velocity dispersion measured in galaxies decreases as the resolution decreases. This is quantified by a ${\sim} 130$ kms$^{-1}$ and ${\sim} 100$ kms$^{-1}$ decrease in the maximum velocity dispersions measured in NGC 1365 and NGC 1068 respectively as the resolution decreases from their respective native resolutions to a resolution of 1 kpc/pix.

For optimal results when calculating the contribution of emission from AGN activity, the highest possible resolution data should be sought. We show that with finer resolution, the relative contribution from AGN in the total emission within the galaxy continuously decreases. Furthermore, the zone of influence of the AGN is also shown to be continually overestimated with coarser resolution. Hence, only the highest possible resolution will provide the most accurate measure of the AGN activity within a galaxy.

We intend to extend this project in the future. Emission in galaxies may arise from sources other than AGN or star formation, such as shocks. Hence, future work will include the separation of emission from shocks in addition to AGN and star formation in IFU data.

\section*{Acknowledgements}

Parts of this research were conducted by the Australian Research Council Centre of Excellence for All Sky Astrophysics in 3 Dimensions (ASTRO 3D), through project number CE170100013. The authors would also like to thank referee Santiago Garc\'ia-Burillo, who provided valuable input and helped improve the condition of the publication. 
%%%%%%%%%%%%%%%%%%%%%%%%%%%%%%%%%%%%%%%%%%%%%%%%%%



\bibliographystyle{mnras}
\bibliography{typhoon1} % if your bibtex file is called example.bib

\appendix

\section{Power-law spectral index}
\label{sec:alpha}

\begin{figure*}
\centering
\begin{subfigure}{0.85\textwidth}
\includegraphics[width=\linewidth]{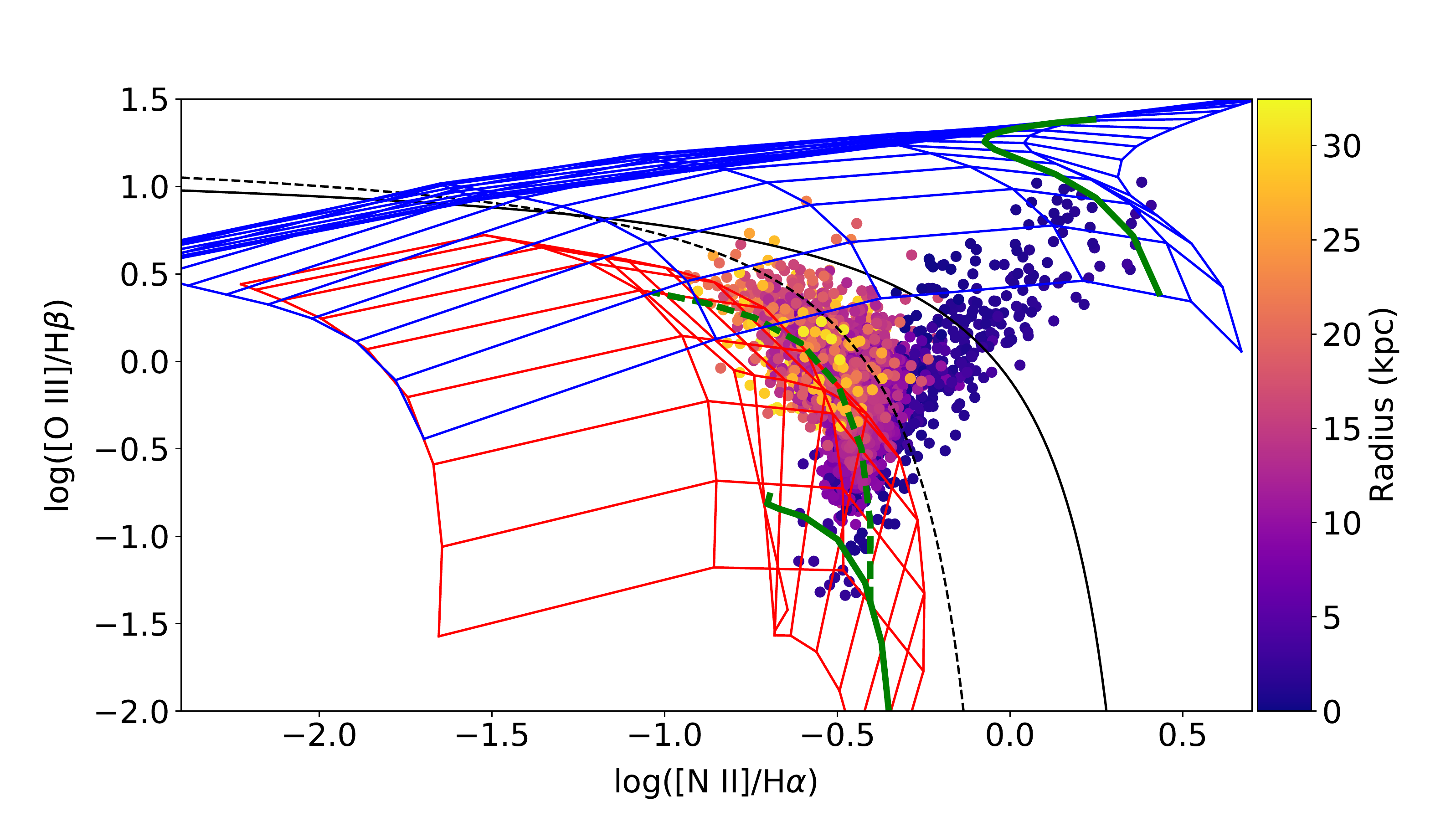}
\caption{$\alpha = -1.4$}
\label{fig:136514}
\end{subfigure}\hspace{0.2\textwidth}
\begin{subfigure}{0.85\textwidth}
\includegraphics[width=\linewidth]{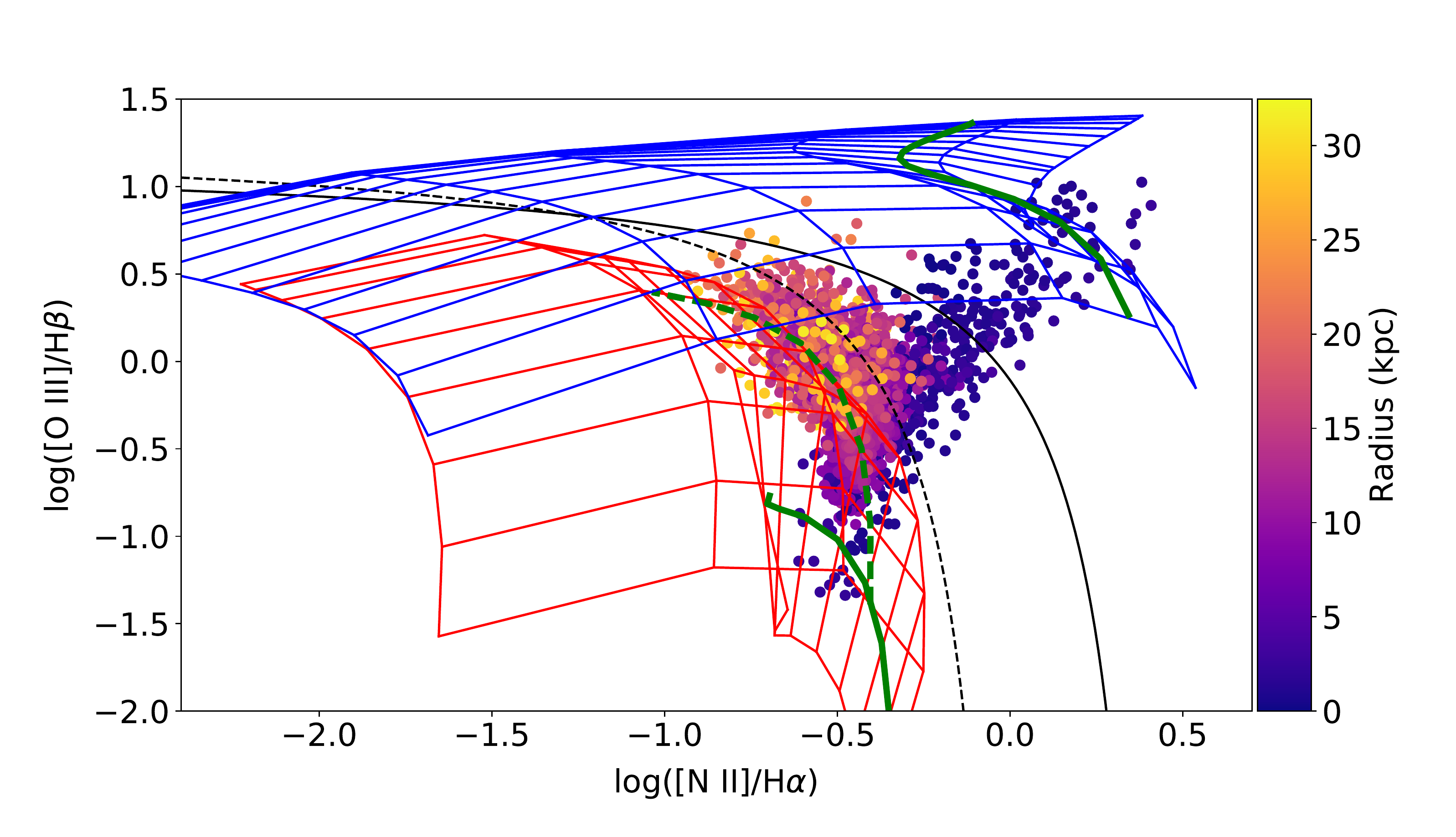}
\caption{$\alpha = -1.7$}
\label{fig:136517}
\end{subfigure}
\caption{NGC 1365, showing $n = 1000\;\mathrm{cm}^{-3}$ AGN model grids computed with power-law spectral indices of $\alpha = -1.4$ and -1.7.}
\label{fig:1365_alpha}
\end{figure*}

Here we show BPT diagrams of NGC 1365 and NGC 1068, with AGN model grids featuring power-law spectral indices of $\alpha = -1.4$ and -1.7. In both $n = 1000\;\mathrm{cm}^{-3}$ and $n = 10000\;\mathrm{cm}^{-3}$ cases for NGC 1365 and NGC 1068 respectively, lowering the spectral index $\alpha$ lowers the [O \textsc{iii}]/H$\beta$ and [N \textsc{ii}]/H$\alpha$ fluxes produced by the models. Several caveats result from using a lower spectral index with these models:

\begin{itemize}
\item The line of constant metallicity representing the central region of the galaxy is moved further down the mixing sequence towards the starburst region of the BPT diagram, and in the case of NGC 1068, does not provide a good fit to the uppermost spaxels in the mixing sequence. The result of using such a model to define the 100\% AGN region would be many spaxels saturated at 100\% AGN. Further, the error in the photoionisation models of ${\sim} 1.5$ dex in emission line ratio (D'Agostino et al. in prep) does not account for the amount of spaxels above the central metallicity line.

\item While a case may be made for the use of $\alpha = -1.4$ for NGC 1365 seen in Figure~\ref{fig:136514}, use of such a model (and models which decrease $\alpha$ even further) results in several off-grid spaxels which are not explained or defined by the model. At $\alpha = -1.7$ and below, increasing the metallicity of the $n = 1000\;\mathrm{cm}^{-3}$ models fails to describe the off-grid spaxels, as the models begin to turn over at metallicities of $Z = 0.060$ and beyond, seen in Figure~\ref{fig:136517}. A turn in the shape of the model is not yet evident when using $\alpha = -1.4$ for the $n = 1000\;\mathrm{cm}^{-3}$ model seen in Figure~\ref{fig:136514}, hence increasing the maximum metallicity shown in the model beyond $Z = 0.060$ may describe the off-grid spaxels. However, it should be noted that the maximum metallicity calculated in each galaxy ($1.42 Z_\odot$ and $1.41 Z_\odot$ for NGC 1365 and NGC 1068 respectively, where $Z_\odot \equiv Z = 0.020$) is far below the maximum already included in the models of $Z = 0.060$. Hence, using higher metallicities to explain the off-grid spaxels is inconsistent with our metallicity calculations, and such a high metallicity is possibly considered unphysical.
\end{itemize}

Hence, a power-law spectral index of $\alpha = -1.2$ is the only value considered which is consistent with the data and our metallicity calculations.

\begin{figure*}
\centering
\begin{subfigure}{0.85\textwidth}
\includegraphics[width=\linewidth]{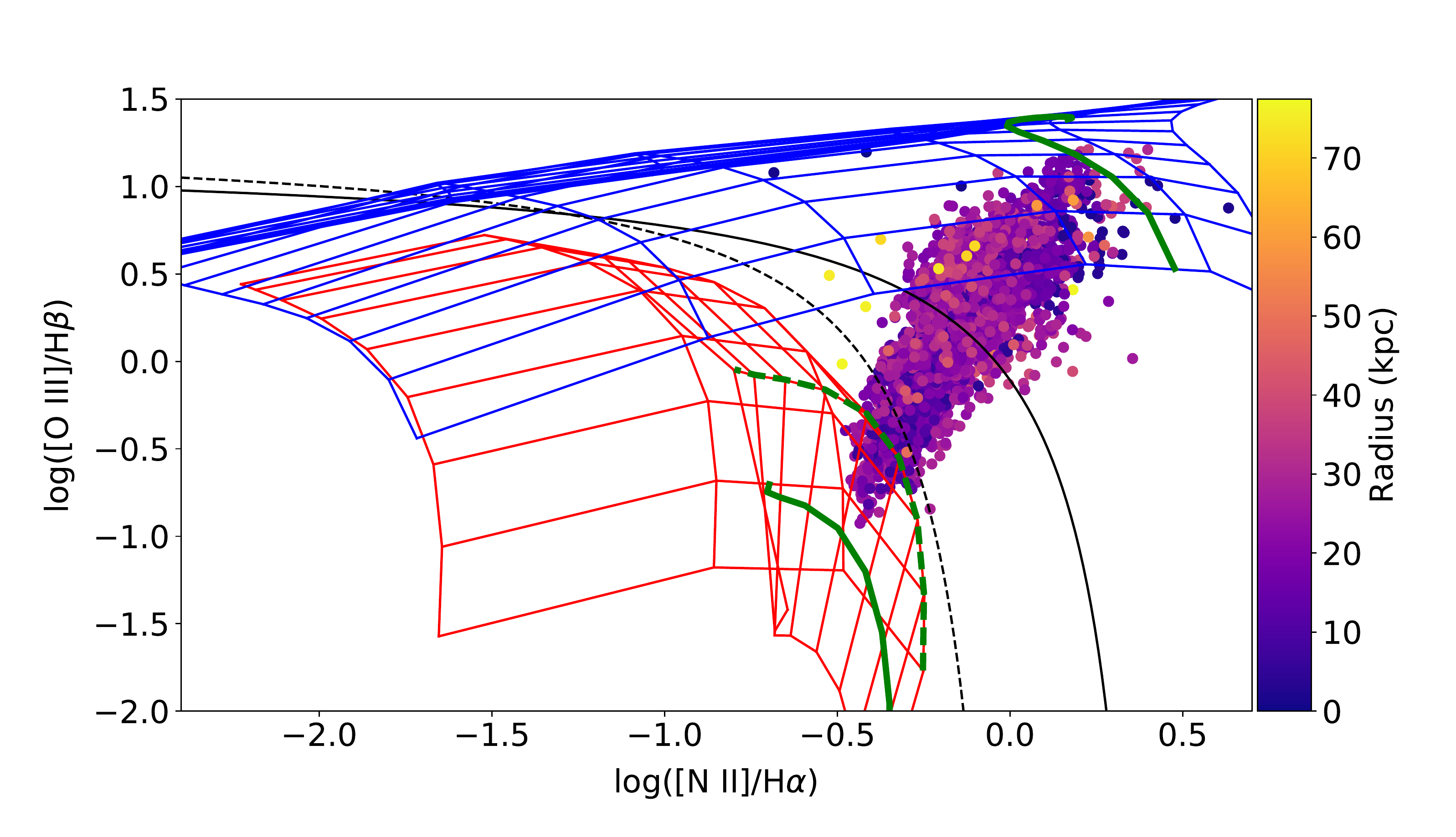}
\caption{$\alpha = -1.4$}
\label{fig:106814}
\end{subfigure}\hspace{0.2\textwidth}
\begin{subfigure}{0.85\textwidth}
\includegraphics[width=\linewidth]{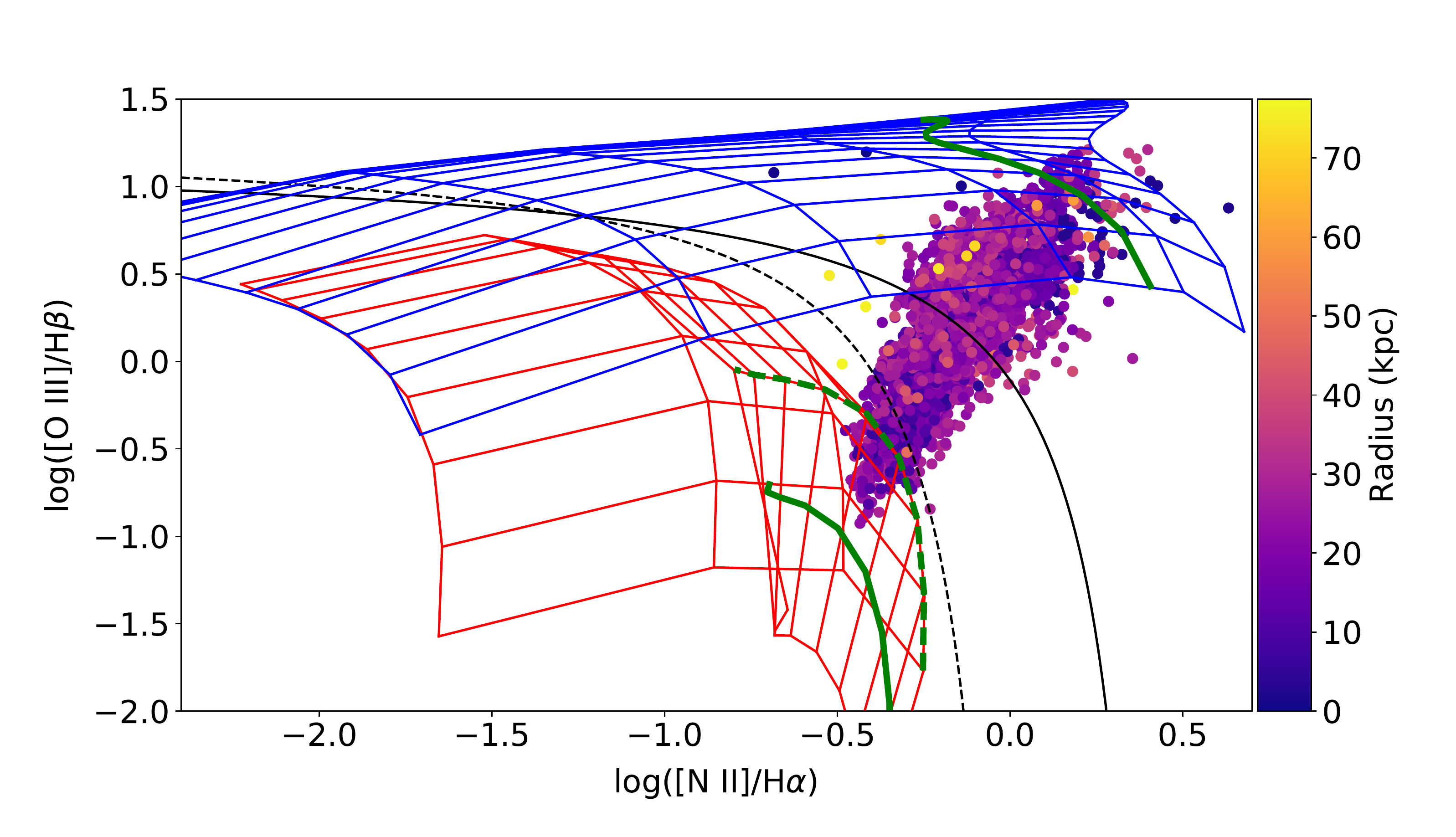}
\caption{$\alpha = -1.7$}
\label{fig:106817}
\end{subfigure}
\caption{NGC 1068, showing $n = 10000\;\mathrm{cm}^{-3}$ AGN model grids computed with power-law spectral indices of $\alpha = -1.4$ and -1.7.}
\label{fig:1068_alpha}
\end{figure*}

\bsp	% typesetting comment
\label{lastpage}
\end{document}